\documentclass[twocolumn]{aastex6}

\usepackage{natbib}

\begin{document}

\title{The FMOS-COSMOS survey of star-forming galaxies at $z\sim 1.6$. IV: Excitation state and chemical enrichment of the interstellar medium}
\author{
D.~Kashino\altaffilmark{1}, 
J.~D.~Silverman\altaffilmark{2},
D.~Sanders\altaffilmark{3},
J.~S.~Kartaltepe\altaffilmark{4},
E.~Daddi\altaffilmark{5},
A.~Renzini\altaffilmark{6},
F.~Valentino\altaffilmark{5,7},
G.~Rodighiero\altaffilmark{8},
S.~Juneau\altaffilmark{5},
L.~J.~Kewley\altaffilmark{9},
H.~J.~Zahid\altaffilmark{10},
N.~Arimoto\altaffilmark{11,12},
T.~Nagao\altaffilmark{13}
J.~Chu\altaffilmark{3}, 
N.~Sugiyama\altaffilmark{14},
F.~Civano\altaffilmark{15,16},
O.~Ilbert\altaffilmark{17},
M.~Kajisawa\altaffilmark{13,18},
O.~Le F{\' e}vre\altaffilmark{17},
C.~Maier\altaffilmark{19},
D.~Masters\altaffilmark{20},
T.~Miyaji\altaffilmark{21},
M.~Onodera\altaffilmark{11,12}, 
A.~Puglisi\altaffilmark{8},
and 
Y.~Taniguchi\altaffilmark{22}}

\email{kashinod@phys.ethz.ch}
\altaffiltext{1}{
Institute for Astronomy, Department of Physics, ETH Z{\" u}rich, Wolfgang-Pauli-strasse 27, CH-8093 Z{\"u}rich, Switzerland
}
\altaffiltext{2}{
Kavli Institute for the Physics and Mathematics of the Universe (WPI), Todai Institutes for Advanced Study, the University of Tokyo, Kashiwanoha, Kashiwa, 277-8583, Japan
}
\altaffiltext{3}{
Institute for Astronomy, University of Hawaii, 2680 Woodlawn Drive, Honolulu, HI 96822, USA
}
\altaffiltext{4}{
School of Physics and Astronomy, Rochester Institute of Technology, 84 Lomb Memorial Dr., Rochester, NY 14623, USA
}
\altaffiltext{5}{
Laboratoire AIM-Paris-Saclay, CEA/DSM-CNRS-Universit{\' e} Paris Diderot, Irfu/Service d’Astrophysique,
CEA-Saclay, Service d'Astrophysique, F-91191 Gif-sur-Yvette, France
}
\altaffiltext{6}{
INAF Osservatorio Astronomico di Padova, vicolo dell'Osservatorio 5, I-35122 Padova, Italy
}
\altaffiltext{7}{
Dark Cosmology Centre, Niels Bohr Institute, University of Copenhagen, Juliane Mariesvej 30, DK-2100 Copenhagen, Denmark
}
\altaffiltext{8}{
Dipartimento di Fisica e Astronomia, Universit{\' a} di Padova, vicolo dell'Osservatorio, 2, I-35122 Padova, Italy
}
\altaffiltext{9}{
Research School of Astronomy and Astrophysics, Australian National University, Cotter Road, Weston Creek, ACT 2611, Australia
}
\altaffiltext{10}{
Smithsonian Astrophysical Observatory, Harvard-Smithsonian Center for Astrophysics, 60 Garden Street, Cambridge, MA 02138, USA
}
\altaffiltext{11}{
Subaru Telescope, National Astronomical Observatory of Japan, National Institutes of Natural Sciences (NINS), 650 North A’ohoku Place, Hilo, Hawaii 96720, USA
}
\altaffiltext{12}{
Department of Astronomical Science, SOKENDAI (The Graduate University for Advanced Studies), 2-21-1 Osawa, Mitaka, Tokyo, Japan
}
\altaffiltext{13}{
Research Center for Space and Cosmic Evolution, Ehime University, Bunkyo-cho 2-5, Matsuyama 790-8577, Japan
}
\altaffiltext{14}{
Division of Particle and Astrophysical Science, Graduate School of Science, Nagoya University, Nagoya, 464-8602, Japan
}
\altaffiltext{15}{
Yale Center for Astronomy and Astrophysics, 260 Whitney Avenue, New Haven, CT 06520, USA
}
\altaffiltext{16}{
Harvard-Smithsonian Center for Astrophysics, 60 Garden Street, Cambridge, MA 02138, USA
}
\altaffiltext{17}{
Aix-Marseille Universit{\'e}, CNRS, LAM (Laboratoire d’Astrophysique de Marseille), UMR 7326, F-13388 Marseille, France
}
\altaffiltext{18}{
Graduate School of Science and Engineering, Ehime University, Bunkyo-cho 2-5, Matsuyama, Ehime, 790-8577, Japan
}
\altaffiltext{19}{
University of Vienna, Department of Astrophysics, T{\"u}rkenschanzstrasse 17, A-1180 Vienna, Austria
}
\altaffiltext{20}{
Infrared Processing and Analysis Center, California Institute for Technology, MC 100-22, 770 South Wilson Ave, Pasadena, CA 91125, USA
}
\altaffiltext{21}{
Instituto de Astronom{\'i}a sede Ensenada, Universidad Nacional Aut{\'o}noma de M{\'e}xico, Km. 103, Carret. Tijunana-Ensenada, Ensenada, 22860, Mexico
}
\altaffiltext{22}{
The Open University of Japan 2-11 Wakaba, Mihama-ku, Chiba 261- 8586 Japan
}

\begin{abstract}

We investigate the physical conditions of ionized gas in high-$z$ star-forming galaxies using diagnostic diagrams based on the rest-frame optical emission lines.  The sample consists of 701 galaxies with an H$\alpha$ detection at \hbox{$1.4\lesssim z \lesssim 1.7$}, from the FMOS-COSMOS survey, that represent the normal star-forming population over the stellar mass range $10^{9.6} \lesssim M_\ast/M_\odot \lesssim 10^{11.6}$ with those at $M_\ast>10^{11}~M_\odot$ being well sampled.  We confirm an offset of the average location of star-forming galaxies in the BPT diagram ($\textrm{[O\,{\sc iii}]}/\mathrm{H\beta}$ vs. $\textrm{[N\,{\sc ii}]}/\mathrm{H\alpha}$), primarily towards higher $\textrm{[O\,{\sc iii}]}/\mathrm{H\beta}$, compared with local galaxies.  Based on the [S\,{\sc ii}] ratio, we measure an electron density ($n_\mathrm{e}=220^{+170}_{-130}~\mathrm{cm^{-3}}$), that is higher than that of local galaxies.  Based on comparisons to theoretical models, we argue that changes in emission-line ratios, including the offset in the BPT diagram, are caused by a higher ionization parameter both at fixed stellar mass and at fixed metallicity with additional contributions from a higher gas density and possibly a hardening of the ionizing radiation field.  Ionization due to AGNs is ruled out as assessed with {\it Chandra}.  As a consequence, we revisit the mass--metallicity relation using $\textrm{[N\,{\sc ii}]}/\mathrm{H\alpha}$ and a new calibration including $\textrm{[N\,{\sc ii}]}/\textrm{[S\,{\sc ii}]}$ as recently introduced by Dopita et al.  Consistent with our previous results, the most massive galaxies ($M_\ast\gtrsim10^{11}~M_\odot$) are fully enriched, while those at lower masses have metallicities lower than local galaxies.  Finally, we demonstrate that the stellar masses, metallicities and star formation rates of the FMOS sample are well fit with a physically-motivated model for the chemical evolution of star-forming galaxies.

\end{abstract}

\newpage

\section{Introduction}
\label{sec:intro}
The physical conditions of the interstellar medium (ISM) provide clues to understanding the current state and past activity of star formation and gas reprocessing in galaxies.  The study of ionized gas in star-forming (i.e., H\,{\sc ii}) regions has been carried out through many early observational efforts (e.g., \citealt{1942ApJ....95...52A,1979MNRAS.189...95P,1980MNRAS.193..219P}), and the underlying physics has been developed theoretically with photoionization models (e.g., \citealt{1985ApJS...58..125E,2000ApJ...542..224D,2002ApJS..142...35K}).  Numerous efforts to study the ISM in low-redshift galaxies ($z\lesssim 0.3$) are based on large spectroscopic data sets such as the Sloan Digital Sky Survey (SDSS; e.g., \citealt{2003MNRAS.346.1055K,2004ApJ...613..898T,2004MNRAS.351.1151B}).  A wide variety of empirical diagnostics of the gas properties have been established, especially based on spectral features in the rest-frame optical window.  

In particular, the Baldwin--Phillips--Terlevich (BPT; \citealt{1981PASP...93....5B,1987ApJS...63..295V}) diagnostic diagram compares line ratios [N\,{\sc ii}]/H$\alpha$ and [O\,{\sc iii}]/H$\beta$ to distinguish star-forming galaxies from those hosting and/or dominated by an active galactic nucleus (AGN).  Local star-forming galaxies form a tight `abundance sequence' in the BPT diagram (e.g., \citealt{2003MNRAS.346.1055K}) with ionization attributed to radiation from young massive stars (type O and B).  In contrast, AGNs deviate from this sequence due to harder radiation from an accretion disk.  Large data sets such as SDSS have established a precise location of star-forming galaxies and AGNs thus facilitating a relatively clean selection of these populations. 

However, the ISM properties of galaxies at higher redshifts ($z\gtrsim1$), where most key rest-frame optical emission lines are redshifted into infrared, are still unclear.  One would expect the properties of ionized gas in high-$z$ galaxies to be dissimilar to local galaxies given their higher star formation rate (SFR) \citep[see][for a review]{2014ARA&A..52..415M} and higher gas fraction \citep[e.g.,][]{2010MNRAS.407.2091G,2012ApJ...758L...9M,2014ApJ...783...84S}. Over the last decade, the universality of the BPT diagram for high-redshift galaxies has been examined through near-infrared spectroscopic campaigns.  Early studies have reported an offset of the distribution of star-forming galaxies at higher redshifts ($z\sim 1$--3) from the local abundance sequence based on relatively small samples (e.g., \citealt{2005ApJ...635.1006S,2006ApJ...644..813E,2008ApJ...678..758L}). More recent, the offset has been reaffirmed using larger samples (e.g., \citealt{2014ApJ...795..165S,2015ApJ...801...88S,2014ApJ...792...75Z,2015ApJ...806L..35K,2015PASJ...67...80H}) observed with multi-object near-infrared spectrographs.  However, the amount of the offset in the BPT diagram and other emission-line diagrams has not been firmly established across the general star-forming population at the various epochs since the location of galaxies on such diagrams can be significantly affected by sample selection (see \citealt{2014ApJ...788...88J,2015ApJ...801...88S,2016ApJ...817...57C}).  These offsets likely reflect more extreme conditions of H\,{\sc ii} gas in high-$z$ galaxies such as having a higher value of the ionization parameter ($q_\mathrm{ion}$), i.e., the ratio of ionizing photon number density to the hydrogen atom density of H\,{\sc ii} regions, or a harder radiation field inferred by the presence of metal-poor, Fe-depleted stars and/or massive binaries \citep{2015ApJ...812L..20K,2016ApJ...817...57C,2014ApJ...795..165S,2016ApJ...826..159S}.  Alternative explanations are an enhancement of the nitrogen-to-oxygen abundance and high gas pressure \citep[e.g.,][]{2014ApJ...785..153M,2014ApJ...787..120S,2016ApJ...816...23S,2016ApJ...817...57C}.  However, it still remains unsettled what physical factors cause the BPT offset and other changes in the emission-line ratios of high-$z$ galaxies.

Directly related to diagnostics of ionization, the gas-phase metallicity (hereafter metallicity) is one of the important probe of the galaxy evolution as a tool to trace past star formation history.  The abundance of an element from nucleosynthesis is further influenced by both inflowing gas that dilutes the metal fraction of the ISM, and outflows that transpose metals into the circumgalactic environment (e.g., \citealt{1999MNRAS.306..317K,2007ApJ...658..941D,2008ApJ...672L.107E,2008ASPC..399..239E,2008MNRAS.385.2181F,2010MNRAS.408.2115M,2010Natur.467..811C,2011MNRAS.417.2962P,2012MNRAS.426..801B,2013ApJ...772..119L}).  In the local universe, a correlation between stellar mass and metallicity has been robustly established based on large data sets such as SDSS (e.g., \citealt{2004ApJ...613..898T,2013ApJ...765..140A,2011ApJ...730..137Z}).  The existence of a mass--metallicity (MZ) relation has been extended up to $z\sim3$ or more (e.g., \citealt{2006ApJ...644..813E,2008A&A...488..463M,2012PASJ...64...60Y,2014MNRAS.437.3647Y,2015PASJ...67..102Y,2014ApJ...791..130Z,2014ApJ...792...75Z,2014ApJ...789L..40W,2014ApJ...792....3M,2015A&A...577A..14M,2015ApJ...799..138S,2016ApJ...822...42O}) with an evolution where the metallicity decreases with redshift at a fixed stellar mass.  We have previously reported a MZ relation at $z\sim1.6$ \citep{2014ApJ...792...75Z}, using a smaller subset of the data from the FMOS-COSMOS program \citep{2015ApJS..220...12S}, that shows the most massive galaxies as being fully mature at a level similar to local massive galaxies while lower mass galaxies are less enriched as compared to local galaxies.  However there still remain some discrepancies between various studies that implement different sample selection methods and/or metallicity determinations.

An anti-correlation between metallicity and SFR at a fixed stellar mass has been seen in local galaxies (e.g., \citealt{2008ApJ...672L.107E}).  In particular, \citet{2010MNRAS.408.2115M} introduced the SFR as a third parameter in the MZ relation, and proposed the existence of a universal relation between these three quantities, referred to as the {\it Fundamental Metallicity Relation} (FMR).  Several studies have found that introducing the FMR reduces the scatter relative to the MZ relation \citep{2010A&A...521L..53L,2013MNRAS.434..451L,2011MNRAS.414.1263M,2012MNRAS.422..215Y,2014ApJ...792...75Z}, although the actual shape of the FMR may differ appreciably from one study to another due to differences in sample selection and/or in the adopted metallicity indicator \citep[e.g.,][]{2013ApJ...765..140A}.  The apparent SFR--$Z$ anti-correlation is interpreted as resulting from upward fluctuations of the infall rate of pristine/metal-poor gas boosting the SFR while diluting the metallicity of the ISM \citep{2008ApJ...672L.107E,2010MNRAS.408.2115M}, as supported by several analytic studies and numerical simulations \citep[e.g.,][]{2012MNRAS.421...98D,2013MNRAS.430.2891D,2013ApJ...772..119L}.  Measuring the $M_\ast$-$Z$-SFR relation at all redshifts is an essential step towards our understanding of star formation and the chemical evolution of galaxies.  However, beyond the local universe, the existence and shape of the FMR has not been firmly established yet (see e.g., \citealt{2014ApJ...789L..40W,2014ApJ...792...75Z,2014MNRAS.437.3647Y,2014ApJ...795..165S,2015ApJ...808...25S,2015ApJ...799..138S,2016ApJ...822..103G}).

To further understand the typical characteristics of high-$z$ H\,{\sc ii} regions, it is highly desirable to construct a well-controlled sample of typical star-forming galaxies at each epoch.  With the availability of multi-object near-infrared spectrographs (i.e., MOSFIRE, FMOS, KMOS) on large telescopes, we can access key rest-frame optical emission lines such as H$\beta$, [O\,{\sc iii}]$\lambda 5007$, H$\alpha$, [N\,{\sc ii}]$\lambda 6584$, and [S\,{\sc ii}]$\lambda\lambda$6717, 6731 for samples consisting of up to $10^3$ galaxies at $z\gtrsim1$ (e.g., \citealt{2012ApJ...750...67R,2015PASJ...67...81T,2015ApJS..218...15K}).  In this paper, we use a sample of H$\alpha$-detected galaxies that trace the star-forming main sequence over a stellar mass range $10^{9.6} \lesssim M_\ast/M_\odot \lesssim 10^{11.6}$, from the FMOS-COSMOS survey \citep{2015ApJS..220...12S} to study the typical properties of H\,{\sc ii} regions of galaxies at $1.43\le z\le1.74$.  The unique advantages of our sample are the large size with 701 galaxies having an H$\alpha$ detection, which is four times larger than that used in our previous study \citep{2014ApJ...792...75Z}, and the high sampling rate of the massive galaxy population ($M_\ast>10^{11}~M_\odot$).  In addition, our target selection is primarily based on the $K$ band photometry, which is a good proxy of stellar mass, thus tends to avoid significant selection biases, as compared to other samples at similar redshifts \citep[e.g.,][]{2014ApJ...795..165S,2015PASJ...67...80H}.  We also highlight that our FMOS observations are much deeper (3--5 hours integration times) than in \citet{2015PASJ...67..102Y} and we have $J$-band spectra for about half of our sample that cover [O\,{\sc iii}] and H$\beta$.  We use the rest-frame optical key emission lines to evaluate the ionization and excitation of heavy elements, gaseous metallicity, and electron density of ionized gas while minimizing the impact of AGNs. 

This paper is organized as follows.  In Section \ref{sec:data}, we give an overview of our FMOS-COSMOS survey and describe our samples and spectral analyses.  We present our measurements of the emission-line properties in Section \ref{sec:results}.  Section \ref{sec:BPTorigin} discusses the characteristics of H\,{\sc ii} regions in high-$z$ galaxies.  We present the metallicity measurements and a reanalysis of the MZ relation in Section \ref{sec:metallicity}.  In Section \ref{sec:M-Z-SFR}, we study the relation between mass, metallicity, and SFR.  We finally summarize our results and conclusions in Section \ref{sec:conclusion}. Throughout this paper, we use a cosmology with $(h, \Omega_M, \Omega_\Lambda)=(0.7, 0.3, 0.7)$ and assume the \citet{1955ApJ...121..161S} initial mass function (IMF; 0.1--100$~M_\odot$).  All magnitudes given in this paper are in the AB magnitude system.

\section{Data}
\label{sec:data}

\subsection{The FMOS-COSMOS survey overview}
\label{sec:survey}

The galaxy sample used in this paper is constructed from data set of the FMOS-COSMOS survey.  Here, we provide an overview of the survey, which is extensively described in \citet{2013ApJ...777L...8K} and \citet{2015ApJS..220...12S}.  The FMOS-COSMOS survey is a completed near-infrared spectroscopic survey carried out between Mar 2012 and Apr 2016, designed to detect the H$\alpha$ and [N\,{\sc ii}] lines from galaxies at $1.43 < z < 1.74$ with the Fiber Multi-Object Spectrograph (FMOS; \citealt{2010PASJ...62.1135K}) in high-resolution mode ($R\sim3000$; $1.6$--$1.8~\mathrm{\mu m}$).  The emission-line sensitivity of the $H$-long grating with an integration time of five hours is $\sim2\times10^{-17}~\mathrm{erg~cm^{-2}~s^{-1}}$ for a $5\sigma$ detection.  In addition to H$\alpha $ and [N\,{\sc ii}], the [S\,{\sc ii}]$\lambda \lambda6717,6731$ doublet lines can be observed within the $H$-long window for galaxies at $1.43< z <1.67$.   At the given spectral resolution, all lines are well separated, thus do not suffer from any blending issues.  The accuracy of the spectroscopic redshift determination is $\Delta z/\left( 1+z \right) =2.2\times10^{-4}$ \citep{2015ApJS..220...12S}.  Galaxies with a positive detection of an H$\alpha$ emission line in the $H$-long spectral window are re-observed with the $J$-long grating (1.11--1.35$\mathrm{\mu m}$; $R\sim2200$) to detect H$\beta$ and [O\,{\sc iii}]$\lambda\lambda4959,5007$, which are essential to determine the excitation states of the ionized gas in star-forming regions.  All data are reduced using the FMOS pipeline FIBRE-pac (FMOS Image-Based Reduction Package; \citealt{2012PASJ...64...59I}).

\subsection{Target selection}
\label{sec:target}

In this paper, we utilize a larger catalog of galaxies with spectroscopic redshifts from the FMOS-COSMOS survey than presented in \citet{2013ApJ...777L...8K} and \citet{2014ApJ...792...75Z}.  This larger sample is the result of additional FMOS observations carried out between Dec 2013 and Feb 2014 that are not reported in \citet{2015ApJS..220...12S}. The characteristics of this sample are statistically equivalent to those used in the aforementioned papers. Here, we give a brief overview of the construction of our galaxy catalog.

Our galaxy sample is based on the COSMOS photometric catalog \citep{2012A&A...544A.156M,2013A&A...556A..55I} that includes the Ultra-VISTA/VIRCAM photometry.  A magnitude limit of $K_\mathrm{S}<23.5$ is imposed for our sample selection, which provides a high level of completeness ($\sim85\%$) in stellar mass of galaxies with $M_\ast\ge10^{10}~M_\odot$ \citep{2013A&A...556A..55I}.  The majority of our sample is selected to have a stellar mass above $10^{9.8}~M_\odot$ and a photometric redshift $1.46\le z_\mathrm{phot} \le 1.72$.  These values are derived for each object by fitting the spectral energy distribution (SED) using Le Phare \citep{2011ascl.soft08009A} with population synthesis models \citep{2003MNRAS.344.1000B} and a \citet{2003PASP..115..763C} IMF.  Hereafter, we convert all stellar masses to a Salpeter IMF by applying a multiplicative factor of 1.7 \citep{2010ApJ...709..644I}.  

To achieve a high success rate of detecting the H$\alpha $ emission line, we calculate the expected H$\alpha $ flux for each galaxy in the photometric catalog and use these values in our target selection.  The prediction of H$\alpha$ flux represents a total flux from each galaxy (without considering flux loss due to the fiber aperture) calculated with Equation 2 of \citet{1998ARA&A..36..189K} from SFR, which is derived from the SED fitting assuming a constant star formation history (see \citealt{2015ApJS..220...12S} for details).  Dust extinction towards the H$\alpha$ emission line is derived via 
\begin{equation}
A(\mathrm{H\alpha})=3.325 E_\mathrm{star}(B-V)/f,
\label{eq:extinction}
\end{equation}
where the color excess $E_\mathrm{star}(B-V)$ is estimated from the SED and a \citet{2000ApJ...533..682C} extinction curve is assumed.  We note that large uncertainties remain in the conversion between the amount of extinction towards stellar emission and that towards nebular emission, which is likely to depend on the geometrical properties of stars, dust, and star-forming regions.  While $f=0.44$ is canonically applied locally \citep{2000ApJ...533..682C}, some studies report on higher values between 0.44 and 1 for high-$z$ galaxies based on measurements of the Balmer decrement (e.g., H$\alpha$/H$\beta$) or comparisons between SED- (or UV-) based SFRs and H$\alpha$-SFRs \citep[e.g.,][]{2013ApJ...777L...8K,2014ApJ...788...86P,2015MNRAS.453..879K,2015ApJ...801..132V,2016A&A...586A..83P}.   Here, we use $f=0.66$ to calculate predicted H$\alpha$ fluxes for target galaxies.  In early pilot observations, a predicted H$\alpha$ flux threshold of $4\times10^{-17}~\mathrm{erg~cm^{-2}~s^{-1}}$ is set without taking into account the $\sim50\%$ aperture loss.  In the subsequent intensive program, the limit of the predicted H$\alpha$ flux was raised to $1\times10^{-16}~\mathrm{erg~cm^{-2}~s^{-1}}$.  This flux limit is equivalent to $\mathrm{SFR}\sim 20~M_\odot ~\mathrm{yr^{-1}}$ for galaxies at $z\sim 1.6$ with typical values of dust attenuation.  The majority (89\%) of our sample have a predicted H$\alpha$ flux higher than $1\times10^{-16}~\mathrm{erg~cm^{-2}~s^{-1}}$.  

\subsection{Spectral fitting}
\label{sec:fitting}
We perform a fitting procedure that utilizes the MPFIT package for IDL \citep{2009ASPC..411..251M} to measure the flux of emission lines and associated error, both on individual and composite spectra.  The fitting of key emission lines (H$\alpha$, [N\,{\sc ii}]$\lambda \lambda$6548, 6584, H$\beta$, [O\,{\sc iii}]$\lambda \lambda$4959, 5007) present in individual spectra are described in detail elsewhere \citep{2013ApJ...777L...8K,2015ApJS..220...12S}.  In this study, we further perform a fit to the [S\,{\sc ii}]$\lambda \lambda$6717, 6731 lines based on redshift determined from H$\alpha$.  

In our emission-line fitting procedure, the value of each pixel is weighted by the noise spectra and the pixels that are impacted by the OH airglow mask are excluded from the fit.  For composite spectra, we use weights based on the variance estimated by jackknife resampling (see Section \ref{sec:stack}).  The continuum is first fit with a linear function to pixels near the emission lines and subtracted from the data.  Each emission line is then modeled with a Gaussian profile.  The [N\,{\sc ii}] and H$\alpha$ lines are simultaneously fit with a single line width.  We further fix the ratio [N\,{\sc ii}]$\lambda$6584/[N\,{\sc ii}]$\lambda$6548 to the laboratory value of 2.86, and the line width of [S\,{\sc ii}] to that of H$\alpha$.  The H$\beta$ and [O\,{\sc iii}] lines are modeled independently from H$\alpha$+[N\,{\sc ii}] with the ratio of the [O\,{\sc iii}] doublet fixed to 2.98.  This avoids any systematic effects due to uncertainties of the wavelength calibration and provides an independent evaluation of key lines (e.g., H$\alpha$, [O\,{\sc iii}]).  The observed line widths of H$\alpha$ and [O\,{\sc iii}]--H$\beta$ system agree with each other on average.  

\subsection{Correction for Balmer absorption}
\label{sec:abcorr}

Stellar atmospheric absorption lowers the observed flux of the Balmer emission lines, in particular H$\beta$ (e.g., \citealt{2004AJ....127.2511N,2012MNRAS.419.1402G}).  Therefore, we correct the observed H$\beta$ flux for this underlying absorption as a function of stellar mass as given in \cite{2014ApJ...792...75Z}:
\begin{equation}
f_\mathrm{corr} = \text{max}\left[1,~1.02+0.30\log\left(M_\ast/10^{10} M_\odot\right)\right].
\label{eq:abcorr}
\end{equation}
This relation has been converted to be used with a Salpeter IMF.  We apply the absorption correction to the observed H$\beta$ flux of individual galaxies in our FMOS sample, and the measurements based on the composite spectra.  The H$\alpha$ flux is not corrected for the stellar absorption since the flux loss is expected to be negligible ($<$ a few percent; see \citealt{2013ApJ...777L...8K}).  

The Balmer absorption correction reduces the [O\,{\sc iii}]/H$\beta$ ratio by $11\%$ on average and maximally by $50\%$ at the high-mass end.  The typical amount of this correction is consistent with \citet{2015ApJ...801...88S} and \citet{2014ApJ...795..165S}.  The application of such a correction does not affect our scientific conclusions.  We note that the Balmer absorption correction is applied for only the FMOS sample, and not for the local galaxies from the MPA/JHU catalog (see Section \ref{sec:sdss}), in which the correction is already taken into account by measuring the emission-line intensities after the stellar continuum subtraction based on a population synthesis model (see \citealt{2004ApJ...613..898T} for details).

\subsection{Sample selection for analysis}
\label{sec:sample}

In this study, we use 701 galaxies (approximately 40\% of observed galaxies) having a detection of the H$\alpha$ emission line with a signal-to-noise ratio (S/N) greater than 3 in the $H$-long spectrum.  The range of spectroscopic redshift is $1.43\le z\le1.74$.  We define two subsamples based on the spectral coverage. {\it Sample-1} consists of all 701 galaxies with a positive H$\alpha$ detection in the $H$-long band, regardless of the presence or absence of the $J$-long coverage.  {\it Sample-2} consists of 310 galaxies (a subset of Sample-1) having additional $J$-long coverage.  The numbers of detections of each emission line are summarized in Table \ref{tb:samples}.  For the [S\,{\sc ii}] doublet, we count the number of objects with both [S\,{\sc ii}] lines detected at either $S/N>1.5$ or $>3$.  The numbers of galaxies with simultaneous detections of multiple emission lines are provided as well.  We further group galaxies by the S/N of their emission-line measurements: high-quality (HQ) if $S/N>5$ for H$\alpha$ and $S/N>3$ for other lines and low-quality (LQ) if $S/N>3$ for H$\alpha$ and $S/N>1.5$ for others.  For stacking analyses (see Section \ref{sec:stack}), galaxies are restricted to those having a redshift measurement between $1.43 \le z \le 1.67$ to cover all the key emission lines including [S\,{\sc ii}]$\lambda \lambda 6717, 6731$.  The numbers of galaxies used for the stacking analysis are given in the bottom row in Table \ref{tb:samples}.

\onecolumngrid
\newpage
\begin{deluxetable*}{llllll}
\tablecaption{Summary of emission-line detections of the FMOS sample\tablenotemark{a}\label{tb:samples}}
\tablehead{\colhead{Emission lines}&\multicolumn{2}{c}{$H$-long}&\multicolumn{2}{c}{$H$ \& $J$-long}\\
\colhead{}&\colhead{Sample-1\tablenotemark{b}}&\colhead{w/o AGNs\tablenotemark{c}}&\colhead{Sample-2\tablenotemark{d}}&\colhead{w/o AGNs}}
\startdata
H$\alpha$ & 701 (500) & 642 (469) & 310 (241) & 283 (224) \\
$\left[ \textrm{N\,{\sc ii}} \right]$ & 436 (278) &  383 (233) & 216 (133) & 190 (110) \\
$\left[ \textrm{S\,{\sc ii}} \right]$ & 77 (17) & 73 (17) & 32 (5) & 31 (5) \\
H$\beta$ & - & - & 138 (99) & 127 (94) \\
$\left[ \textrm{O\,{\sc iii}} \right]$ & - & - & 171 (160) & 158 (147) \\
$\mathrm{H\alpha} + \left[ \textrm{N\,{\sc ii}} \right]$ \tablenotemark{e} & 436 (246) & 383 (217) & 216 (115) & 190 (99) \\
$\mathrm{H\alpha} + \left[ \textrm{S\,{\sc ii}} \right]$                  & 77 (17)    & 73 (17)      & 32 (5)      & 31 (5) \\
$\mathrm{H\alpha} + \left[ \textrm{N\,{\sc ii}} \right]+\left[ \textrm{S\,{\sc ii}} \right]$          & 61 (13)    & 57 (13)      & 28 (2)       & 27 (2) \\
$\left[ \textrm{O\,{\sc iii}} \right]+\mathrm{H\beta}$                 & - & - & 121 (70) & 110 (67) \\
$\mathrm{H\alpha} + \left[ \textrm{N\,{\sc ii}} \right] + \left[ \textrm{O\,{\sc iii}} \right]+\mathrm{H\beta}$        & - & - & 87 (40) & 76 (37) \\
$\mathrm{H\alpha} + \left[ \textrm{S\,{\sc ii}} \right] + \left[ \textrm{O\,{\sc iii}} \right]+\mathrm{H\beta}$         & - & - & 19 (5) & 18 (5)\\
$N_\mathrm{stack}$\tablenotemark{f} & - & 554 & - & 246
\enddata
\tablenotetext{a}{The threshold S/N is 3 for H$\alpha$ and 1.5 for other lines.  In parentheses, the numbers of detections with a higher S/N ($>5$ for H$\alpha$ and $>3$ for other lines) are listed.}
\tablenotetext{b}{Sample-1 consists of 701 galaxies with a positive H$\alpha$ detection.}
\tablenotetext{c}{The numbers of galaxies after the AGN exclusion (see Section \ref{sec:AGNrem}.)}
\tablenotetext{d}{Sample-2 consists of 310 galaxies with both an positive H$\alpha$ detection and an additional $J$-long spectrum.}
\tablenotetext{e}{The numbers of galaxies with multiple emission-line detections are listed in the 6th-11th rows (see Section \ref{sec:sample}).}
\tablenotetext{f}{The numbers of objects that are used for stacking ($1.43\le z \le1.67$; AGNs are excluded; see Section \ref{sec:stack}).}
\end{deluxetable*}

\subsection{Identification of AGN}
\label{sec:AGNrem}

AGNs are excluded for a clean investigation of the conditions of H\,{\sc ii} regions of star-forming galaxies.  In our sample, we identify 22 objects (3\%) associated with an X-ray point source in the catalog provided by the {\it Chandra} COSMOS Legacy survey \citep{2009ApJS..184..158E,2016ApJ...819...62C}, which covers the entire FMOS survey area.  These X-ray detected galaxies are likely to host an AGN because such luminous X-ray emission ($L_\mathrm{X-ray}\gtrsim 10^{42}~\mathrm{erg~s^{-1}}$ at $0.5$--$7~\mathrm{keV}$) is expected to arise from a hot accretion disk. The fraction of the objects with an X-ray detection (3\%) is roughly similar to that reported by \citet[2\% at $z \sim1$]{2009ApJ...696..396S}, who utilized a galaxy sample from the 10k catalog of the zCOSMOS-bright survey \citep{2007ApJS..172...70L}.  The X-ray source fraction of our sample achieves 10\% at $M_\ast>10^{11}~M_\ast$.  Such an X-ray AGN fraction at the high-mass end is lower than reported at a similar redshift range ($\sim 25\text{--}30\%$; e.g., \citealt{2005ApJ...633..748R,2008ApJ...681..931B,2009A&A...507.1277B,2009ApJ...699.1354Y,2015MNRAS.450..763M}).  This is likely due to the prior selection on the predicted H$\alpha$ flux applied for galaxies in our sample.

We also flag FMOS sources as AGN based on their emission-line ratios.  While the BPT diagram is commonly used to separate the pure star-forming population and AGNs for low redshift galaxies, the same boundary is unlikely to be applicable for high-$z$ galaxies \citep[e.g.,][]{2013ApJ...774L..10K}.  Here, we use a boundary between star-forming galaxies and AGN in the BPT diagram at $z=1.6$ derived by \citet{2013ApJ...774L..10K} (see Section \ref{sec:BPT}).  We identify 39 objects (5.5\%) as AGNs based on their rest-frame optical emission-line properties that are located above this boundary or have either a line ratio of $\log\text{[N\,{\sc ii}]/H}\alpha > -0.1$ or $\log\text{[O\,{\sc iii}]/H}\beta>0.9$.  Of these, four galaxies are detected in the X-ray.  The majority of AGNs, identified by their narrow line ratios, are likely obscured (type-II) AGNs.   In addition, four objects (less than 1 per cent) have an emission-line width (full width at half maximum) greater than $1000~\mathrm{km~s^{-1}}$, usually H$\alpha$, that are taken to be unobscured (type-I) AGNs; one of these is included in the X-ray point source catalog and another object is flagged as AGN based on the line ratios.

In total, 59 (8\%) objects are identified as AGN.  The total AGN fraction is roughly consistent with that of \citet[6\%]{2014MNRAS.437.3647Y}, who use a sample of star-forming galaxies at $z\sim1.4$, although they do not have X-ray observations thus probably miss some AGNs.  The AGN fraction increases with stellar mass and achieves 25\% at $M_\ast>10^{11}~M_\ast$.  Such a trend is consistent with previous studies \citep[e.g.,][]{2010ApJ...720..368X}.  Table \ref{tb:samples} lists the numbers of galaxies in each sample before and after the exclusion of those hosting AGNs.  For stacking analysis, we only use galaxies without AGN.  In Appendix \ref{sec:AGNcontami}, we address the additional contribution of AGN photoionization to the stacked emission-line measurements.

\subsection{Stellar masses, SFRs, and the main sequence}
\label{sec:SFRdet}

Stellar masses and SFRs are derived from SED fitting, based on the spectroscopic redshift measured with FMOS, using Le Phare \citep{2011ascl.soft08009A} while assuming a constant star formation history and a Chabrier IMF.  Masses and SFRs are converted to a Salpeter IMF by multiplying them by a factor of 1.7.    The statistical errors of the derived stellar masses and SFRs are typically 0.05~dex and 0.2~dex, respectively.  The sample spans a range of stellar mass $9.6\lesssim \log~M_\ast/M_\odot \lesssim 11.6$ with a median mass of $\log M^\mathrm{med}_\ast/M_\odot = 10.4$.  We highlight that our sample includes a large number of very massive galaxies ($M_\ast \ge 10^{11}~M_\odot$; 120 in total, 92 after removing AGNs; see Section \ref{sec:AGNrem}).

\begin{figure*}[htbp]
   \centering
   \includegraphics[width=6.5in]{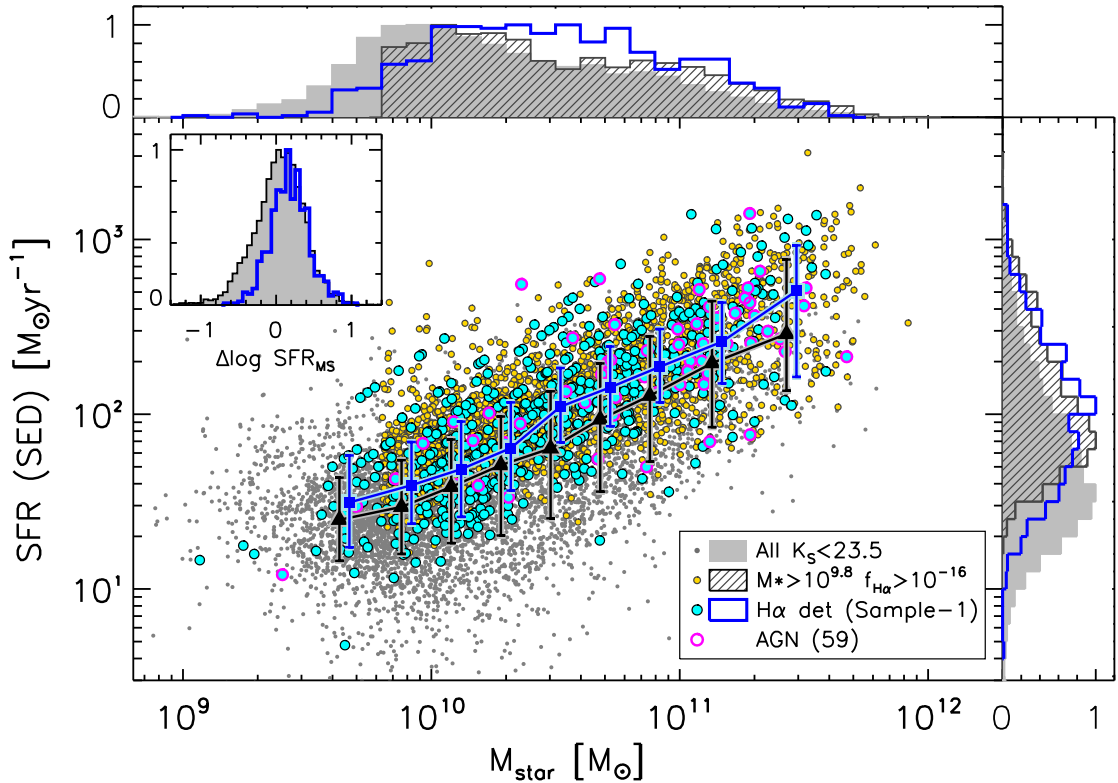} 
   \caption{SFR (SED) versus $M_\ast$ for galaxies at $1.43 \le z\le1.74$: all galaxies with $K_\mathrm{S}<23.5$ and $1.46\le z_\mathrm{photo}\le 1.72$ in the FMOS survey area (gray dots, filled histograms normalized by their peak values), galaxies with $M_\ast \ge 10^{9.8}~M_\odot$ and a predicted H$\alpha$ flux $f(\mathrm{H\alpha})\ge 1\times10^{-16}~\mathrm{erg~s^{-1}~cm^{-2}}$ (yellow circles, hatched histograms), galaxies with an H$\alpha$ detection (cyan circles, blue thick-line histograms) are shown.  AGN candidates are marked as magenta circles.  The median SFRs in eight stellar mass bins for both the $K_\mathrm{S}$-selected and H$\alpha$-detected samples are indicated by large symbols (triangles and squares respectively) with error bars indicating the 68th percentiles.  The H$\alpha$-detected galaxies trace well the star-forming main sequence, although there is a slight bias towards higher SFRs as compared with the underlying $K_\mathrm{S}$-selected sample (see inset panel).}
   \label{fig:sample}
\end{figure*}

Figure \ref{fig:sample} shows the SFR as a function of $M_\ast$ for our sample, compared with the distribution of $K_\mathrm{S}$-selected ($K_\mathrm{S}<23.5$) galaxies in the equivalent redshift range ($1.46\le z_\mathrm{photo}\le1.72$; 7987; gray dots).  The $K_\mathrm{S}$-selected sample shows a clear correlation between stellar mass and SFR (i.e., star-forming main sequence) with a standard deviation around the relation of $0.37~\mathrm{dex}$.  Of these, galaxies with $M_\ast\ge10^{9.8}$ and predicted H$\alpha$ flux $f(\mathrm{H\alpha})\ge1\times10^{-16}~\mathrm{erg~s^{-1}~cm^{-2}}$ (2324) are marked as small yellow circles; those with an H$\alpha$ detection (Sample-1) are shown as cyan circles.  Objects identified as AGN are marked as magenta circles.  The Sample-1 galaxies apparently follow the distribution of the parent $K_\mathrm{S}$-selected sample.  To quantify a potential bias in our sample, we fit a power-law relation ($\mathrm{SFR}\propto M_\ast^\alpha$) to the $K_\mathrm{S}$-selected and H$\alpha$-detected samples separately for galaxies with $M_\ast\ge 10^{9.8}~M_\odot$.  While the slope is consistent between the two samples ($\alpha=0.66\pm0.01$ for the former and $0.68\pm0.02$ for the latter), the H$\alpha$-detected sample is slightly biased towards higher SFRs by $\sim0.15~\mathrm{dex}$ over the entire stellar mass range due to the self-imposed limit on the predicted H$\alpha $ flux.  However, such a bias is less than half of the scatter of the main sequence.  The inset panel shows the normalized distribution of the difference in SFR from the fit to the main sequence for both $K_\mathrm{S}$-selected (filled gray histogram) and H$\alpha$-detected (blue solid histogram) galaxies.  As clearly evident, our final sample traces well the star-forming main sequence at $z\sim1.6$.

We also measure SFRs from the H$\alpha$ flux measured by FMOS following Equation 2 of \citet{1998ARA&A..36..189K}.  As described in \citet{2013ApJ...777L...8K}, the observed H$\alpha$ fluxes are corrected for the flux falling outside the FMOS 1\arcsec.2-diameter fiber.  This aperture correction factor is evaluated for each object from a comparison between the observed continuum flux from FMOS spectra and the Ultra-VISTA $H$ and $J$-band photometry \citep{2012A&A...544A.156M}.  The observed H$\alpha$ flux is also corrected for dust extinction.  The amount of extinction (i.e., color excess $E_\mathrm{star}(B-V)$) is estimated from the SED of each galaxy, then it is converted to the extinction towards the H$\alpha$ emission line by using Equation (\ref{eq:extinction}) with $f=0.59$.  This $f$ value is adjusted to bring observed H$\alpha$-based SFRs into agreement with the SED-based SFRs.  While we assumed $f=0.66$ in our target selection (Section \ref{sec:target}), the application of a different $f$-factor does not change any of our conclusions.

\subsection{Spectral stacking}
\label{sec:stack}

We make use of co-added spectra to account for galaxies with faint emission lines and avoid a bias induced by galaxies with the strongest lines (i.e., highest SFR and/or less extinction).  Individual spectra of galaxies at $1.43\le z \le1.67$ are stacked in bins of stellar mass.  This redshift range ensures that all the emission lines used in this study fall within the spectral coverage of FMOS.  A fair fraction ($30\%$) of the pixels in individual spectra that are strongly impacted by the OH mask and residual sky lines are removed by identifying them in the noise spectra as regions with relatively large errors as compared with the typical noise level of $\sim5\times10^{-19}~\mathrm{erg~cm^{-2}~s^{-1}~\textrm{A}^{-1}}$ (see Figure 11 in \citealt{2015ApJS..220...12S}).  We transform all individual spectra to the rest-frame wavelength based on their redshift and resample them to a common wavelength grid with a spacing of $0.5~\text{\AA}$ per pixel.  This wavelength sampling is equivalent to the observed-frame spectral resolution of FMOS ($1.25~\text{\AA/pix}$) for galaxies at $z=1.5$.

\begin{figure*}[htbp]
   \centering
   \includegraphics[width=7in]{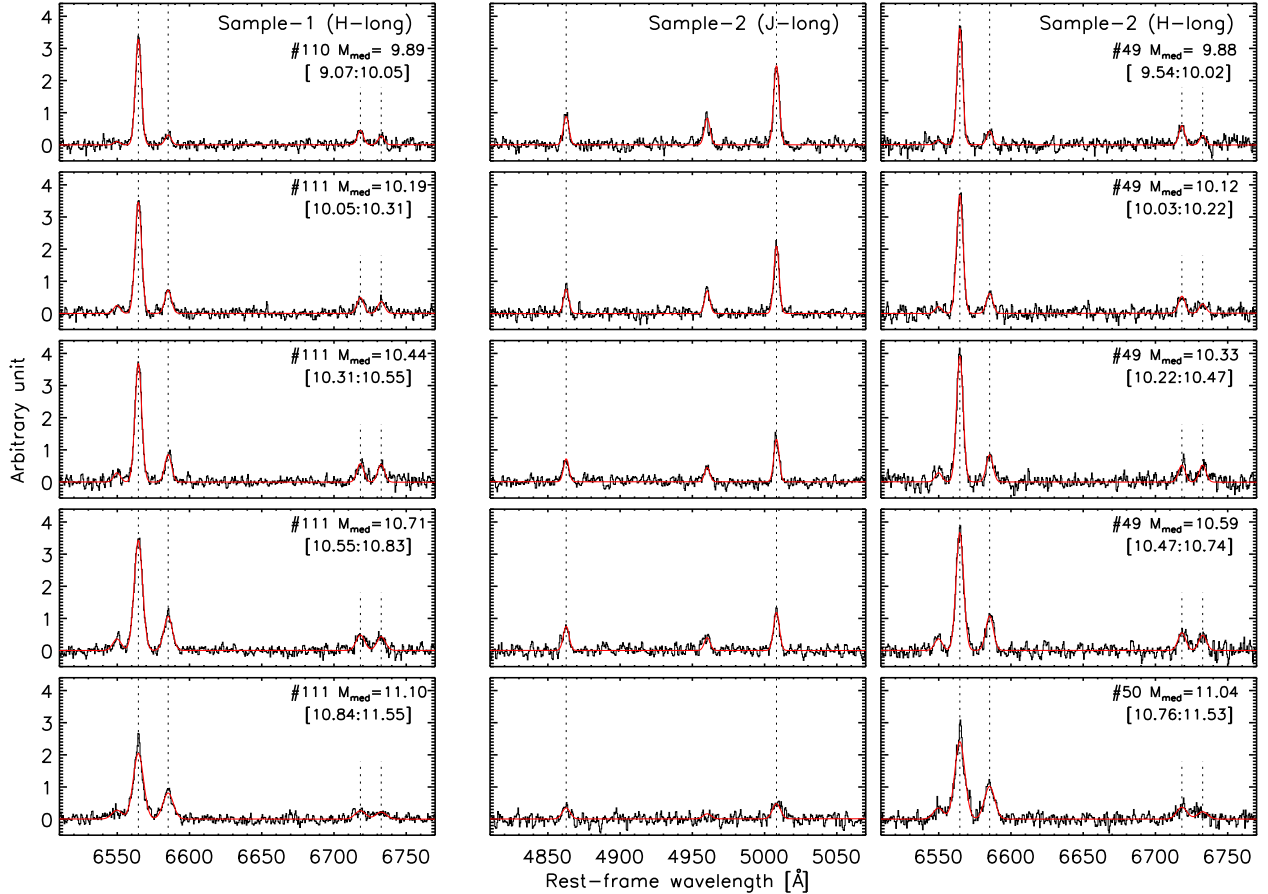} 
   \caption{Composite spectra in five bins of stellar mass of Sample-1 (left column) and Sample-2 (middle and right columns).  The number of individual spectra is indicated for each bin (e.g., \#110).  The stellar mass increases from top to bottom, as indicated by their median value ($\log M_\ast/M_\odot$) and the interval shown in each panel.  The observed spectra and the best-fit models are shown by black thin solid lines and red thick curves.  Vertical dotted lines show the positions of H$\alpha$, [N\,{\sc ii}]$\lambda 6584$, [S\,{\sc ii}]$\lambda \lambda 6717,6731$, H$\beta$, and [O\,{\sc iii}]$\lambda 5007$.}
   \label{fig:spec}
\end{figure*}

After subtracting the continuum, the individual de-redshifted spectra are averaged by using the {\it resistant\_mean.pro}, an IDL routine available in the  Astronomy User's Library.  We apply a $5\sigma$ clipping, which removes data that deviates from the median by more than five times the median absolute deviation at each pixel.  We do not apply any weighting scheme to avoid possible biases.  The associated noise spectra are computed using a jackknife resampling method.  The variance at each pixel is given by the following equation
\begin{eqnarray}
	\sigma^2_\mathrm{jack}  = \frac{N-1}{N}\sum^{N}_{i=1} \left( F_i - \frac{1}{N}\sum^N_{i=1} F_i\right)^2
	\label{eq:jack}
\end{eqnarray}
where $N$ is the sample size used for stacking (after a 5 $\sigma$ clip) and $F_i$ is stacked spectra composed of $N-1$ spectra by removing the $i$-th spectrum.  Figure \ref{fig:spec} shows the composite spectra of Sample-1 and Sample-2 in five bins of stellar mass.  We ensure that the reduced $\chi^2$ of the fit to each composite spectrum is approximately unity (0.8--1.2).  The errors on the flux measurements and flux ratios for the stacked spectra are calculated from Equation (\ref{eq:jack}) with $F_i$ replaced by an arbitrary quantity (i.e., flux or line ratio) measured on the $i$-th-removed stacked spectrum.

We note that individual spectra are averaged without any renormalization; this may result in a flux-weighted co-added spectrum that is not representative of galaxies with the faintest emission lines.  To examine such effects, we normalize each spectrum by the observed amplitude of the H$\alpha$ line before co-adding.  We then find that the results presented in this study do not depend on whether such a scaling is applied or not.  All results presented hereafter are based on a stacking analysis without any such renormalization of the individual spectra.  Throughout, we measure average emission-line ratios using co-added spectra of Sample-1 and Sample-2 split into eight and five bins of stellar mass, respectively.  In Table \ref{tb:linemeasure}, we list the emission-line ratios and their associated jackknife errors.

To further check the potential impact of residual OH emission and the suppression mask in the stacking, we construct a sample that consists of only galaxies for which the redshifted wavelength of [N\,{\sc ii}]$\lambda6584$ is more than 9{\AA} from every OH line\footnote{The list of the OH lines in the FMOS coverage are available at http://www.subarutelescope.org/Observing/Instruments/FMOS/index.html} ($\sim 50\%$ of the full sample) to ensure that the [N\,{\sc ii}] line is free from OH contamination (see also \citealt{2013MNRAS.436.1130S}). We have checked that the results do not change when restricting the analysis to this smaller sample.

\onecolumngrid
\begin{deluxetable*}{llcccc}
\tablecaption{Emission line ratios from co-added spectra\label{tb:linemeasure}}
\tablehead{\colhead{Median $M_\ast$\tablenotemark{a}}& $M_\ast$ range & \colhead{$N2$\tablenotemark{b}}&\colhead{$S2$\tablenotemark{c}}&\colhead{$N2S2$\tablenotemark{d}}&\colhead{$O3$\tablenotemark{e}}} 
\startdata
\multicolumn{6}{l}{Sample-1}\\
 9.84 & 9.07--9.94 & $  -1.09\pm   0.12$ & $  -0.61\pm   0.07$ & $  -0.48\pm   0.13$ & -- \\
10.04 & 9.95--10.11 & $  -0.87\pm   0.07$ & $  -0.61\pm   0.07$ & $  -0.25\pm   0.09$ & -- \\
10.21 & 10.12--10.28 & $  -0.64\pm   0.04$ & $  -0.60\pm   0.06$ & $  -0.04\pm   0.07$ & -- \\
10.35 & 10.29--10.44 & $  -0.70\pm   0.05$ & $  -0.55\pm   0.05$ & $  -0.15\pm   0.07$ & -- \\
10.51 & 10.44--10.59 & $  -0.55\pm   0.04$ & $  -0.52\pm   0.06$ & $  -0.03\pm   0.06$ & -- \\
10.69 & 10.60--10.78 & $  -0.52\pm   0.04$ & $  -0.62\pm   0.07$ & $   0.10\pm   0.08$ & -- \\
10.85 & 10.78--11.05 & $  -0.46\pm   0.04$ & $  -0.57\pm   0.06$ & $   0.12\pm   0.07$ & -- \\
11.16 & 11.06--11.55 & $  -0.37\pm   0.05$ & $  -0.61\pm   0.09$ & $   0.24\pm   0.10$ & -- \\
\hline
\multicolumn{6}{l}{Sample-2}\\
  9.88 & 9.54--10.02 & $  -0.93\pm   0.08$ & $  -0.60\pm   0.09$ & $  -0.34\pm   0.11$ & $   0.46\pm   0.06$ \\
10.12 & 10.03--10.22 & $  -0.77\pm   0.06$ & $  -0.64\pm   0.06$ & $  -0.13\pm   0.07$ & $   0.43\pm   0.06$ \\
10.33 & 10.22--10.47 & $  -0.67\pm   0.04$ & $  -0.56\pm   0.07$ & $  -0.11\pm   0.07$ & $   0.23\pm   0.07$ \\
10.59 & 10.47--10.74 & $  -0.53\pm   0.04$ & $  -0.54\pm   0.05$ & $   0.01\pm   0.05$ & $   0.14\pm   0.08$ \\
11.04 & 10.76--11.53 & $  -0.37\pm   0.04$ & $  -0.60\pm   0.07$ & $   0.22\pm   0.08$ & $   0.03\pm   0.16$ 
\enddata
\tablenotetext{a}{Median $\log M_\ast / M_\odot$ of each mass bin.}
\tablenotetext{b}{$N2=\log (\textrm{[N\,{\sc ii}]} \lambda 6584 / \mathrm{H\alpha})$.}
\tablenotetext{c}{$S2=\log (\textrm{[S\,{\sc ii}]} \lambda\lambda 6717,6731 / \mathrm{H\alpha})$.}
\tablenotetext{d}{$N2S2=\log (\textrm{[N\,{\sc ii}]} \lambda 6584 / \textrm{[S\,{\sc ii}]} \lambda\lambda 6717,6731)$.}
\tablenotetext{e}{$O3=\log (\textrm{[O\,{\sc iii}]} \lambda 5007 / \mathrm{H\beta})$, corrected for Balmer absorption (see Section \ref{sec:abcorr}).}
\end{deluxetable*}

\subsection{Local comparison sample}
\label{sec:sdss}
We extract a sample of local galaxies from the Sloan Digital Sky Survey (SDSS) Data Release 7 \citep{2009ApJS..182..543A} to compare with the ISM properties of our high-$z$ galaxies.  The emission-line flux measurements are from the MPA/JHU catalog \citep{2003MNRAS.341...33K,2004MNRAS.351.1151B,2004ApJ...613..898T} based on Data Release 12 \citep{2015ApJS..219...12A} for which the stellar absorption is taken into account by subtracting a stellar component as determined from a population synthesis model.  We use {\it total} SFRs from the MPA/JHU catalog, for which SFRs within and outside the SDSS fiber aperture are combined.  The {\it in-fiber} SFRs are based on the extinction-corrected H$\alpha$ luminosities falling within the fiber \citep{2004MNRAS.351.1151B}, and SFRs outside the fiber are estimated from the SDSS photometry \citep[see][]{2007ApJS..173..267S}.  We convert SFRs to a Salpeter IMF from a \citet{2001MNRAS.322..231K} IMF by multiplying by a factor of 1.5 \citep{2004MNRAS.351.1151B}.

Stellar masses are derived from {\it ugriz} photometry with Le Phare (see \citealt{2011ApJ...730..137Z} for details) instead of using the JHU/MPA values.  Stellar masses from Le Phare are based on a Chabrier IMF and are smaller than those provided by MPA/JHU by approximately 0.2~dex on average with a dispersion between the two estimates of 0.14~dex.  For this study, the stellar masses are converted to a Salpeter IMF.

The SDSS galaxies are selected over a redshift range of $0.04<z<0.1$ to reduce the effects of redshift evolution.  \citet{2005PASP..117..227K} report that line measurements are highly biased towards the central area that may be more quenched when a covering fraction is less than 20\%.   To avoid such aperture effects, we impose a lower redshift limit of 0.04.  Furthermore, we require 5$\sigma$ detections for H$\alpha$ and $3\sigma$ for all other emission lines used in this paper ([N\,{\sc ii}]$\lambda 6584$, [S\,{\sc ii}]$\lambda \lambda$ 6717, 6731, H$\beta$, and [O\,{\sc iii}]$\lambda 5007$).  We classify galaxies as either star-forming or AGN as those below or above the classification line of \citet{2003MNRAS.346.1055K} in the BPT diagram (see Section \ref{sec:BPT}).  

\begin{figure}[htbp]
   \centering
   \includegraphics[width=3.5in]{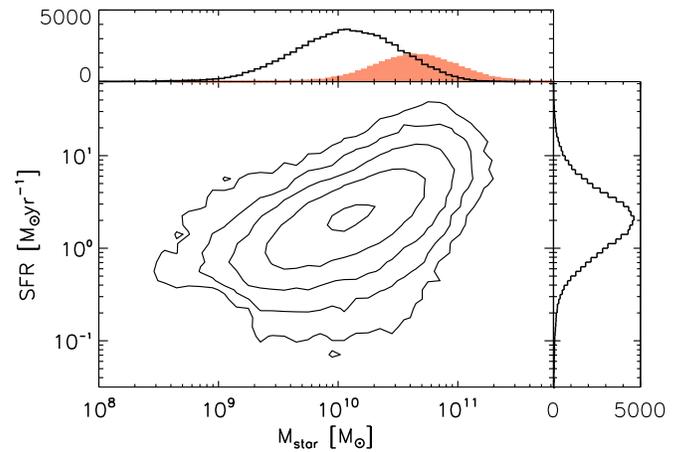} 
   \caption{$M_\ast$ vs. total SFR for local star-forming galaxies from SDSS.  Their distributions of stellar mass and SFR are shown by empty histograms.  Filled histogram indicates the stellar mass distribution of local AGNs.} 
   \label{fig:SDSS_MxSFR}
\end{figure}

The final sample of local sources consists of $80,003$ star-forming galaxies and $31,899$ AGNs.  In Figure \ref{fig:SDSS_MxSFR}, we show the distribution of $M_\ast$ and total SFR for the SDSS star-forming galaxies.  For AGNs, the stellar mass distribution is shown by the filled histogram.  The central 99 percentiles of the star-forming galaxies and AGNs include objects with $8.7 < \log M_\ast/ M_\odot < 11.1$ and $9.7 < \log M_\ast/ M_\odot < 11.6$, respectively.  We note that our results and conclusions do not change when a different selection is implemented, in which a higher limit is applied for only H$\alpha$ ($S/N>20$) without any S/N limit on the other lines to avoid possible biases concerning the detection of faint lines (i.e., H$\beta$, [N\,{\sc ii}], and [S\,{\sc ii}]).

To illustrate the average relations between line ratios and stellar mass of local star-forming galaxies in various diagrams, we split the local star-forming galaxies into the same eight/five stellar mass bins as the FMOS sample or into smaller, equally-spaced twenty-four bins of stellar mass for $10^{8.9} \le M_\ast/M_\odot \le 10^{11.3}$ (bin size $\Delta \log M_\ast=0.1~\mathrm{dex}$).  For fair comparisons with the FMOS stacked measurements, we define the average line ratios of the SDSS sample as follows.  In each bin, we derive the mean luminosities of individual emission lines and calculate the line ratios from these values.  To exclude extreme objects, we apply a $3\sigma$ clipping in log scales of the luminosities of both H$\alpha$ and [O\,{\sc iii}] lines in each bin, which removes approximately $2\%$ of the sample.  For the remainder of the paper, we refer to these average line ratios as the {\it stack} or {\it stacked line ratios} of the SDSS sample.

The FMOS stacked measurements include objects with no detection of some emission lines, while the SDSS sample consists of those having measurements of all line ratios of interest.  Hence it is potentially not straightforward to compare the stacked line ratios of our high-$z$ sample to the {\it median} or the distribution (e.g., ridge lines of the contours) of all the individual measurements in the reference sample.  In fact, the SDSS stacked line ratios are slightly offset from the median values of the individual line ratios or the ridge lines of the contours in various diagrams shown in this paper.  These offsets arise from the use of the mean luminosities which are biased towards more luminous, thus higher SFR objects, while the similar effects are potentially expected for the stacked measurements of our FMOS sample.  Therefore, this caveat should be kept in mind as long as one relies on the stacking analysis including undetected objects.

\section{Emission line properties}
\label{sec:results}
\subsection{BPT diagnostic diagram}
\label{sec:BPT}

\begin{figure*}[t]
   \centering
   \includegraphics[width=5.5in]{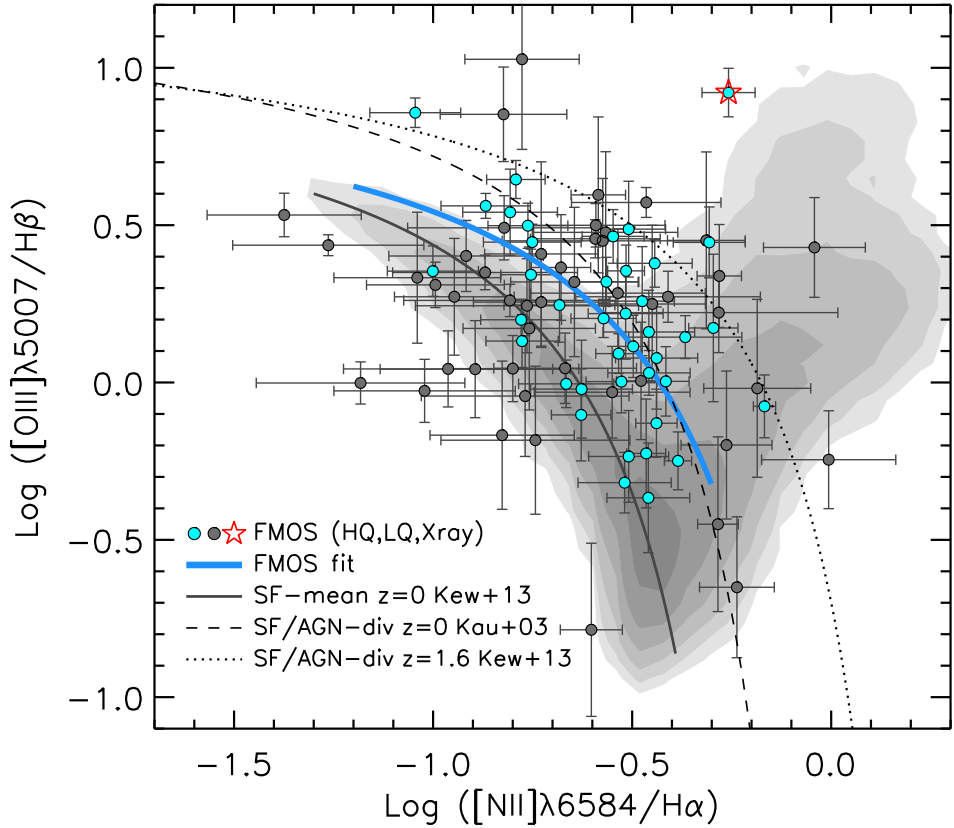} 
      \caption{The BPT diagnostic diagram: $\log ( [\text{O\sc iii}]/\mathrm{H\beta})$ versus $\log ([\text{N\sc ii}]/\mathrm{H\alpha})$.  Individual FMOS galaxies (87) are shown with circles for two groups: high- (HQ, cyan) and low- (LQ, dark gray) quality.  A single object detected in the X-ray band is indicated by a star.  Thin curves indicate the empirical abundance sequence of local galaxies given by Equation (\ref{eq:kew13}) (solid line: \citealt{2013ApJ...774..100K}), the empirical separation between star-forming galaxies and AGNs for the local SDSS sample (dashed line: \citealt{2003MNRAS.346.1055K}), and the empirical prediction of the division at $z=1.6$ (dotted line: \citealt{2013ApJ...774L..10K}).  Shaded contours show the distribution of the local sample (star-forming galaxies plus AGNs) in log scale.  Individual data points from FMOS show an offset from the local sequence, as illustrated by the best-fit (thick blue solid line; Equation \ref{eq:BPT}).}
   \label{fig:BPT}
\end{figure*}

We present in Figure \ref{fig:BPT} the BPT diagram for high-redshift galaxies from our FMOS sample (Sample-2).  The line ratios for 87 individual galaxies that have detections of all four lines are indicated by the data points.  We illustrate the quality of the measurements by splitting the sample into two groups: high-quality (HQ, $S/N>5$ for H$\alpha$ and $S/N>3$ for other relevant lines, i.e., [N\,{\sc ii}]$\lambda$6584, H$\beta$, and [O\,{\sc iii}]$\lambda$5007) and low-quality measurements (LQ, $S/N>3$ for H$\alpha$ and $S/N>1.5$ for others).  An absorption correction is applied to the observed H$\beta$ fluxes (see Section \ref{sec:abcorr}).  To aid in our interpretation of the high-redshift data, we plot the distribution of line ratios of local SDSS galaxies (star-forming population plus AGN; see Section \ref{sec:AGNrem}; gray contours).  We highlight the local abundance sequence of star-forming galaxies by a commonly used functional form \citep{2013ApJ...774..100K}:
\begin{equation}
\log ( [\text{O\sc iii}]/\mathrm{H\beta}) = \frac{0.61}{\log ([\text{N\sc ii}]/\mathrm{H\alpha})+0.08} + 1.10.
\label{eq:kew13}
\end{equation}

More than half of our FMOS galaxies deviate from the above local abundance sequence towards higher [O\,{\sc iii}]/H$\beta$ and/or higher [N\,{\sc ii}]/H$\alpha$ ratios, and a large number of objects are beyond a local demarcation between star-forming galaxies and AGNs \citep{2003MNRAS.346.1055K}.  Here it is important to highlight that our sample maps the distribution in the BPT diagram below $\log\left( \text{[O\,{\sc iii}]/H}\beta\right)=0$ down to $\sim -0.5$ as our sample includes a fairly large number of very massive galaxies (see Figure \ref{fig:sample}).  \citet{2013ApJ...774L..10K} derive a redshift evolution of the boundary to distinguish star-forming galaxies and AGNs.  We show this boundary at $z=1.6$ (thin dotted line) that is consistent with the location of the majority of the FMOS sources.  We remove FMOS sources located above this line as AGN candidates, and perform a fit to our sample with the same functional form as Equation (\ref{eq:kew13}), yielding
\begin{equation}
\log ( [\text{O\sc iii}]/\mathrm{H\beta}) = \frac{0.61}{\log ([\text{N\sc ii}]/\mathrm{H\alpha})-0.1336} + 1.081.
\label{eq:BPT}
\end{equation}
Here, the coefficient is fixed at the same value (0.61) as the local relation (Equation \ref{eq:kew13}).  The fit is shown as a blue thick line in Figure \ref{fig:BPT}.

\begin{figure*}[htbp]
   \centering
   \includegraphics[width=5.5in]{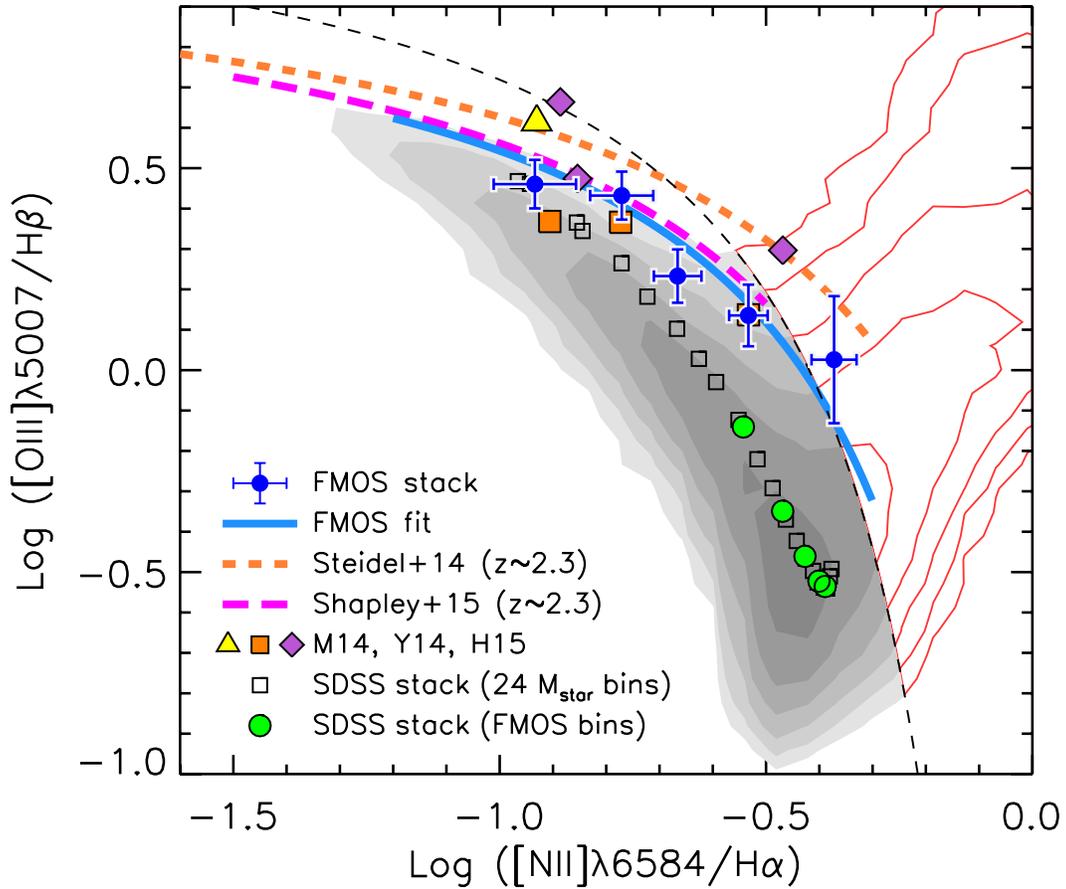} 
      \caption{BPT diagram with average measurements. FMOS stacked measurements (Sample-2; blue circles) are shown in five mass bins (with the median value increasing from left to right) with their associated errors.  The best-fit relation for the FMOS sample is reproduced from Figure \ref{fig:BPT} (blue thick line).  Shaded and red-line contours indicate local star-forming galaxies and AGNs, classified with the \citet{2003MNRAS.346.1055K} division line (thin dashed line), respectively.  For comparison, we show the stacked line ratios of the SDSS star-forming galaxies using the same mass bins as the FMOS sample (green circles) and smaller, equally-spaced mass bins (open squares) to illustrate the average locus of the local sample.  High-redshift samples are indicated as well (\citealt{2014ApJ...785..153M} -- yellow triangle; \citealt{2014MNRAS.437.3647Y} -- orange squares; \citealt{2015PASJ...67...80H} -- purple diamonds).  In addition, the best-fit relations at $z\sim2.3$ (\citealt{2014ApJ...795..165S} -- orange short-dashed line; \citealt{2015ApJ...801...88S} -- magenta long-dashed line) are shown.  The FMOS points are in good agreement with the fit to the individual points and indicate an offset of similar magnitude to other studies (see Section \ref{sec:BPT} for comparisons with other studies)}.
   \label{fig:BPTst}
\end{figure*}

Figure \ref{fig:BPTst}, the average measurements, based on the stacked spectra of Sample-2 in five bins of stellar mass (see Table \ref{tb:linemeasure}), are consistent with the individual data points and in good agreement with the best-fit relation thus confirming a clear offset from the local sequence.  For local SDSS galaxies, the stacked line ratios in twenty-four stellar mass bins for $10^{8.9} \le M_\ast/M_\odot \le 10^{11.3}$ (open squares) and in the same bins as the FMOS sample (green circles) are shown.  We note that the local points are slightly off from the ridge line of the contours due to the effects of taking the mean line luminosities (see Section \ref{sec:sdss}).  At the same stellar mass, the FMOS sample exhibits [N\,{\sc ii}]/H$\alpha$ ratios lower than those of local galaxies except for the most massive bin, while having much higher [O\,{\sc iii}]/H$\beta$ ratios ($>1$) over the entire stellar mass range.  \citet{2015ApJ...808...25S} find a similar behavior for galaxies with similar stellar masses ($M_\ast\gtrsim10^{10}~M_\ast$) when comparing local and $z\sim2$ galaxies.  Conversely, in comparison with the sequence of the SDSS stacked points (squares), the FMOS data point in the lowest mass bin ($M_\ast\sim10^{9.6\textrm{--}10}~M_\ast$) and the low-mass part of the best-fit relation are very close to local loci at low masses ($M_\ast\sim10^{9}~M_\ast$; see Section \ref{sec:BPTorigin} for further discussion).  In addition, we note that while the stacked measurements shown in Figure \ref{fig:BPTst} have $\log ([\text{O\sc iii}]/\mathrm{H\alpha}) >0$, a number of individual FMOS galaxies exist with lower [O\,{\sc iii}]/H$\beta$ values that are more consistent with the locus of local massive galaxies in the BPT diagram (Figure \ref{fig:BPT}).  Unbiased individual measurements of the line ratios are obviously essential to constrain the lower portion of the sequence of high-$z$ galaxies.

To further assess the robustness of the high-$z$ abundance sequence, we apply a more strict limit on the [N\,{\sc ii}]/H$\alpha$ ratio to remove potential AGNs.  Following \citet{2015ApJ...801..132V}, we decrease the limit on $\log \text{[N\,{\sc ii}]/H}\alpha$ from $-0.1$ (see Section \ref{sec:AGNrem}) to $-0.4$ \citep{2010MNRAS.403.1036C}.  With this restricted sample ($N_\mathrm{stack}=225$), stacked measurements also show a significant offset from the local abundance sequence, thus confirming our results.  Therefore, we conclude that the offset of the high-$z$ star-forming population cannot be explained by the effects of AGNs.

The offset in the BPT diagram has been previously reported in our FMOS studies \citep{2014ApJ...792...75Z,2015ApJ...806L..35K} and other works (e.g., \citealt{2005ApJ...635.1006S,2006ApJ...644..813E,2008ApJ...678..758L,2014ApJ...781...21N,2016ApJ...817...57C}) based on stacked and/or individual measurements.  \citet{2014ApJ...795..165S} find an offset using rest-frame UV-selected galaxies at $z\sim2.3$ from the KBSS-MOSFIRE survey, as shown in Figure \ref{fig:BPT} (orange dashed line), which lies above our relation as a whole.  \citet{2015ApJ...801...88S} also confirm the locus of a rest-frame optical selected galaxy sample at $z\sim2.3$ from the MOSDEF survey (magenta long-dashed line), although it is more similar to our FMOS sample than the KBSS results.  The difference between the results from these studies at $z\sim2$ is likely due to the difference in the adopted sample selection that results in varying typical ionization and excitation states.  The larger offset of the \citet{2014ApJ...795..165S} sample may be caused by the selection based on the rest-frame UV emission, which likely leads to higher [O\,{\sc iii}]/H$\beta$ ratios (see Section \ref{sec:MEx}).  \citet{2014ApJ...785..153M} measure the line ratios for highly star-forming galaxies at $z\sim2$ ($M_\ast\sim10^{8.5\text{--}9.5}~M_\odot$), which are consistent with the \citet{2014ApJ...795..165S} sample, as shown in Figure \ref{fig:BPTst}.

At $z\sim1.5$, three groups provide BPT measurements using Subaru/FMOS, including our program.  \citet{2014MNRAS.437.3647Y} measure the line ratios using the stacked, low-resolution ($R\sim600$) spectra of star-forming galaxies at $z\sim1.4$.  They use a sample selected based on stellar mass and H$\alpha$ flux, similar to our study, thus likely to be representative of the normal star-forming population at this epoch.  Their results are in good agreement with our stacked measurements as shown in Figure \ref{fig:BPTst} (orange squares).  \citet{2015PASJ...67...80H} construct a sample of [O\,{\sc ii}]$\lambda\lambda$3727, 3729 emitters at $z\sim1.5$ with an H$\alpha$ detection with FMOS.  Their sample shows a larger offset as compared to the Yabe et al., Shapley et al., and our samples, although it is likely to be more consistent with the Steidel et al. sample (purple diamonds in Figure \ref{fig:BPTst}).  The Hayashi et al. sample follows a $M_\ast$--SFR relation with a slope of 0.38, which is shallower than reported by other studies ($\sim0.7$--0.9, e.g., \citealt{2013ApJ...777L...8K,2014ApJ...795..104W}).  Consequently, the sample is biased towards a population having high specific SFR ($\mathrm{sSFR}=\mathrm{SFR}/M_\ast$) at $\log (M_\ast/M_\odot) \sim 9$, which is possibly responsible for their high [O\,{\sc iii}]/H$\beta$ ratios (i.e., a large BPT offset) in their sample.  The authors also mention the possibility that their [O\,{\sc ii}] emitter selection causes such a bias.  

After these studies, it remains to be understood whether the offset in the BPT diagram keeps increasing beyond $z\sim1.5\textrm{--}2$, or saturates.  In this respect, it is important to properly account for all possible section biases \citep[see e.g.,][]{2016ApJ...817...57C}.  Further investigation on the selection biases and the redshift evolution will be presented in a companion paper (S.~Juneau et al, in preparation).  Obviously, the presence of an offset indicates that high-$z$ star-forming galaxies have ionized gas with physical properties dissimilar to local galaxies.  In Section \ref{sec:BPTorigin}, we discuss several possible origins of the offset of high-$z$ galaxies from the local sequence, including variations in $q_\mathrm{ion}$, shape of the UV radiation from the ionizing sources, and gas density \citep[e.g.,][]{2014ApJ...785..153M,2015ApJ...801...88S,2015PASJ...67...80H,2015ApJ...812L..20K,2016ApJ...817...57C}.  

\subsection{Mass--Excitation (MEx) diagram}
\label{sec:MEx}

\begin{figure*}[tbp]
   \centering
   \includegraphics[width=5.5in]{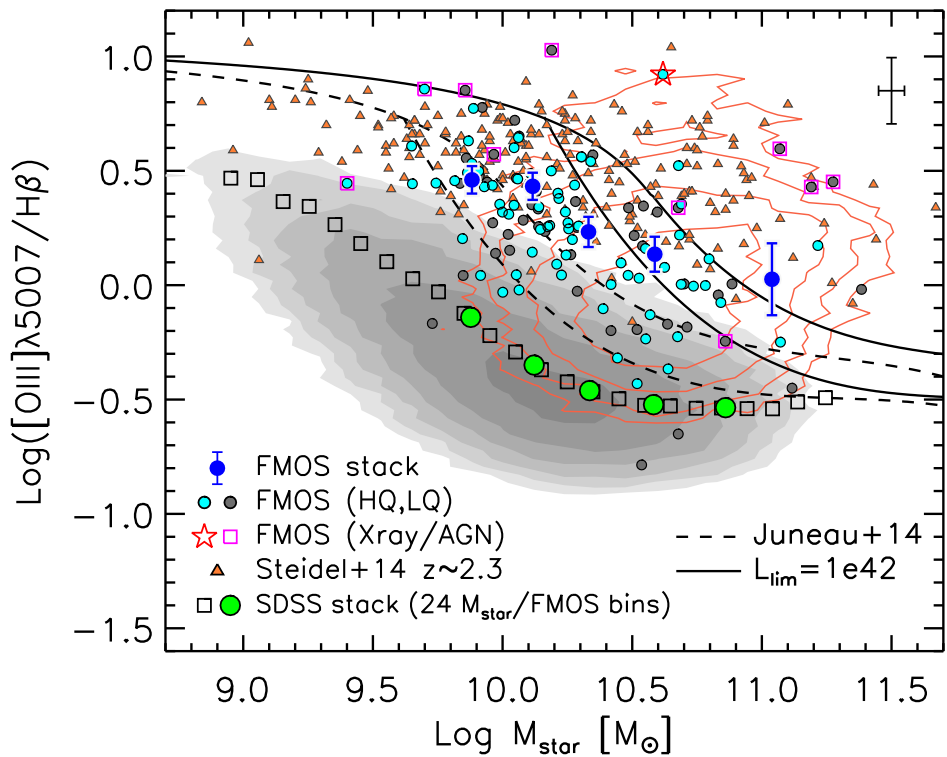} 
      \caption{The mass--excitation (MEx) diagram: $M_\ast$ vs. [O\,{\sc iii}]$\lambda5007$/H$\beta$.  Individual FMOS galaxies are shown with filled circles for two groups: high- (HQ; cyan) and low- (LQ; gray) quality.  Typical error bars of FMOS galaxies are shown in the upper-right corner.  AGN candidates are indicated by a star (X-ray detected source) and magenta squares (others, see Section \ref{sec:AGNrem}).  Large blue circles are average measurements of Sample-2 based on the stacked spectra in five mass bin.  Contours and symbols, representing the local sample, are the same as in Figure \ref{fig:BPTst}.  In addition, a sample at $z\sim 2.3$ from \citet{2014ApJ...795..165S} is shown with orange triangles.  Dashed curves indicate a demarcation between star-forming, composite objects, and AGNs, derived by \citet{2014ApJ...788...88J}.  The solid lines indicate these same boundaries for a luminosity-limited sample with $L_\mathrm{H\alpha}>10^{42}~\mathrm{erg~s^{-1}}$.  Our individual and stacked points show much higher [O\,{\sc iii}]/H$\beta$ ratios than those of local star-forming galaxies, and agree with the luminosity-adjusted boundary.}
   \label{fig:MEx}
\end{figure*}

To further illustrate the magnitude of the [O\,{\sc iii}]/H$\beta$ offset, we show in Figure \ref{fig:MEx} the mass-excitation (MEx) diagram ($M_\ast$ vs. [O\,{\sc iii}]/H$\beta$).  This diagram was introduced by \citet{2011ApJ...736..104J} to identify galaxies hosting an AGN at intermediate redshifts.  Local star-forming galaxies show a decline of the [O\,{\sc iii}]/H$\beta$ ratio with increasing stellar mass because more massive galaxies are more metal-rich and metal-line cooling becomes more efficient in such systems.  In contrast, AGNs present much higher ratios as compared to star-forming population, especially at $M_\ast \gtrsim 10^{10}~M_\odot$ (red contours in Figure \ref{fig:MEx}).

Figure \ref{fig:MEx} shows the [O\,{\sc iii}]$\lambda$5007/H$\beta$ ratios as a function of $M_\ast$ for 121 galaxies in our Sample-2 (filled circles).  For comparison, local star-forming galaxies are shown by shaded contours, with their median line ratios in stellar mass bins.  We highlight the divisions between star-forming galaxies, composite objects and AGNs, as derived by \citet{2014ApJ...788...88J} for local galaxies (dashed line).  It is clear that FMOS galaxies show higher [O\,{\sc iii}]/H$\beta$ ratios on average as compared to local galaxies at a fixed stellar mass, and that many individual measurements fall within regions classified as composite objects (between the two dashed lines) or AGN (above the upper line) although the error bars are typically too large to constrain the true category to which each galaxy belongs.   The average measurements, based on the stacked spectra of Sample-2 in five mass bins (see Table \ref{tb:linemeasure}), confirm a substantial offset towards higher [O\,{\sc iii}]/H$\beta$ ratios by $\sim0.5$~dex.  In addition, we note that the FMOS sample shows a clear negative correlation between stellar mass and line ratio, similar to local galaxies.

\citet{2014ApJ...788...88J} also derive the offset of the boundary as a function of luminosity threshold of the H$\alpha$ emission line (Equation B1 of \citealt{2014ApJ...788...88J}) to maximize the fraction of objects successfully classified as either star-forming or AGNs.  We shift the boundary by $\Delta \log M_\ast = +0.54~\mathrm{dex}$ (solid lines in Figure \ref{fig:MEx}), assuming an H$\alpha$ luminosity limit of $L_\mathrm{H\alpha}=10^{42}~\mathrm{erg~s^{-1}}$, which corresponds to $\mathrm{SFR}\sim 10~M_\ast~\mathrm{yr^{-1}}$ for our sample although the amount of the offset is less constrained at high luminosities ($L_\mathrm{H\alpha}>10^{40.7}~\mathrm{erg~s^{-1}}$; \citealt{2014ApJ...788...88J}).  The luminosity-adjusted boundary shows a good agreement with our FMOS galaxies.  The majority of the FMOS sources fall below the demarcation line thus classified as star-forming galaxies, while most of the remaining ones are classified as composite galaxies.  At the same time, half of potential AGNs (star and squares in Figure \ref{fig:MEx}) falls within the AGN-dominant region.  While our sample is not purely luminosity-limited, it is likely that the luminosity-dependent boundary derived based on local galaxies provides a good classification even for high redshift galaxies, as originally reported by \citet{2014ApJ...788...88J}.  A companion paper further investigates the locus of star-forming galaxies in the BPT and MEx diagrams as a function of the detection limits of emission-line luminosity (S.~Juneau et al. in preparation).

We also plot the measurements for $z\sim 2.3$ galaxies from \citet{2014ApJ...795..165S}.  They show significantly higher line ratios compared to local galaxies and even to our FMOS sample over the entire stellar mass range.  These higher [O\,{\sc iii}]/H$\beta$ ratios are essentially responsible for the larger offset in the BPT diagram (orange dashed line in Figure \ref{fig:BPT}).  These elevated ratios may be due to the redshift evolution of typical conditions (e.g., an enhanced amount of hotter stars; see \citealt{2014ApJ...795..165S}) and possibly to the UV-bright selection favoring such higher [O\,{\sc iii}]/H$\beta$ ratios.

\subsection{Diagnostics with [S\,{\sc ii}]/H$\alpha$}
\label{sec:BPT_S2}

An alternative BPT diagram compares the line ratios [S\,{\sc ii}]$\lambda\lambda$6717,6731/H$\alpha$ and [O\,{\sc iii}]$\lambda$5007/H$\beta$ to separate the star-forming population from the AGNs (hereafter ``[S\,{\sc ii}]-BPT'' diagram; \citealt{1987ApJS...63..295V}), although it does not provide as clear a division as that based on [N\,{\sc ii}]/H$\alpha$ \citep{2006MNRAS.372..961K,2009MNRAS.398..949P}.  Contrary to offsets seen in the standard BPT diagram at high redshift (Figure \ref{fig:BPT}), no systematic difference of the [S\,{\sc ii}]-BPT diagram for high-redshift galaxies has been reported \citep[e.g.,][]{2013ApJ...763..145D,2014ApJ...785..153M}.

\begin{figure*}[tbp]
   \centering
   \includegraphics[width=5.5in]{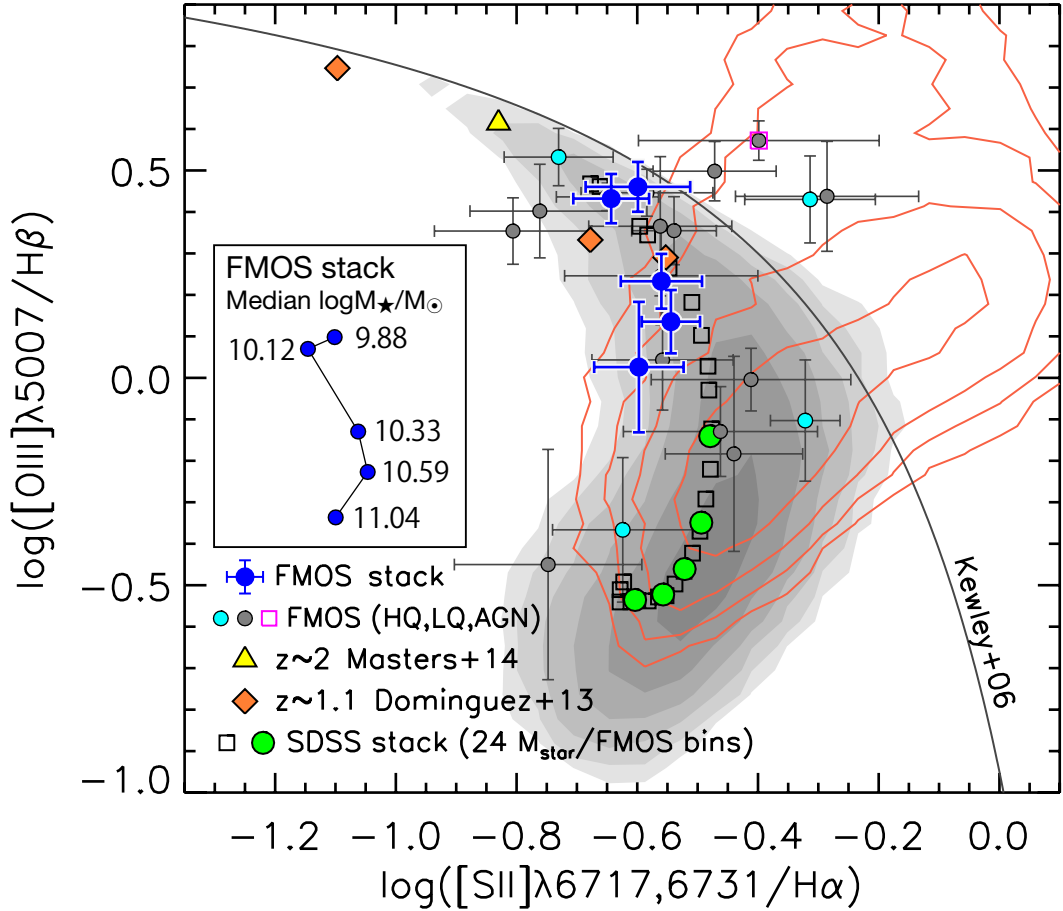} 
   \caption{[S\,{\sc ii}]-BPT diagram: [O\,{\sc iii}]$\lambda5007$/H$\beta$ vs. [S\,{\sc ii}]$\lambda \lambda 6717,6731$/H$\alpha$.  The FMOS sample is shown as filled circles for two groups: high- (HQ; cyan) and low- (LQ; gray) quality.  A magenta square marks an AGN candidate.  Large blue circles are stacked points in five mass bins (Sample-2) with the median masses given in the inset box.  Data from \citet{2013ApJ...763..145D} (diamond) and \citet{2014ApJ...785..153M} (triangle) are presented.  Contours and symbols, representing the local sample, are the same as in Figure \ref{fig:BPTst}.  A thin solid curve is a demarcation between local star-forming galaxies and AGNs \citep{2006MNRAS.372..961K}.  Our FMOS sample shows an offset towards lower [S\,{\sc ii}]/H$\alpha$ ratios at a fixed [O\,{\sc iii}]/H$\beta$ from the local average relation at $M_\ast\gtrsim10^{10.5}~M_\odot$, which suggests an enhancement in $q_\mathrm{ion}$ in high-$z$ galaxies (see Section \ref{sec:BPTorigin}).}
   \label{fig:BPT_S2}
\end{figure*}

In Figure \ref{fig:BPT_S2} we show 19 individual galaxies of Sample-2 and the average line ratios based on stacked spectra in five bins of stellar mass (Table \ref{tb:linemeasure}).  Four objects fall within the AGN region, above the classification line derived by \citet[thin solid line in Figure \ref{fig:BPT_S2}]{2006MNRAS.372..961K}.  One of them is identified as an AGN based on the [N\,{\sc ii}]-BPT plot (magenta square).  For comparison, we show the distribution of SDSS galaxies by contours and the stacked line ratios in mass bins (green circles and squares).  Similarly to Figure \ref{fig:BPTst}, the stacked points are slightly off from the ridge line of the contours.  From the individual measurements, the distribution appears to follow that of local star-forming galaxies; although, we are hampered by the limited sample size and errors on individual measurements.  On the other hand, the FMOS stacked data points indicate a departure towards lower [S\,{\sc ii}]/H$\alpha$ ratios compared to average loci of local galaxies (both the ridge line of the contours and the sequence of the stacked points) with increasing stellar mass at $M_\ast\gtrsim10^{10.5}~M_\odot$, while showing a similarity to local galaxies at lower masses.  We note that \citet{2015ApJ...801...88S} find similar [S\,{\sc ii}]/H$\alpha$ and [O\,{\sc iii}]/H$\beta$ ratios, indicative of a shift towards lower [S\,{\sc ii}]/H$\alpha$ values at high stellar masses, although the authors do not comment on it.  We further discuss the origins of the offset in the [S\,{\sc ii}]-BPT diagram in Section~\ref{sec:BPTorigin}.

\begin{figure}[tbp]
   \centering
   \includegraphics[width=3.6in]{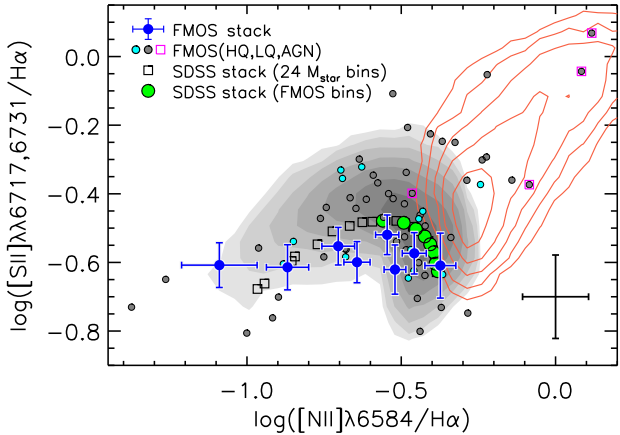} 
   \caption{[N\,{\sc ii}]$\lambda 6584$/H$\alpha$ vs. [S\,{\sc ii}]$\lambda \lambda 6717,6731$/H$\alpha$.  The FMOS sample is shown as filled circles for two groups: high- (HQ, cyan) and low- (LQ, gray) quality.  Typical error bars are shown in the lower-right corner.  Magenta squares mark AGN candidates.  The stacked points are based on the spectra of Sample-1 in eight stellar mass bins (large blue circles).  Contours and symbols, representing the local sample, are the same as in Figure \ref{fig:BPTst}, except that the green circles show the stacked ratios in the eight $M_\ast$ bins as the FMOS sample.  Our sample shows no specific trend between these line ratios, unlike the local sample (see Section \ref{sec:BPT_S2}).}
   \label{fig:N2xS2}
\end{figure}

An additional diagnostic relation for emission-line galaxies is based on the line ratios [S\,{\sc ii}]$\lambda \lambda 6717,6731$/H$\alpha$ and [N\,{\sc ii}]$\lambda$6584/H$\alpha$.  This diagram was first introduced by \citet{1977A&A....60..147S} to distinguish between planetary nebulae, H\,{\sc ii} regions, and supernova remnants.  In past years, this diagram has also been used to separate star-forming galaxies from AGNs \citep{2009A&A...495...53L,2010A&A...519A..31L}.  In Figure \ref{fig:N2xS2}, we show 61 FMOS galaxies and average measurements based on the stacked spectra in eight mass bins (Sample-1; Table \ref{tb:linemeasure}).  Individual galaxies are in general agreement with the distribution of local star-forming galaxies (shaded contours).  In comparison with the local average points (open squares), the FMOS stacked measurements have slightly lower [S\,{\sc ii}]/H$\alpha$ ratios at a given [N\,{\sc ii}]/H$\alpha$ with the exception of those at both the highest and lowest masses.  The FMOS stacked points occupy a narrow range of $\log$[S\,{\sc ii}]/H$\alpha\sim-0.6$, and have no specific trend with [N\,{\sc ii}]/H$\alpha$ or stellar mass, as seen in local galaxies.  Our results differ slightly from \citet{2015PASJ...67..102Y} that have lower [S\,{\sc ii}]/H$\alpha$ ratios over $-1.1< \log(\text{[N\,{\sc ii}]}/\mathrm{H\alpha}) <-0.6$.  Finally, we see that there is minimal contribution of AGNs to our stacked measurements while a few individual sources are possibly present (see Appendix \ref{sec:AGNcontami} for a full assessment on AGNs within our sample).

\subsection{Electron density}
\label{sec:eleden}

The electron density $n_\mathrm{e}$ can be measured from the intensity ratio [S\,{\sc ii}]$\lambda $6717/[S\,{\sc ii}]$\lambda$6731.  Here, we convert the [S\,{\sc ii}] doublet ratio into $n_\mathrm{e}$ using the TEMDEN routine from the NEBULAR package of STSDAS/IRAF, assuming a fixed electron temperature of $T_\mathrm{e}=10^4~\mathrm{K}$.  While this temperature is commonly assumed for typical H\,{\sc ii} regions, the electron temperature is sensitive to the gas-phase metallicity and can vary in the range $\sim 5000$--$20000~\mathrm{K}$ (e.g., \citealt{1983A&A...127..211W,2013ApJ...765..140A}).  We note that the temperature dependence of the electron density, derived from [S\,{\sc ii}] lines, is weak around this choice (maximally $\lesssim 0.2$~dex; \citealt{2000A&A...357..621C}).  The line ratio is sensitive to the electron density over the range $10 \lesssim n_\mathrm{e}/\mathrm{cm^3} \lesssim 5\times 10^3$, while it saturates at upper ([S\,{\sc ii}]$\lambda $6717/[S\,{\sc ii}]$\lambda$6731$\sim 1.45$) and lower ($\sim 0.45$) limits for lower and higher electron densities, respectively.  The doublet lines are well resolved with the high-resolution of FMOS (see Figure \ref{fig:spec}).  Due to the close spacing of these lines in wavelength, the measurements are free from uncertainties in the absolute flux calibration and not impacted by dust extinction.  However, no single galaxy in the FMOS sample has a significant measurement of the line ratio to constrain the electron density.  

\begin{figure}[tbp]
   \centering
   \includegraphics[width=3.6in]{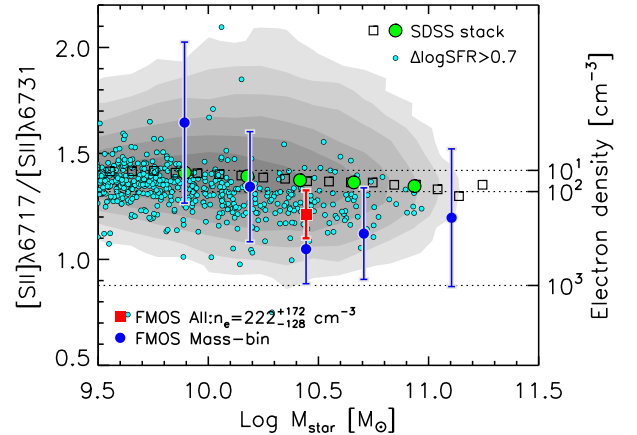} 
   \caption{The [S\,{\sc ii}]$\lambda 6717$/[S\,{\sc ii}]$\lambda 6731$ ratio as a function of stellar mass.  The corresponding electron densities are given on the right vertical axis and by the horizontal dashed lines ($n_\mathrm{e}=10,10^2$ and $10^3~\mathrm{cm^{-3}}$).  Blue filled circles indicate average measurements based on the stacked spectra in five bins of stellar mass (Sample-1), and a red square marks the measurement based on the entire sample ($n_\mathrm{e}=222^{+172}_{-128} ~\mathrm{cm^{-3}}$).  Contours and symbols, representing the local sample, are the same as in Figure \ref{fig:BPTst}.  Cyan circles indicate local sources with an offset from the local star-forming main sequence $\Delta \log(\mathrm{SFR}) \ge 0.7~\mathrm{dex}$ (see Section \ref{sec:eleden}).}
   \label{fig:Sii}
\end{figure}

In Figure \ref{fig:Sii}, we present the [S\,{\sc ii}]$\lambda6717$/[S\,{\sc ii}]$\lambda6731$ ratios based on the stacked spectra (Sample-1) in five mass bins (blue circles) and of the entire sample (red square), and the median values of the local sample (squares and green circles).  The average electron density of the full sample is $n_\mathrm{e}=222^{+172}_{-128} ~\mathrm{cm^{-3}}$, which is higher than the average electron density, derived from [S\,{\sc ii}] lines, in local galaxies as shown in Figure \ref{fig:Sii} ($n_\mathrm{e} = 10\text{--}100~\mathrm{cm^{-3}}$; \citealt{2008MNRAS.385..769B}).  We split the sample into mass bins, but do not find any significant trend with mass due to large uncertainties on our measurements.

Moreover, we compare our measurement with local galaxies having high SFRs that match those of our FMOS sample.  We select local star-forming galaxies with $\Delta \log \mathrm{SFR}>0.7~\mathrm{dex}$, where $\Delta \log \mathrm{SFR}$ is defined as the difference between the observed SFR and the typical value of main-sequence galaxies at a given stellar mass \citep{2007A&A...468...33E}.  Figure \ref{fig:Sii} shows that such high-SFR local galaxies are biased towards a lower [S\,{\sc ii}] doublet ratio (i.e., a higher electron density) that is more similar to that of our high-$z$ sample, as compared to the entire local sample.  

Our result is consistent with findings reported in the literature.  For example, \citet{2014ApJ...787..120S} find that the electron density of ionized gas in high-$z$ star-forming galaxies ($z\approx 2.4\text{--}3.7$) is enhanced as compared to matched local galaxies, selected to have the same stellar mass and sSFR.  \citet{2015MNRAS.451.1284S} measure the electron densities of 14 H$\alpha$ emitters at $z=2.5$ by the line ratio [O\,{\sc ii}]$\lambda$3729/$\lambda$3726, and find a median electron density to be $291~\mathrm{cm^{-3}}$.  \citet{2016ApJ...816...23S} find a median $n_\mathrm{e}=225_{-4}^{+119}~\mathrm{cm^{-3}}$ from the [O\,{\sc ii}] doublet and $n_\mathrm{e}=290_{-169}^{+88}~\mathrm{cm^{-3}}$ from the [S\,{\sc ii}] doublet for a sample at $z\sim 2.3$ from the MOSDEF survey.  \citet{2016ApJ...822...42O} find $\sim10^{2\textrm{--}3}~\mathrm{cm^{-3}}$ for star-forming galaxies at $z\approx3.3$.  Therefore, we can safely conclude that H\,{\sc ii} regions in high-$z$ galaxies have an electron density a few to several times higher than that of local galaxies on average.

\subsection{The line ratio [N\,{\sc ii}]/[S\,{\sc ii}]}
\label{sec:MxN2S2}

\citet{2002ApJS..142...35K} investigate the feasibility of the line ratio [N\,{\sc ii}]$\lambda$6584/[S\,{\sc ii}]$\lambda\lambda$6717,6731 as an abundance diagnostic.  Sulphur is one of the $\alpha$-elements, which include e.g., O, Ne, Si, produced through {\it primary} nucleosynthesis in massive stars and supplied to the ISM through type-II supernovae.  In contrast, nitrogen is generated through both the primary process as well as a {\it secondary} process, where $^{12}$C and $^{16}$O initially contained in stars are converted into $^{14}$N via the CNO cycle.  Therefore, the [N\,{\sc ii}]/[S\,{\sc ii}] ratio is sensitive to the total chemical abundance, particularly in a regime where secondary nitrogen production is predominant, while this ratio is almost constant if most of nitrogen has a primary origin (see e.g., Figure 4 of \citealt{2002ApJS..142...35K}).  The advantage of the use of [N\,{\sc ii}]/[S\,{\sc ii}] is that their wavelengths are separated by only $\sim140~\textrm{\AA}$, hence they are able to be observed simultaneously and the line ratio is nearly free from the effect of dust extinction (typically $<0.03~\mathrm{dex}$).  Here we neglect dust extinction, and define
\begin{equation}
N2S2 = \log \left( [\text{N\,{\sc ii}}]\lambda6584/[\text{S\,{\sc ii}}]\lambda \lambda6717,6731 \right).
\label{eq:N2S2}
\end{equation}

\begin{figure}[htbp]
   \centering
   \includegraphics[width=3.6in]{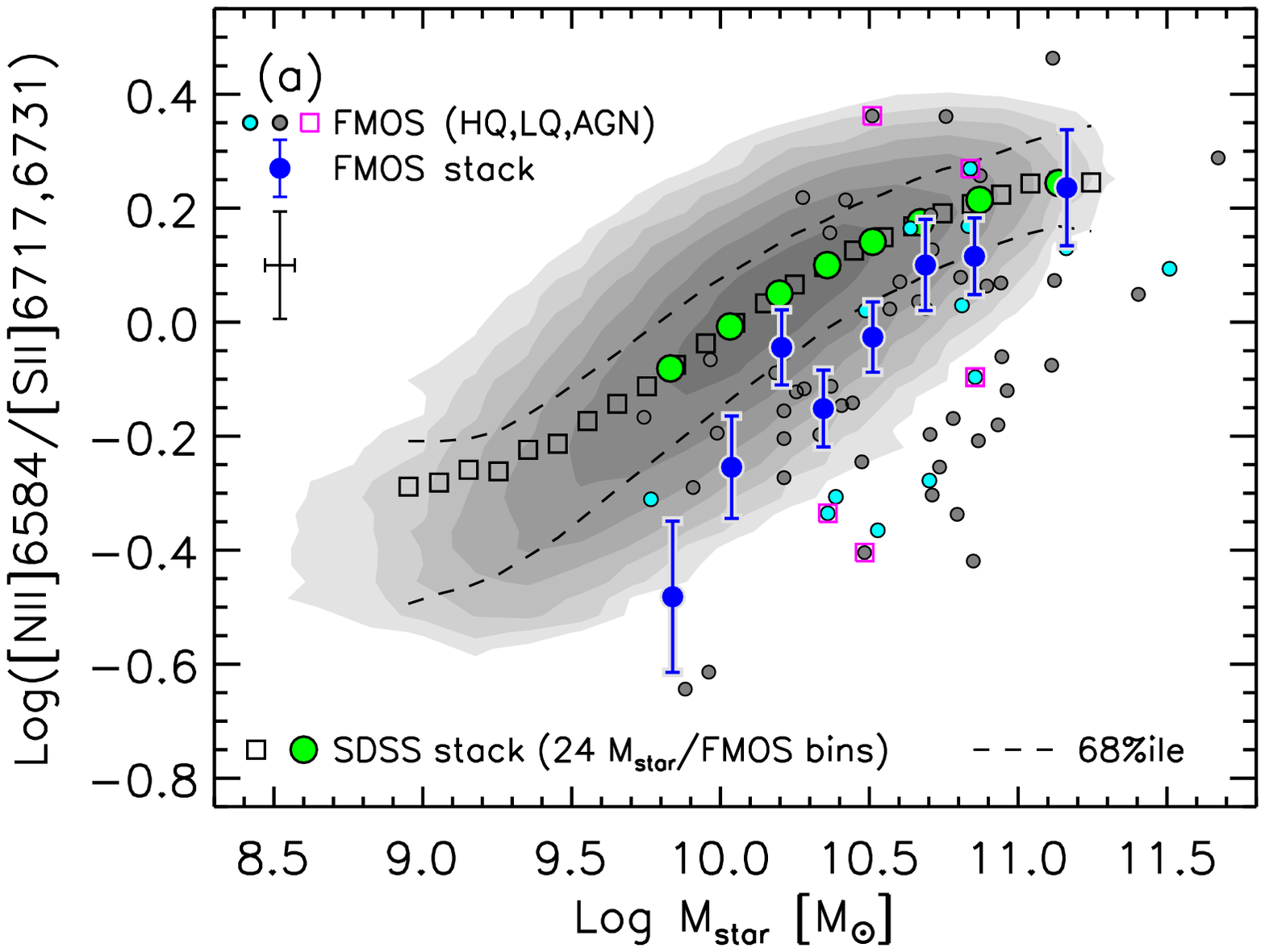} 
   \includegraphics[width=3.6in]{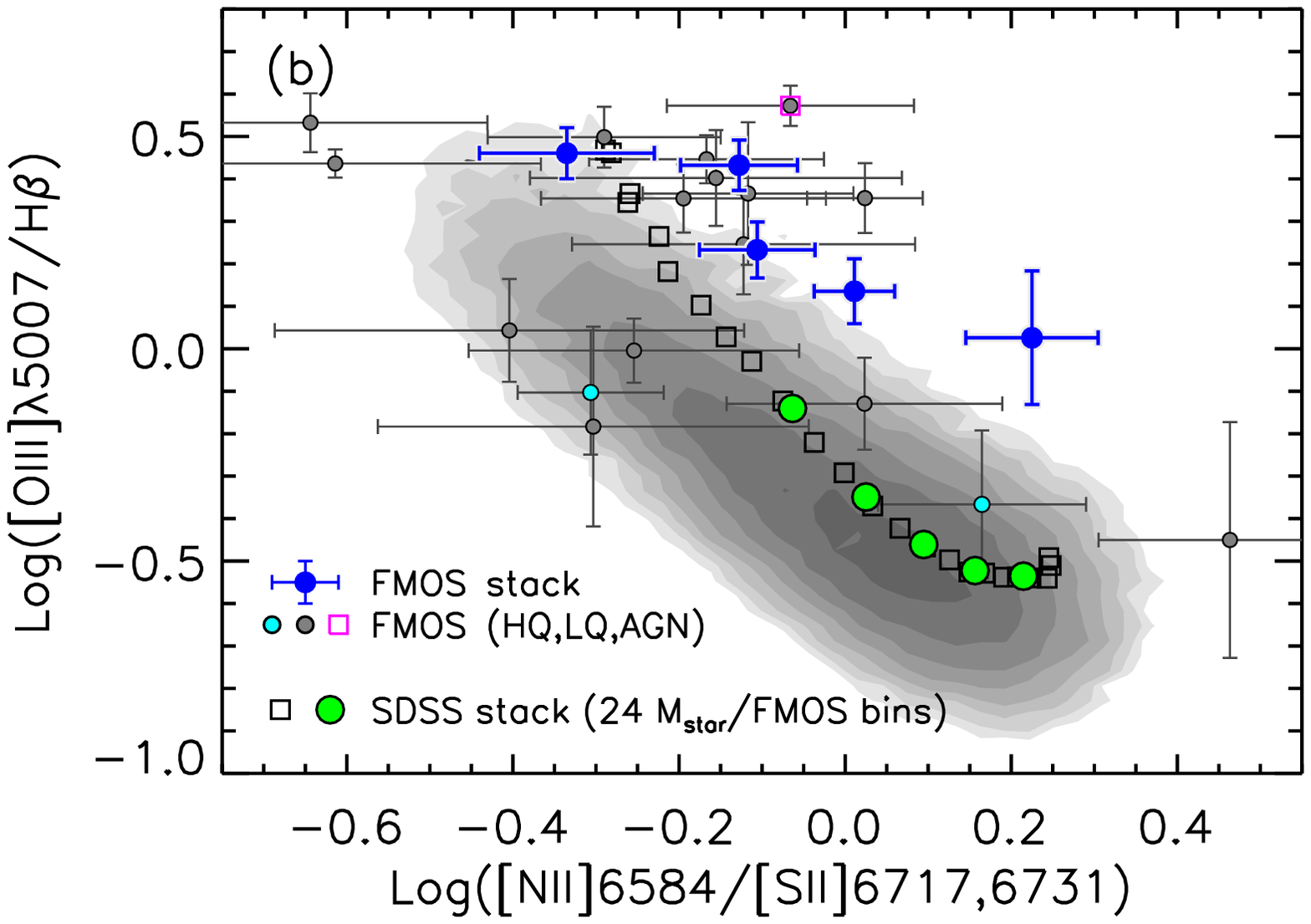} 
   \caption{{\it Panel (a)}: The [N\,{\sc ii}]$\lambda 6584$/[S\,{\sc ii}]$\lambda \lambda 6717,6731$ ratio, a proxy of the N/O ratio (thus the chemical enrichment; see Section \ref{sec:MxN2S2}), as a function of stellar mass.  The FMOS galaxies are shown with filled circles separately for high- (HQ; cyan) and low- (LQ; gray) quality groups.  Typical error bars are shown in the upper-left corner.  The stacked data points are based on co-added spectra of Sample-1 in eight mass bins (blue circles).  {\it Panel (b)}: [N\,{\sc ii}]$\lambda 6584$/[S\,{\sc ii}]$\lambda \lambda 6717,6731$ vs. [O\,{\sc iii}]$\lambda 5007/\mathrm{H\alpha}$.  The FMOS galaxies are shown as filled circles.  A magenta square indicates an AGN candidate.  The average measurements are based on the stacked spectra of Sample-2 in five mass bins (blue circles).  The stellar mass increases from the left to right.  The offset of high-$z$ galaxies to higher [O\,{\sc iii}]/H$\beta$ at fixed [N\,{\sc ii}]/[S\,{\sc ii}] indicates that they have a higher excitation state than local galaxies with a similar degree of chemical enrichment (see Section \ref{sec:MxN2S2}).  In both panels, contours and symbols, representing the local sample, are the same as in Figure \ref{fig:BPTst}, except that the green circles show the values in the eight mass bins in Panel (a).}
   \label{fig:MxN2S2}
\end{figure}

Figure \ref{fig:MxN2S2}{\it a} shows the [N\,{\sc ii}]/[S\,{\sc ii}] ratio as a function of stellar mass for local star-forming galaxies and for the FMOS sample.  The local sample shows a clear correlation between the line ratio and stellar mass above $M_\ast \sim10^{9.5} M_\odot$ and reaches $N2S2\sim0.3$ at the massive end ($\gtrsim10^{11}~M_\odot$), as illustrated by the median points (squares).  This tight correlation reflects the regime where secondary nitrogen production is dominant.  We find that 61 individual FMOS galaxies with a $N2S2$ measurement show a significant correlation with stellar mass at $>99\%$ confidence level with a Spearman's rank correlation coefficient $\rho=0.41$.  The trend of these individual measurements are well represented by the stacked measurements based on the Sample-1 co-added spectra in eight mass bins (blue circles; see Table \ref{tb:linemeasure}).  Therefore, it is likely that the ISM is enriched to a level that nitrogen is dominantly produced by the secondary process for the majority of our sample.  The $N2S2$ reaches to the same level as local galaxies at the massive end ($\gtrsim10^{11}M_\odot$) while, on average, galaxies with $M_\ast<10^{11}~M_\odot$ have lower $N2S2$ values, thus lower secondary-to-primary element abundance ratios, than found in local galaxies at the same stellar mass.  This trend is analogous to the mass--metallicity relation as discussed in Section \ref{sec:metallicity}.

We note that the [N\,{\sc ii}]/[S\,{\sc ii}] ratio is relatively insensitive to the change of $q_\mathrm{ion}$, although the values of [N\,{\sc ii}]/H$\alpha$ and [S\,{\sc ii}]/H$\alpha$ depend on $q_\mathrm{ion}$.  According to calculations of photoionization by \citet[][see Figure 20]{2013ApJS..208...10D}, the [N\,{\sc ii}]/[S\,{\sc ii}] ratio is nearly invariant of $q_\mathrm{ion}$ over $\log q_\mathrm{ion} / (\mathrm{cm~s^{-1}}) \sim 7.0\textrm{--}8.5$ at lower ratios ($N2S2\sim-0.5$; i.e., at lower masses), while the ratio increases maximally by $\lesssim 0.1~\mathrm{dex}$ at at higher ratios ($N2S2\sim0$; i.e., at high masses) for a 0.5~dex increase in $q_\mathrm{ion}$ (from $\log q_\mathrm{ion} / (\mathrm{cm~s^{-1}}) = 7.0$ to $7.5$).  While an enhancement of $q_\mathrm{ion}$ of $\lesssim0.5~\mathrm{dex}$ is expected to arise in high-$z$ galaxies (see Section \ref{sec:BPTorigin}), such a weak dependence of $N2S2$ on $q_\mathrm{ion}$ does not affect the trend seen in Figure \ref{fig:MxN2S2}.

Figure \ref{fig:MxN2S2}{\it b} compares the line ratios [O\,{\sc iii}]/H$\beta$ and [N\,{\sc ii}]/[S\,{\sc ii}].  The stacked measurements, based on the Sample-2 co-added spectra in five mass bins, clearly show higher [O\,{\sc iii}]/H$\beta$ ratios at a given [N\,{\sc ii}]/[S\,{\sc ii}] (and {\it vice versa}) as compared to local galaxies.  The majority of the individual points also fall above the local relation, in agreement with the stacked data points.  This characteristic of high-$z$ star-forming galaxies is very similar to that found in the \citet{2015ApJ...801...88S} sample at $z\sim2.3$.  As discussed extensively in the following sections, this fact qualitatively indicates that high-$z$ star-forming galaxies tend to maintain higher excitation state for their stage of chemical enrichment, than local galaxies.

We also highlight that the local SDSS galaxies have [N\,{\sc ii}]/[S\,{\sc ii}] ratios that begins to flatten below $\sim10^{9.5} M_\odot$ (Figure \ref{fig:MxN2S2}a).  This flattening may arise from the transition between primary and secondary processes as being the predominant mechanism.  However, it is possible that selection effects, concerning the detection of faint emission lines (i.e., [N\,{\sc ii}], [S\,{\sc ii}]), may affect the [N\,{\sc ii}]/[S\,{\sc ii}] ratio at lower masses.  Further investigation of this feature is beyond the scope of this work.

\section{Physical origin of changes in emission-line properties}
\label{sec:BPTorigin}

An explanation of the changes in emission-line properties of high-$z$ galaxies is essential to understand the characteristics of their ionized gas in star-forming regions.  In particular, it is important to specify the primary cause(s) of the offset in the BPT diagram, which has been reported by many studies \citep[e.g.,][]{2005ApJ...635.1006S,2006ApJ...644..813E,2008ApJ...678..758L,2014ApJ...781...21N,2014ApJ...785..153M,2014ApJ...792...75Z,2014MNRAS.437.3647Y,2014ApJ...795..165S,2015ApJ...806L..35K,2015PASJ...67...80H,2016ApJ...817...57C}.  Several factors have been suggested and discussed, including the metal abundance, ionization parameter, hardness of the EUV radiation field, and gas density (pressure).  A higher $q_\mathrm{ion}$ facilitates the ionization of singly-ionized elements to a doubly-ionized state, thus increasing the [O\,{\sc iii}]/H$\beta$ ratio, while decreasing [N\,{\sc ii}]/H$\alpha$ and [S\,{\sc ii}]/H$\alpha$ \citep{2002ApJS..142...35K}.   Alternatively, a harder ionizing radiation field has also been considered as an origin of the BPT offset that causes an increase of the three line ratios given above \citep{2010AJ....139..712L,2015ApJ...812L..20K}.  Recently, \citet{2016ApJ...826..159S} suggest the presence of harder stellar spectra due to the existence of metal-poor and massive binary systems.  In addition, a higher H\,{\sc ii} gas density will increase the three line ratios due to the higher rate of collisional excitation of metal ions.  As a result, the BPT abundance sequence will be shifted towards the upper right (e.g., \citealt{2008MNRAS.385..769B}).  \citet{2013ApJ...774..100K} summarize the effects of varying ISM properties on the star-forming locus in the BPT diagram.

\subsection{Comparison with theoretical models}

\begin{figure}[tbp]
   \centering
   \includegraphics[width=3.5in]{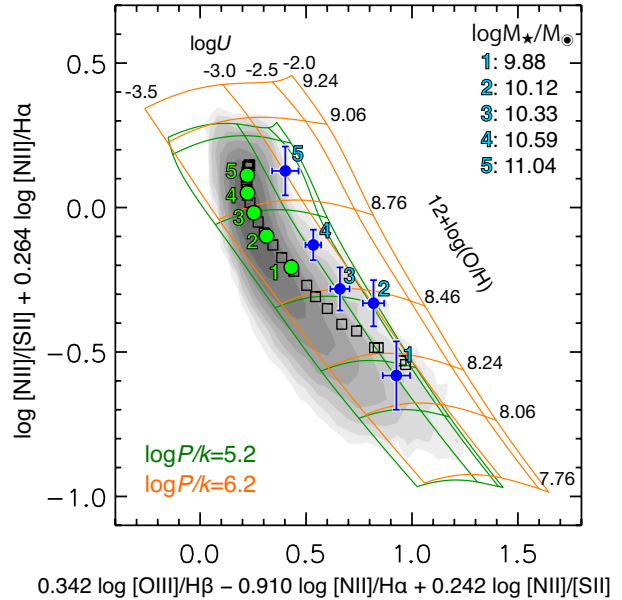} 
   \caption{Comparison of combinations of line ratios, as presented in Figure 2 of \citet{2016Ap&SS.361...61D}, with theoretical model grids (\texttt{MAPPINGS V}) for varying metallicity $Z$ ($12+\log(\mathrm{O/H})$), ionization parameter ($U=q_\mathrm{ion}/c$), and gas pressure ($P$).  Average measurements of the FMOS sample are based on stacked spectra in five stellar mass bins (Sample-2; blue circles, labeled as \#1--5).  The median $M_\ast$ are given in the upper right corner.  Shaded contours show local star-forming galaxies, with their stacked points in equally-spaced 24 mass bins ($10^{8.9} \le M_\ast/M_\odot \le 10^{11.3}$; squares) and in the same five mass bins as the FMOS sample (green circles, labeled as \#1--5).  Based on the theoretical model grids, our FMOS sample shows a $\sim0.3\textrm{--}0.4~\mathrm{dex}$ enhancement in $q_\mathrm{ion}$ as compared to local galaxies over the entire stellar mass range.}
   \label{fig:D16Fig2}
\end{figure}

\begin{figure*}[tbp]
   \centering
   \includegraphics[width=6.6in]{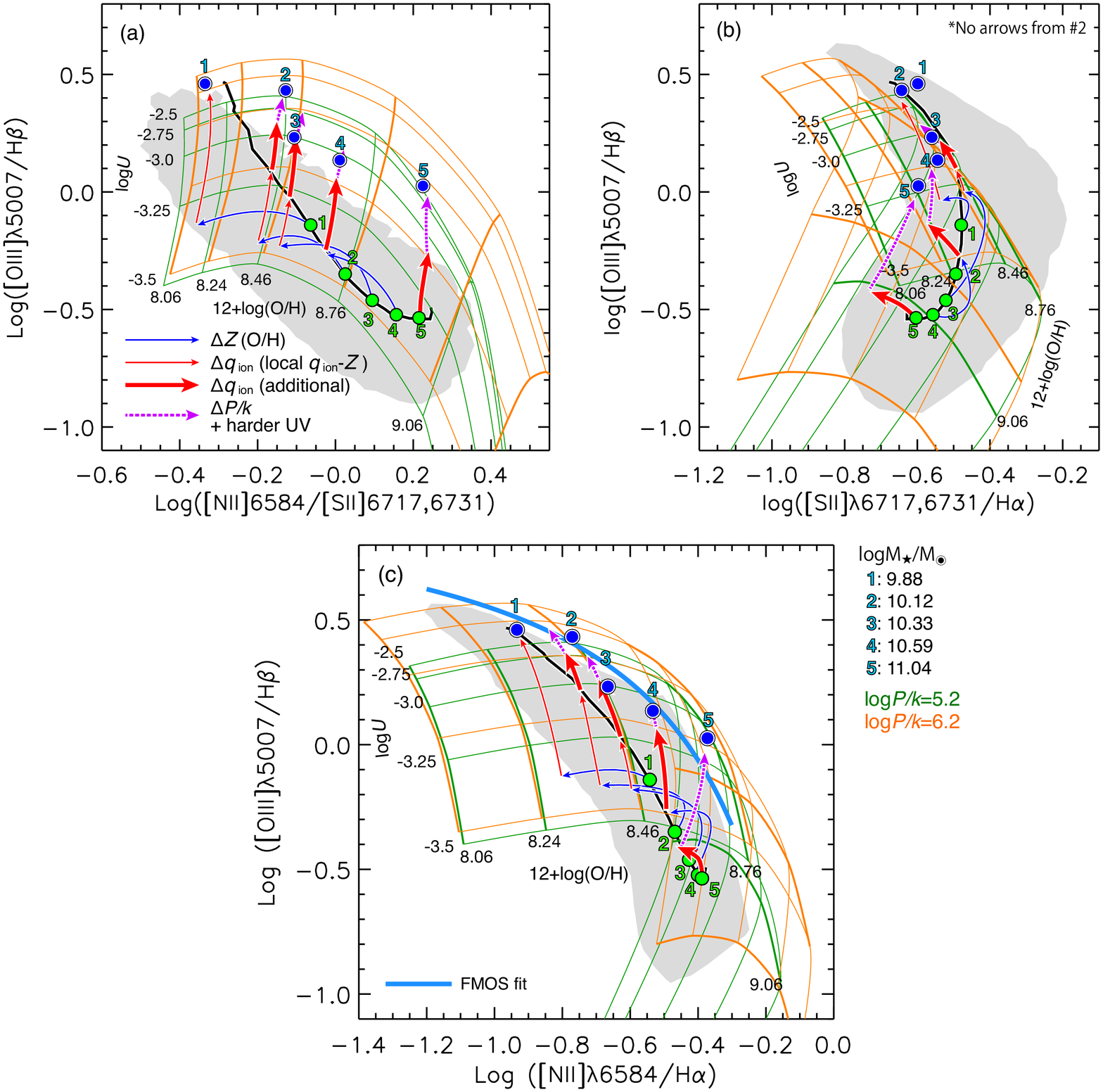} 
   \caption{Illustration of the magnitude of effects produced by changes in each physical factor in the [N\,{\sc ii}]/[S\,{\sc ii}] vs. [O\,{\sc iii}]/H$\beta$ (a), the [S\,{\sc ii}]-BPT (b), and the BPT (c) diagrams.  The average data points of our FMOS and local samples, and the theoretical model are the same as in Figure \ref{fig:D16Fig2} except that the sequence of the stacked points in 24 $M_\ast$ bins (squares in Figure \ref{fig:D16Fig2}) is simply expressed by a black solid line.  The median $M_\ast$ and the gas pressures of the models are shown outside the frame.  Arrows indicate corresponding changes in the line ratios for the each factor separately (blue thin arrows -- decreases in metallicity; red thin arrows -- enhancement of $q_\mathrm{ion}$ due to the local $q_\mathrm{ion}$--$Z$ anti-correlation; red thick arrows -- excessive enhancement of $q_\mathrm{ion}$; magenta dotted arrows -- remaining effects possibly attributed to high gas pressure and harder radiation field).  A shaded region indicates an outline of distribution of local star-forming galaxies. }
   \label{fig:model}
\end{figure*}

We aim to determine the impact of each physical effect on changes in the location of high-$z$ galaxies in emission-line diagrams by comparing our observational data with theoretical models.  We use the photoionization code \texttt{MAPPINGS V}\footnote{https://miocene.anu.edu.au/mappings}.  In Figure \ref{fig:D16Fig2}, we make use of a new diagnostic tool, recently proposed by \citet{2016Ap&SS.361...61D}, using combinations of [N\,{\sc ii}]$\lambda$6584/H$\alpha$, [N\,{\sc ii}]$\lambda$6584/[S\,{\sc ii}]$\lambda\lambda$6717,6731, and [O\,{\sc iii}]$\lambda$5007/H$\beta$ to clearly separate the dependence on metallicity ($Z$) from that on the ionization parameter ($U=q_\mathrm{ion}/c$) at a given gas pressure ($P$) (see Figure 2 of \citealt{2016Ap&SS.361...61D}).  The line ratios are computed as a function of $U$ and $Z$ at gas pressures of $\log P/k=5.2$ or $6.2~\mathrm{cm^{-3}~K}$.  In Figure \ref{fig:model}, we compare the local and high-$z$ line ratios with the model grids in the [N\,{\sc ii}]/[S\,{\sc ii}] vs. [O\,{\sc iii}]/H$\beta$ (panel a), [S\,{\sc ii}]-BPT (b), and the BPT (c) diagrams.  In each panel, we consider the individual physical effects that can be responsible for the differences in the line ratios between low-$z$ and high-$z$ galaxies at a given stellar mass.  The travels of the line ratios from low-$z$ to high-$z$ due to different physical factors are schematically illustrated by various arrows.  In each diagram in Figure \ref{fig:model}, their direction and magnitude are prescribed based on the model grid by comparing the loci, i.e., the coordinates ($\log(\textrm{O/H}), \log U$), of the stacked data points at both epochs in the same mass bins (low-$z$ -- green; high-$z$ -- blue circles).  Although the theoretical calculations do not fully reproduce the observed line ratios with precision, they do provide qualitative insight into the relation between the line ratios and physical parameters.  In the following subsections, we describe the effects of each physical factor in detail.

\subsubsection{Metallicity}

First, we consider the changes in line ratios due to changes in metallicity.  In Figure \ref{fig:D16Fig2}, a comparison between the local and our FMOS samples shows that high-$z$ galaxies have the same metallicity ($y$-axis) as the local sample at the massive end ($M_\ast>10^{10.8}~M_\odot$), while metallicities of lower mass galaxies are lower than those of local galaxies of the same stellar mass\footnote{We further investigate the mass--metallicity relation of our sample based on the [N\,{\sc ii}]/H$\alpha$ ratio and the theoretical calibration from \citet{2016Ap&SS.361...61D} in Section \ref{sec:metallicity}.}.  Such a decline in $Z$ for high-$z$ galaxies with $M_\ast\lesssim10^{10.8}~M_\odot$ (binned data labelled as \#1--4 in the panel) contributes significantly to changes in line ratios.   In Figure \ref{fig:model}, shifts, corresponding to a decline in $Z$, are shown by thin blue arrows (pointing in the direction of decreasing $Z$) while the ionization parameter is fixed.  As evident, a decrease in metallicity does not move galaxies towards high-$z$ loci, hence we conclude that metallicity cannot be the primary cause of the offset between local and our high-$z$ galaxies. 

\subsubsection{Ionization parameter}

We investigate the hypothesis that the ionization parameter is higher in star-forming galaxies at high redshifts as compared to those at low redshift.  As shown in the MEx diagram (Figure \ref{fig:MEx}), high-$z$ galaxies have elevated [O\,{\sc iii}]/H$\beta$ ratios as compared to local galaxies over the full range of stellar mass.  Higher [O\,{\sc iii}]/H$\beta$ ratios are also seen at a given stage of chemical enrichment, as indicated by [N\,{\sc ii}]/[S\,{\sc ii}] (Figure \ref{fig:MxN2S2}{\it b}).  These higher [O\,{\sc iii}]/H$\beta$ ratios can be produced by a higher $q_\mathrm{ion}$.  Evidence of an enhancement in $q_\mathrm{ion}$ comes from a shift towards lower [S\,{\sc ii}]/H$\beta$ in the [S\,{\sc ii}]-BPT diagram (Figure \ref{fig:BPT_S2}); a higher $q_\mathrm{ion}$ leads to the ionization of $\mathrm{S}^+ \rightarrow \mathrm{S}^{++}$ that decreases [S\,{\sc ii}]/H$\alpha$.  Therefore, an increase in $q_\mathrm{ion}$ likely plays an important role in causing the offset in the BPT diagram.  

In Figure \ref{fig:D16Fig2}, we see an indication, based on a comparison with theoretical models, that our FMOS galaxies have ionization parameters $\log q_\mathrm{ion}/c \sim -3$, approximately $0.3\textrm{--}0.4~\mathrm{dex}$ higher, on average, than that of local galaxies of the same stellar mass, even if a higher gas pressure is considered.  This is consistent with previous studies (e.g., \citealt{2014ApJ...787..120S,2014MNRAS.442..900N,2015PASJ...67...80H,2016ApJ...822...42O}).  Figure \ref{fig:D16Fig2} also suggests that the FMOS galaxies have higher $q_\mathrm{ion}$ as compared to local galaxies at fixed metallicity except for the lowest mass bin (labelled as \#1) which shows the level in the ionization parameter similar to the low-mass local galaxies ($M_\ast\sim10^9~M_\odot$).

To evaluate the effects of varying $q_\mathrm{ion}$ on emission-line ratios, it is important to consider the anti-correlation between $q_\mathrm{ion}$ and $Z$ seen in local galaxies at lower stellar masses ($\lesssim 10^{10}~M_\odot$) \citep[e.g.,][]{2000ApJ...542..224D,2016arXiv160503436K}.  Based on model grids, the local average measurements (squares) show an increase of $q_\mathrm{ion}$ with decreasing $Z$ at $12+\log(\mathrm{O/H})\lesssim 8.6$ (Figure \ref{fig:model}a,c).  Following a change in line ratios due to $Z$, we show in Figure \ref{fig:model} a shift corresponding to a change in $q_\mathrm{ion}$ due to the local $q_\mathrm{ion}$--$Z$ anti-correlation (narrow red arrows), with a magnitude set to match the local relation at a given metallicity (squares).  There is no shift for the two highest mass bins (circles labelled as \#4--5) due to small or no change in metallicity.  It is clear that these changes in $q_\mathrm{ion}$ accompanying the metallicity evolution are not fully responsible for the offsets of high-$z$ galaxies.  The remaining offsets evidently require a further enhancement in $q_\mathrm{ion}$ (thick red arrows) that moves galaxies towards higher [O\,{\sc iii}]/H$\beta$ and constitutes the most important origin of the offset.  However, the size of this increase in $q_\mathrm{ion}$ cannot be fully constrained as other effects may also cause an additional shift in a similar direction (see the following subsection).  Here the thick red arrow corresponds to a scaling up of about $0.3~\mathrm{dex}$ at fixed O/H for all mass bins, except for the lowest one (\#1), as being estimated from Figure \ref{fig:D16Fig2}.  The contribution from this additional $q_\mathrm{ion}$ excess seems to be relatively more important for higher mass bins.  In contrast, we note that, for the lowest mass bin (labelled as \#1), the entire enhancement in $q_\mathrm{ion}$ could be explained by the one coincident with the metallicity change (i.e., due to the local $q_\mathrm{ion}$--$Z$ anti-correlation), thus does not need the additional excess.

In panel (b), the shift towards both lower [S\,{\sc ii}]/H$\alpha$ and higher [O\,{\sc iii}]/H$\beta$ due to higher $q_\mathrm{ion}$ is clearly shown, especially, in the most massive bin; it is apparent that the only factor that can produce a shift towards the left is an enhancement in $q_\mathrm{ion}$.  Over the stellar mass range of the FMOS sample, the magnitude of the shift is not fully accounted for by only an enhancement of $q_\mathrm{ion}$ that is expected from the $q_\mathrm{ion}$--$Z$ relation (narrow red arrows).  Therefore, an additional enhancement of the ionization parameter is required to fully produce the offset of high-$z$ galaxies in the emission-line diagrams (thick red arrows).  

\subsubsection{Additional effects (ISM gas density and hardness of the radiation field)}
\label{sec:additionaleffects}

While a higher [O\,{\sc iii}]/H$\beta$ ratio likely indicates an enhancement of $q_\mathrm{ion}$, additional factors may contribute to higher ratios since [O\,{\sc iii}]/H$\beta$ increases with a higher gas density and a harder ionizing radiation field.  The significance of these additional effects is shown in the [S\,{\sc ii}]-BPT diagram (Figure \ref{fig:model}b).  While a higher $q_\mathrm{ion}$ will shift the line ratios towards the upper left (as illustrated by the thick red arrows, mass bins \#4 and \#5 in particular), an additional offset towards the upper right is needed (as highlighted by the magenta dotted arrows) to match the line ratios of high-$z$ galaxies.  We attribute the latter effect to a higher gas pressure, supported by the measurements indicative of a higher electron density (see Section \ref{sec:eleden}), as demonstrated by the model grids of different $P/k$.  However, Figure \ref{fig:model} suggests that the remaining shift in line ratios (in particular, the most massive bin labelled as \#5) is unlikely to be fully explained even by a ten times higher gas pressure.  Hence we surmise that there may be additional help from a harder ionizing radiation field to fully match the offset, although the current data cannot ascertain the size of such a contribution.  Here, for mass bins \#2 and \#3, we indicate a shift corresponding to about ten times higher $P/k$, as expected from higher electron density we measure (Section \ref{sec:eleden}).  Meanwhile, there is no additional shift required for the lowest mass bin (\#1) as its line-ratio offset can fully explained by the $q_\mathrm{ion}$ excess due to the lower O/H (see the previous subsection).  We note that it is challenging to quantitatively distinguish the contribution from a higher $q_\mathrm{ion}$ (red thick arrows) and additional factors (magenta dotted arrows) especially at lower stellar masses because the direction of changes in emission-line ratios are roughly similar for different physical effects.  In addition, we caution the reader that the model line ratios and arrows in Figure \ref{fig:model} may be affected by systematic uncertainties.
\\

To conclude, we find that it is likely that all three factors contribute to changes in the emission-line ratios at high redshift.  Based on our observations and comparisons with the models, we argue that the emission-line properties of high-$z$ galaxies are a result of the following: (1) a higher $q_\mathrm{ion}$ as compared to local galaxies of the same stellar mass with an enhancement in $q_\mathrm{ion}$ being higher than expected from the $q_\mathrm{ion}$--$Z$ anti-correlation seen in local galaxies, (2) higher gas density (pressure), and (3) an additional effect possibly attributed to a hardening of the ionizing radiation field.  While arguing for the existence of multiple effects, we conclude that an elevation of the ionization parameter {\it at fixed metallicity} is the dominant factor influencing the emission-line ratios of high-$z$ galaxies.  In addition, we highlight that the emission-line ratios of FMOS galaxies in the lowest mass bin ($M_\ast\sim10^{9.6\textrm{--}10}~M_\odot$) can be explained by only the enhancement in $q_\mathrm{ion}$ accompanying the change in metallicity without invoking additional effects.  These facts likely indicate that the excitation properties of the ISM in such low-mass high-$z$ galaxies can be commonly seen in local galaxies with about ten times lower masses ($M_\ast\sim10^9~M_\odot$), while the situation in higher-mass high-$z$ galaxies ($M_\ast>10^{10}~M_\odot$) are dissimilar from ordinary local galaxies at any stellar mass.

\subsection{What causes an enhancement of ionization parameter?}

Following the above discussion, we consider how the characteristics of H\,{\sc ii} regions in high-$z$ star-forming galaxies are different from those in local galaxies.  If a fixed inner radius of an ionized gas sphere is assumed, the ionization parameter is inversely proportional to the gas density.  With the higher gas density of our sample (Section \ref{sec:eleden}), a higher $q_\mathrm{ion}$ requires an increase in the production rate of ionizing photons.  An enhancement of the ionizing photon flux in individual H\,{\sc ii} regions could be simply expected to occur when a larger number of stars form in each gas cloud as compared to local galaxies \citep{2015ApJ...812L..20K}.  This corresponds to a higher star formation efficiency (SFE), such that $\mathrm{SFR}=\mathrm{SFE}\times M_\mathrm{gas}$ (where $M_\mathrm{gas}$ is the mass of molecular gas) and indeed the SFE is higher at higher redshift \citep[e.g.,][]{2015ApJ...800...20G}.  Furthermore, a higher ionizing photon flux could also arise due to harder stellar spectra, which induces additional changes in line ratios, as discussed above (Section \ref{sec:additionaleffects}).  A hardening of the overall stellar radiation field could result from a top-heavy IMF that increases the relative amount of hot massive stars, a metal-poor stellar population \citep[see e.g.,][]{2010AJ....139..712L}, or from massive binaries, which have a lifetime on the main sequence that is longer than that of single stars \citep[see][and references therein]{2016ApJ...826..159S}.  \citet{2016ApJ...826..159S} suggest that, at high redshifts, the stellar atmosphere of ionizing massive stars have a lower Fe/O abundance ratio than the solar value, as a result of chemical enrichment being dominated by Type II supernovae.  Since the photospheric line blanketing is mainly due to iron, the Fe depletion results in a lower opacity for ionizing photons at fixed O/H, thus leading to a higher ionizing parameter and/or harder ionizing radiation field for the circumstellar environment.  Examination of these various possible origins needs not only measuring the ISM properties, but also further understanding the properties of stellar component, for which very deep spectroscopy in optical (rest-frame UV) is especially important.

The geometrical properties of gas clouds and star-forming regions could also affect the physical properties of the ionized gas, thus the emission-line ratios.  Given high sSFRs of high-$z$ galaxies, it is considered likely that an H\,{\sc ii} region is close to, or overlapping with, another ionized gas region.  In such cases, ionizing photons from adjacent H\,{\sc ii} regions effectively increase $q_\mathrm{ion}$.  The model photoionization calculations used above assume an `ionization-bounded' H\,{\sc ii} region, for which the radius is determined by the ionization equilibrium (St{\"o}mgren radius).  In contrast, `density-bounded' H\,{\sc ii} regions are entirely ionized and have a radius determined by the size of the gas cloud.  In such a situation, the sizes of a singly-ionized (i.e., [N\,{\sc ii}], [S\,{\sc ii}]) and the hydrogen recombination regions are smaller than the ionization-bounded cases, while the doubly-ionized region (i.e., [O\,{\sc iii}]) is less affected.  As a consequence, the [O\,{\sc iii}]/H$\beta$ ratio will increase, while the [N\,{\sc ii}]/H$\alpha$ and [S\,{\sc ii}]/H$\alpha$ ratios decrease, as compared with the ionization-bounded case.  These changes in the line ratios will be interpreted as a result of an enhancement of $q_\mathrm{ion}$ \citep[e.g.,][]{2013ApJ...774..100K,2014MNRAS.442..900N}.  However, it remains completely unclear how common density-bounded H\,{\sc ii} regions are in both local and higher redshift star-forming galaxies.

\begin{figure}[tbp]
   \centering
   \includegraphics[width=3.5in]{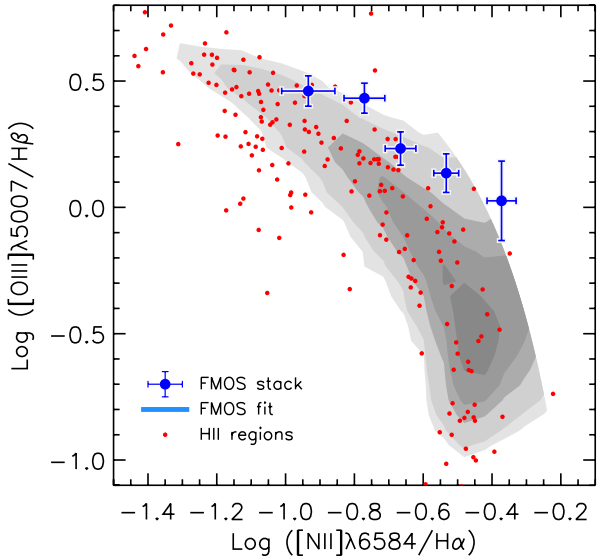} 
   \caption{Comparison of line ratios (BPT diagram) for FMOS galaxies (average values as in Figure \ref{fig:BPTst}) and local H\,{\sc ii} regions (\citealt{1998AJ....116.2805V}; red circles).  Local star-forming galaxies are shown by gray contours.}
   \label{fig:vanZee}
\end{figure}

Alternatively, we consider the effects from low density ($\sim0.1~\mathrm{cm^{-3}}$) diffuse ionized gas (DIG) in galaxies, which is widely distributed in the interstellar space and is ionized by radiation from hot massive stars \citep[e.g.,][]{1984ApJ...282..191R,1992ApJ...392L..35R,1994ApJ...428..647D}.  In a galaxy-wide spectrum, the nebular emission consists of the combination of emission from individual H\,{\sc ii} regions and the DIG component.  The DIG is known to have a very low $q_\mathrm{ion}$, as compared to denser H\,{\sc ii} regions, as indicated by weak [O\,{\sc iii}] lines \citep{1994ApJ...428..647D}.  Therefore, the contribution from DIG may decrease an effective value of $q_\mathrm{ion}$ estimated from the galaxy-wide flux, although such effects have not been well understood yet.  Low-$z$ galaxies could be expected to have a larger DIG contribution to the integrated flux than high-$z$ galaxies, which may be more similar to pure H\,{\sc ii} regions owing to their high SFRs.  We test the hypothesis that the changes in the emission-line ratios are caused by the change of the DIG contribution in the galaxy-wide emission.  In Figure \ref{fig:vanZee}, our FMOS galaxies are compared with individual H\,{\sc ii} regions in local spiral galaxies \citep{1998AJ....116.2805V}.  The data points of H\,{\sc ii} regions trace well the BPT abundance sequence of local galaxies based on the galaxy-wide spectra, and thus the offset of our FMOS galaxies remains even in comparison with pure H\,{\sc ii} regions.  This apparently means that a contribution from DIG does not explain the changes in the observed emission-line ratios between local and high-$z$ galaxies, thus rejecting the above working hypothesis.

\subsection{Enhancement of N/O}
\label{sec:NOenhance}
Previous studies have reported a lack of any significant offset between local and high-$z$ galaxies in the [S\,{\sc ii}]-BPT diagram, as an indication that the offset in the BPT diagram is caused by an effect other than a higher $q_\mathrm{ion}$ or a higher ionizing radiation field \citep{2014ApJ...785..153M,2015ApJ...801...88S}.  An elevated nitrogen-to-oxygen (N/O) abundance ratio at a given O/H has been suggested as such a possible cause of the BPT offset towards higher [N\,{\sc ii}]/H$\alpha$ at a fixed [O\,{\sc iii}]/H$\beta$ (e.g., \citealt{2014ApJ...785..153M,2016ApJ...828...18M,2014ApJ...795..165S,2015ApJ...801...88S,2015PASJ...67..102Y,2016ApJ...816...23S,2016ApJ...817...57C}).  In contrast to these studies, we find an offset in the [S\,{\sc ii}]-BPT diagram and explain the emission-line ratios of high-$z$ galaxies without invoking a change in the N/O--O/H relation.  A quantitative constraint would require a metallicity determination that is independent of the N abundance (e.g., direct $T_\mathrm{e}$ or $R_\mathrm{23}$ index) and a reliable measurement of N/O for high-$z$ galaxies.  While such studies are not feasible with the primary FMOS spectroscopic sample, future [O\,{\sc ii}]$\lambda \lambda 3726,3729$ follow-up observations will permit us to further investigate these issues.

We highlight a recent study by \citet{2016ApJ...826..159S} that find by using a composite spectrum of $z\sim2.4$ galaxies that an average N/O and $T_\mathrm{e}$-based O/H are in excellent agreement with the N/O-O/H relation of local extragalactic H\,{\sc ii} regions \citep{2012MNRAS.421.1624P}.  Therefore, it is conceivable that an enhancement of N/O is not the main factor of the BPT offset at high redshifts.  However, we caution that the N/O ratio measured by \citet{2016ApJ...826..159S} is higher at fixed O/H when compared to a `galaxy-wide' relation derived by \citet{2013ApJ...765..140A}.  Therefore, it still remains unsettled whether the N/O--O/H relation evolves with redshift or not, though to some extent it should, given the different timescales of nitrogen and oxygen release by stellar nucleosynthesis.

\section{Revisiting the mass--metallicity relation}
\label{sec:metallicity}

We previously reported on the mass--metallicity (MZ) relation based on our FMOS program \citep{2014ApJ...792...75Z} using a smaller subset of galaxies (162) than presented in this work.  Our MZ relation indicated that the most massive galaxies at $z\sim1.6$ are fully mature similar to local massive galaxies while the lower mass galaxies are less enriched as compared to local galaxies at a fixed stellar mass.  Here, we revisit the MZ relation using a sample four times larger than presented in \citet{2014ApJ...792...75Z}.  In particular, the number of massive ($M_\ast>10^{11}~M_\odot$) galaxies in the sample is larger by an order-of-magnitude.  

\subsection{Empirical metallicity determination using [N\,{\sc ii}]/H$\alpha$}
\label{sec:MxN2}

We use the [N\,{\sc ii}]/H$\alpha$ ratio to evaluate the gas-phase metallicity, i.e., the oxygen abundance, $12+\log(\text{O/H})$.  The advantage of using this ratio is the close spectral proximity of the lines thus not requiring any correction for extinction.  The high-resolution mode of FMOS cleanly separates the two lines.  However, the [N\,{\sc ii}]/H$\alpha$ ratio is sensitive to the ionization parameter and hardness of the ionizing radiation field.  A $0.5~\mathrm{dex}$ increase in $q_\mathrm{ion}$ produces a decrease in [N\,{\sc ii}]/H$\alpha$, thus leading to a $\sim0.4~\mathrm{dex}$ underestimate of metallicity \citep{2002ApJS..142...35K}.  Furthermore, metallicities based on the [N\,{\sc ii}] line depend on the relation between the N/O and O/H ratios.  Any empirically or theoretically calibrated metallicity indicator involving a nitrogen line implicitly rests on the assumption of a universal (locally calibrated) N/O vs. O/H relation.  However, the universality of the relation is still under debate both at low and high redshifts.  For instance, the accretion of a substantial amount of metal-poor gas could reduce a galaxy's O/H while leaving its N/O largely unchanged, causing a deviation from the local relation \citep[see][]{2016ApJ...823L..24K,2016ApJ...828...18M}.  Indeed, an enhancement of N/O at fixed O/H has been advocated to explain the offset in the BPT diagram of high-$z$ galaxies (see Section \ref{sec:NOenhance}), although we do not invoke the change of the N/O vs. O/H relation as an explanation for the emission-line properties of high-$z$ galaxies \citep[see also][]{2016Ap&SS.361...61D}.  It is clear that a direct calibration of the relation between N/O and O/H is required to improve upon metallicity determinations of high-$z$ galaxies.

With these caveats in mind, we estimate the metallicities of both local galaxies and the FMOS sample based on a locally-calibrated relation.  The line ratio is converted to metallicity as given in \citet{2008A&A...488..463M}:
\begin{equation}
N2 = -0.7732 + 1.2357x -0.2811x^2 -0.7201x^3 -0.3330x^4
\label{eq:Z_N2}
\end{equation}
where $N2=\log(\text{[N\,{\sc ii}]}\lambda6584/\mathrm{H\alpha})$ and $x=12+\log (\mathrm{O/H}) - 8.69$.  This relation is nearly linear over the metallicity range of interest ($8<12+\log (\mathrm{O/H})<9$).  On the other hand, the line ratio begins to saturate at higher metallicity (above $N2\sim-0.3$) as a result of efficient metal cooling, which leads to a decrease in the collisional excitation rate of N$^+$ \citep{2006agna.book.....O}.

\begin{figure}[htbp]
   \centering
   \includegraphics[width=3.6in]{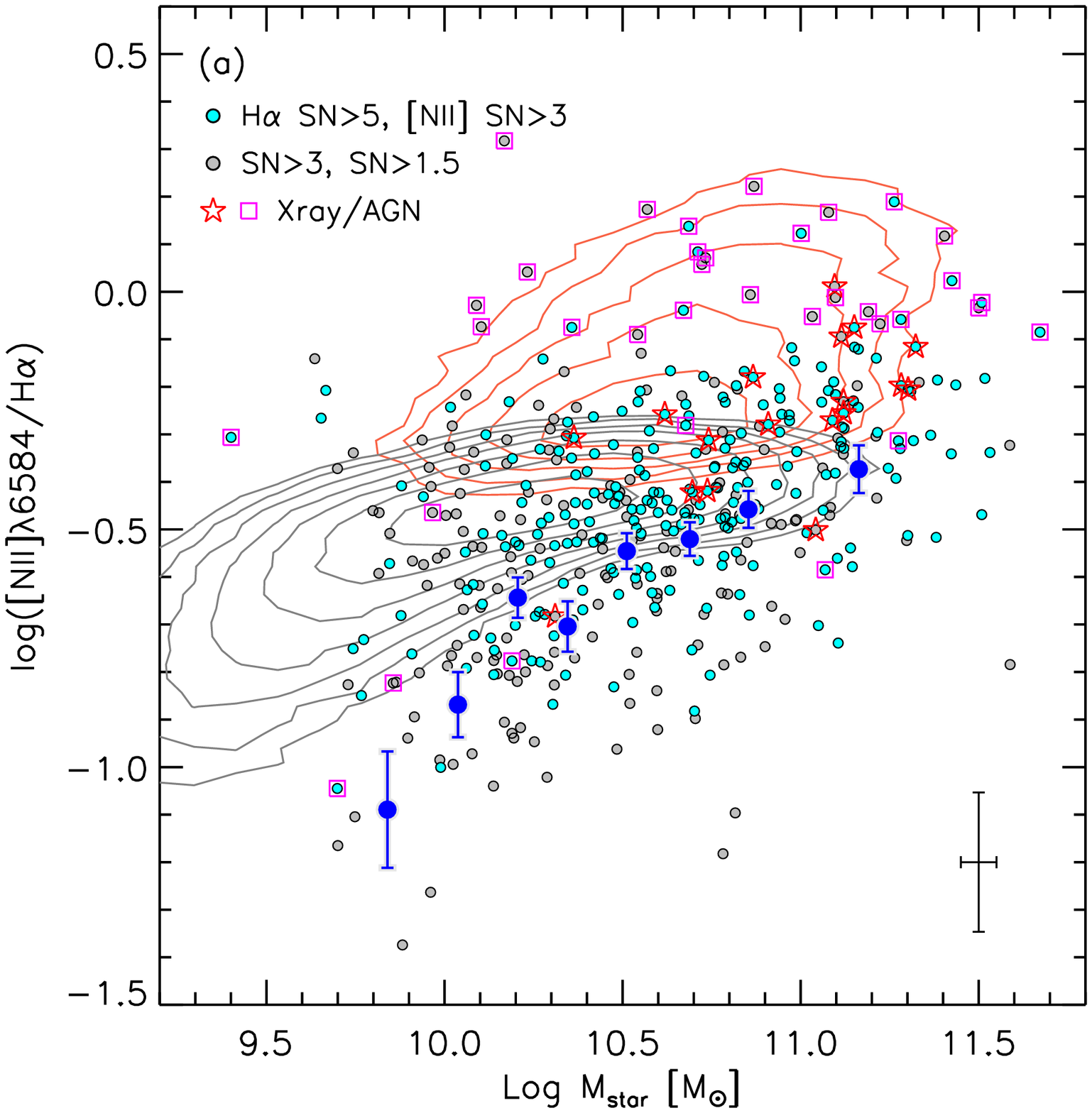} 
   \includegraphics[width=3.6in]{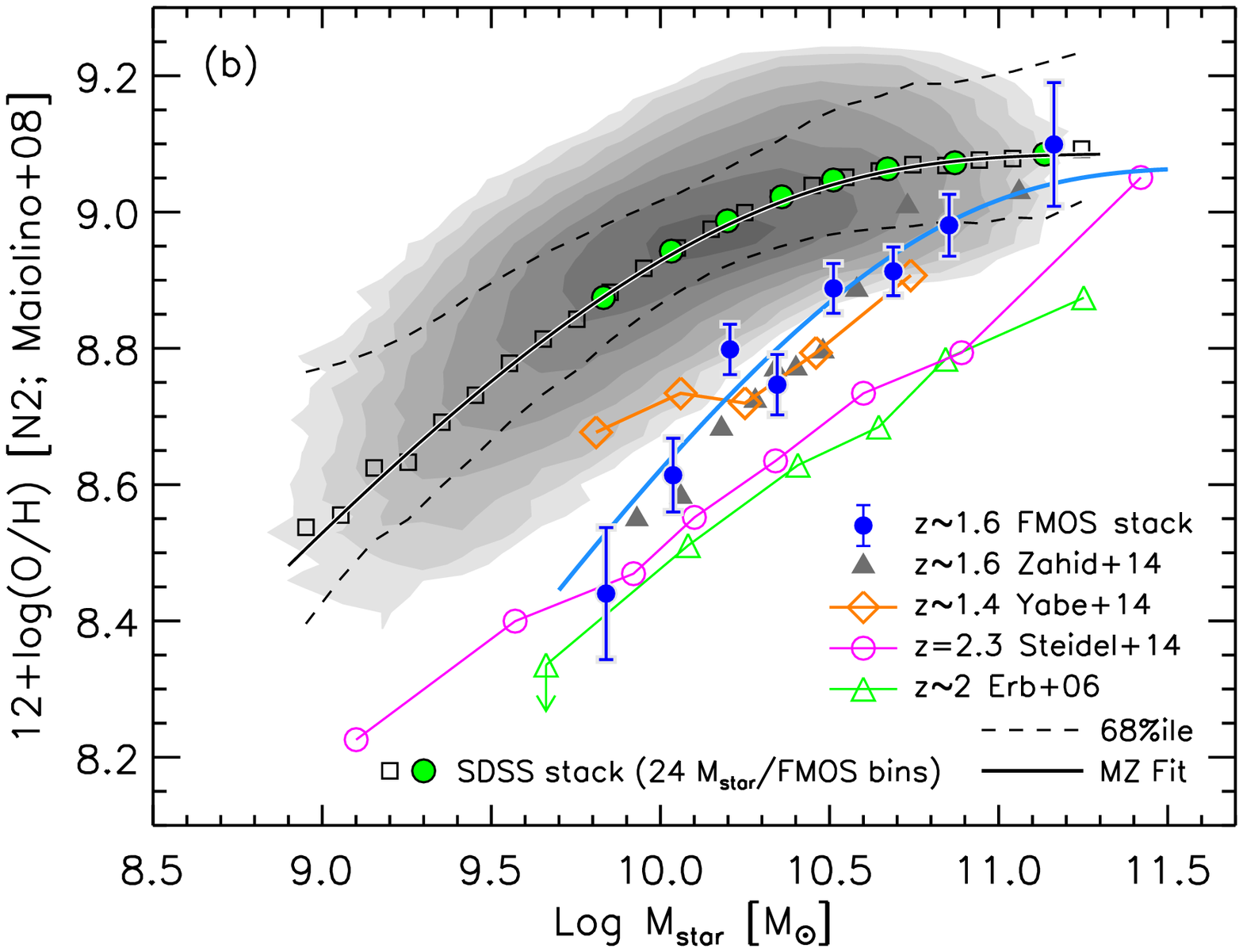} 
   \caption{Mass--metallicity relation; {\it Panel (a)}: [N\,{\sc ii}]/H$\alpha$ as a function of stellar mass.  Individual FMOS galaxies are shown with the quality of measurement as labelled (cyan -- $S/N(\mathrm{H}\alpha)>5$ and $S/N(\text{[N\,{\sc ii}]})>3$; gray -- $S/N(\mathrm{H}\alpha)>3$ and $S/N(\text{[N\,{\sc ii}]})>1.5$).  Typical error bars are shown in the lower-right corner.  AGN candidates are highlighted by stars (X-ay detected sources) and magenta squares (others).  Contours show the distribution of the local star-forming galaxies (gray) and AGNs (red).  The stacked points in the eight $M_\ast$ bins (Sample-1) are shown by blue filled circles, for which AGNs are removed.  {\it Panel (b)}:  Mass--metallicity relation.  The oxygen abundance is calculated from the average [N\,{\sc ii}]/H$\alpha$ ratio with Equation (\ref{eq:Z_N2}).  The FMOS stacked points are the same as in the upper panel (filled circles).  Filled triangles indicate our previous measurement from \citet{2014ApJ...792...75Z}.  Empty symbols show the measurements in the literature as labelled.  Shaded contours show local star-forming galaxies with the stacked data points in equally-spaced 24 mass bins (squares) or in the same eight bins as the FMOS sample (green circles).  Dashed lines indicate the central 68th percentile of the local SDSS galaxies.  Solid curves show the best-fit relations parametrized in Equation (\ref{eq:MZ}), for the FMOS (blue) and the local (black) samples.}
   \label{fig:MZ}
\end{figure}

Figure \ref{fig:MZ}{\it a} shows the [N\,{\sc ii}]$\lambda$6584/H$\alpha$ ratios as a function of stellar mass.  We plot 436 galaxies with H$\alpha$ and [N\,{\sc ii}] detections (Sample-1).  The FMOS sample has a broad distribution of [N\,{\sc ii}]/H$\alpha$ that spans much of the SDSS locus (see also \citealt{2014ApJ...792...75Z}).  Many galaxies with less secure measurements ($1.5<S/N(\textrm{[N\,{\sc ii}]})<3$) have a low $N2$.  This is expected because the intensity of the [N\,{\sc ii}] line is generally much weaker than H$\alpha$ and difficult to detect with FMOS for a reasonable amount ($\sim$a few hours) of integration time.  It is also shown that 54 galaxies have $N2>-0.2$ (12\% of sources with a $N2$ measurement).  The locus of such a population is in agreement with the distribution of the local AGN (red contours in Figure \ref{fig:MZ}{\it a}).  In contrast, roughly half of the X-ray detected FMOS sources are located at $N2<-0.2$, consistent with local star-forming galaxies, while the others show higher $N2$ values, especially at high masses.  This suggests that AGNs are present in our sample if there is no prior exclusion, especially, at high masses \citep{2015MNRAS.450..763M}, and that the availability of the $X$-ray observations is important to construct a sample of pure star-forming galaxies.

To overcome the low S/N of individual detections of the [N\,{\sc ii}] line, we measure the average [N\,{\sc ii}]/H$\alpha$ ratios from the stacked spectra of Sample-1 split into eight bins of stellar mass (see Table \ref{tb:linemeasure}; blue circles in Figure \ref{fig:MZ}ab).  The stacked points show a clear trend with stellar mass while lie below the average of the individual sources with the detection (Figure \ref{fig:MZ}a).  Co-adding spectra reduces the bias with respect to detecting the [N\,{\sc ii}] line since galaxies are identified only by the presence of an H$\alpha$ detection.  In Figure \ref{fig:MZ}b, the metallicities converted from the stacked [N\,{\sc ii}]/H$\alpha$ ratios are shown, which are consistent with the previous measurements from \citet[gray triangles]{2014ApJ...792...75Z}.  For local galaxies, we show the stacked data points in equally-spaced twenty-four mass bins (squares) and in the same bins as the FMOS sample (green circles).  Both the local and the FMOS samples are well fit with a model by \citet{2014ApJ...791..130Z}.  The model fit and the interpretations are fully described in Section \ref{sec:MZev}.  Our data show a level of metallicity similar to local galaxies at the massive end ($M_\ast>10^{11}~M_\odot$), while the average measurements in less massive galaxies are lower than those in local galaxies.  The difference in metallicity between the local and the FMOS sample increases as the stellar mass decreases.  Such a behavior of the MZ relation is consistent with an evolutionary scenario, so-called ``downsizing''\citep[e.g.,][]{2008A&A...488..463M}, where massive galaxies form earlier, thus chemically mature at high redshift, while lower mass systems evolve more slowly.

As described in Section \ref{sec:AGNrem}, we remove galaxies identified as AGN based on the X-ray detection, the BPT diagram (or $N2>-0.1$ if no [O\,{\sc iii}]/H$\beta$ measurement is available), or FWHM$>1000~\mathrm{km/s}$.  We further assess the effects of excluding potential AGN by their [N\,{\sc ii}]/H$\alpha $ values, particularly at the massive end.  With an upper limit on $N2$ decreased from $-0.1$ to $-0.3$ , the stacked measurements are still consistent with local massive galaxies, although the values are lower by $\Delta \log \mathrm{(O/H)}\sim0.15$.  Therefore, we conclude that the effects of AGNs do not impact our measurement of the MZ relation at $z\sim1.6$.

In Figure \ref{fig:MZ}{\it b}, we also show MZ relations based on [N\,{\sc ii}]/H$\alpha$ from the literature at $z\sim1.4$--2.3 \citep{2006ApJ...644..813E,2014MNRAS.437.3647Y,2014ApJ...795..165S}, which are converted here to a Salpeter IMF and the \citet{2008A&A...488..463M} metallicity calibration.  Our measurement is nearly in agreement with \citet{2014MNRAS.437.3647Y}, which is based on a low resolution FMOS program.  The MZ relations at $z\sim2$ seem to indicate a further decline of metallicity as compared to our relation at $z\sim1.6$.  In addition, the slope seems to be shallower than that of our measurement at $z\sim1.6$ (see Section \ref{sec:MZev}).

Given the sensitivity of the [N\,{\sc ii}]/H$\alpha$ ratio to ionization parameter, gas pressure, N/O, and hardness of the ionizing radiation field, some caution is advised with interpreting the shifts seen between the low and high-$z$ samples in Figure \ref{fig:MZ}b as purely due to metallicity changes. \citet{2016ApJ...817...57C} find that the [N\,{\sc ii}]/H$\alpha$ vs. O/H relation depends weakly on H$\beta$ luminosity.  The luminosity-adjusted relation for the typical $\log L(\mathrm{H\beta})/(\mathrm{erg~s^{-1}})\sim41.5\textrm{--}42$ of our sample reduces the difference between the MZ relations at $z\sim0$ and $z\sim1.6$ by $\sim0.15\textrm{--}0.2~\mathrm{dex}$.  Even though, that the metallicities of high-$z$ galaxies with $M_\ast\sim10^{10}~M_\odot$ are still smaller than those of local galaxies.  We recognize that there may be systematic biases between the metallicity determination of different samples at these two epochs.  However, we use locally- and/or theoretically-calibrated metallicity indicators without any specific correction for high redshift for the remainder of this study.

\subsection{New calibration with [N\,{\sc ii}]/[S\,{\sc ii}]}
\label{sec:Dopita16}

Recently, \citet{2016Ap&SS.361...61D} have introduced a new metallicity calibration using both the line ratios [N\,{\sc ii}]/H$\alpha$ and [N\,{\sc ii}]/[S\,{\sc ii}], that is almost independent of the ionization parameter and gas pressure, as described in Section \ref{sec:BPTorigin} (see Figure \ref{fig:D16Fig2}).  This calibration is effective over the metallicity range of $8<12+\log(\mathrm{O/H})<9$, expressed as follows:
\begin{equation}
12+\log (\mathrm{O/H}) = 8.77 + N2S2 +0.264N2,
\label{eq:Dopita16}
\end{equation}
where $N2S2$ is defined as Equation (\ref{eq:N2S2}).  Inclusion of the [N\,{\sc ii}]/[S\,{\sc ii}] term ($N2S2$) in the determination of the metallicity provides an assessment of how nucleosynthesis has proceeded through the ratio of secondary-to-primary elements.  This metallicity indicator assumes a relation between the N/O ($\approx $N/S) ratio and the O/H ratio based on a set of local sources (see Figure 1 of \citealt{2016Ap&SS.361...61D}).  The authors argue that changes in the N/O ratio at a fixed O/H are not required to explain the emission-line properties of high-$z$ galaxies.  However, it is not proven whether such N/O vs. O/H relation is universal, hence applicable at higher redshfits (or even for local galaxies other than those used for the calibration). This calibration also assumes a one-to-one relation between N/S and N/O, i.e., a constant S/O ratio.  We note that the variation in S/O are expected to be small ($\sim0.1~\mathrm{dex}$), and indeed the S/O ratio is almost independent of metallicity \citep{2006A&A...448..955I}, which supports the use of the [N\,{\sc ii}]/[S\,{\sc ii}] ratio as a proxy of the N/O ratio \citep{2009MNRAS.398..949P}.

\begin{figure}[htbp]
   \centering
   \includegraphics[width=3.6in]{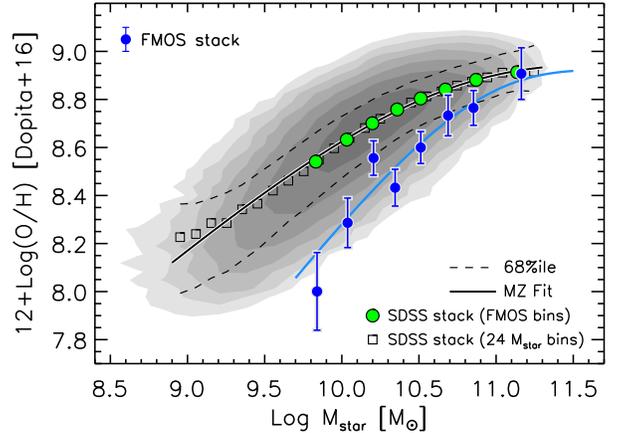} 
   \caption{Mass--metallicity relation based on a calibration introduced by \citet[Equation \ref{eq:Dopita16}]{2016Ap&SS.361...61D}.  Symbols are the same as those in Figure \ref{fig:MZ}.  Solid curves show the best-fit relation parametrized in Equation (\ref{eq:MZ}) for the FMOS (blue) and the local (black) samples.}
   \label{fig:Dopita16}
\end{figure}

With these considerations and caveats in mind, we present in Figure \ref{fig:Dopita16} the MZ relation with metallicity determined from Equation (\ref{eq:Dopita16}) for both local and our FMOS galaxies.  Similar to the [N\,{\sc ii}]/[S\,{\sc ii}] ratio shown in Figure \ref{fig:MxN2S2}{\it a}, the metallicity in local star-forming galaxies begins to flatten at low masses ($M_\ast \lesssim 10^{9.2}~M_\odot$) (see Section \ref{sec:MxN2S2}).  Except for the low metallicity tail, the average points of local galaxies and those of the FMOS sample are well fit with the model of \citet[see Section \ref{sec:MZev}]{2014ApJ...791..130Z}.  The MZ relation for the FMOS sample is qualitatively equivalent to the results based on the [N\,{\sc ii}]/H$\alpha$ ratios (shown in Figure \ref{fig:MZ}{\it b}) thus indicating that the metallicities of high-$z$ galaxies are, on average, smaller than those in local galaxies at low masses, and increase with stellar mass, reaching a level similar to local galaxies at $M_\ast\gtrsim10^{11}~M_\odot$.  However, we note that the metallicity range spanned by our sample ($8\lesssim12+\log(\mathrm{O/H})\lesssim8.9$) is slightly different from that of the $N2$-based metallicity with the \citet{2008A&A...488..463M} calibration  ($8.4\lesssim12+\log(\mathrm{O/H})\lesssim9.1$).  We remind readers that there are likely systematic uncertainties in the absolute metallicity calibrations based on different techniques and/or indicators \citep[e.g.,][]{2008ApJ...681.1183K}.  It is worth noting that the $N2$-based MZ relation at $z\sim1.6$ shows a slightly larger offset from the local relation (Figure \ref{fig:MZ}b) than that based on the \citet{2016Ap&SS.361...61D} calibration.  Comparing the deviations of the high-$z$ sample to the scatter of the local MZ relation (dashed lines in Figures \ref{fig:MZ}b and \ref{fig:Dopita16}), we find that the $N2$-based metallicity of the FMOS galaxies is offset from the local average by about five times the local scatter at the lowest mass bin, while the  deviation is about three times the scatter based on the other indicator.  This is most likely explained by the different sensitivities of these metallicity indicators on the ionization parameter as described above and in Section \ref{sec:MxN2}.

In this analysis, we remove those with $N2>-0.1$ to exclude AGNs (see Section \ref{sec:AGNrem}).  However, there is still a possibility that the average [N\,{\sc ii}]/H$\alpha$ ratios are enhanced by weak AGNs.  Therefore, we examine if the metallicity measurements are affected by the changes in selecting which galaxies to stack.  We lower the threshold to $N2=-0.3$ thus excluding all galaxies with values above this limit.  We find that galaxies at the high-mass end still have metallicities in agreement with local massive galaxies.  Therefore, we conclude that the most massive galaxies are fully mature, as reported when using the $N2$-based metallicity.

\subsection{Evolution of the mass--metallicity relation}
\label{sec:MZev}

\citet{2014ApJ...791..130Z} introduced the concept of a universal relation between metallicity and the stellar-to-gas mass ratio, $M_\ast/M_\mathrm{gas}$, as the underlying origin of the MZ relation.  In this framework, an observed MZ relation can be parametrized as
\begin{equation}
12+\log (\mathrm{O/H}) = Z_0 + \log \left[ 1-\exp \left( -\left[ \frac{M_\ast}{M_0}\right]^\gamma \right) \right].
\label{eq:MZ}
\end{equation}
Here, $Z_0$ is the asymptotic metallicity at the massive end.  $M_0$ is the characteristic stellar mass where the relation begins to flatten, and $\gamma$ is the power-law slope of the relation at $M_\ast \ll M_0$.  Given the assumption of a power-law relation between stellar and gas mass, $M_\mathrm{gas} \propto M_\ast^\delta$ \citep{2014ApJ...786...54P}, the term $(M_\ast / M_0)^\gamma$ approximately equals $M_\ast/M_\mathrm{gas}$, and the slope $\gamma$ is prescribed by the slope $\delta$ as $\gamma=1-\delta$ \citep{2014ApJ...791..130Z}.

We model the observed MZ relations based on both the two metallicity indicators described in Sections \ref{sec:MxN2} and \ref{sec:Dopita16} with Equation (\ref{eq:MZ}).  This fitting function differs from the function used in our previous study \citep{2014ApJ...792...75Z} and has a more physically motivated form.  We note that fitting the current measurements with the model given in \citet{2014ApJ...792...75Z} yields results consistent with those found in our past study.  The best-fit parameters in Equation (\ref{eq:MZ}) are listed in Table \ref{tb:MZ}.  For our sample, the stacked measurements in eight stellar mass bins are used for the fit.  To derive the local MZ relation, we fit the same formulation to the median values in twenty-four stellar mass bins at $10^{8.9}\le M_\ast/M_\odot \le 10^{11.3}$.

\onecolumngrid
\begin{deluxetable}{lcccc}
\tablecaption{Best-fit parameters of the mass--metallicity relation\tablenotemark{a}\label{tb:MZ}}
\tablehead{\colhead{Sample}&\colhead{Redshift\tablenotemark{b}}&\colhead{$Z_0$}&\colhead{$\log (M_0/M_\odot)$}&\colhead{$\gamma$}}
\startdata
\multicolumn{5}{l}{Indicator: [N\,{\sc ii}]/H$\alpha$ (Maiolino et al. 2008; Eq. \ref{eq:Z_N2})} \\
SDSS & 0.07 & $9.086\pm0.003$ & $9.87\pm0.02$ & $0.56\pm0.01$ \\
FMOS & 1.55 & $9.07\pm0.15$ & $10.50\pm0.34$ & $0.71\pm0.45$ \\
FMOS & 1.55 & $[9.086]\tablenotemark{c}$ & $10.59\pm0.05$ & $[0.56]$ \\
\hline
\multicolumn{5}{l}{Indicator: Dopita et al. (2016; Eq. \ref{eq:Dopita16})} \\
SDSS & 0.07 & $8.95\pm0.01$ & $10.35\pm0.05$ & $0.55\pm0.01$ \\
FMOS & 1.55 & $8.92\pm0.22$ & $10.71\pm0.49$ & $0.83\pm0.30$ \\
FMOS & 1.55 & $[8.95]$ & $10.94\pm0.08$ & $[0.55]$
\enddata
\tablenotetext{a}{The MZ relation is parametrized in Equation (\ref{eq:MZ}).}
\tablenotetext{b}{Median redshift of each sample.}
\tablenotetext{c}{Values given in square brackets have been fixed to the SDSS values.}
\end{deluxetable}

In Figure \ref{fig:MZ}{\it b}, the model represents well the observed $N2$-based MZ relations at $z\sim0$ and $z\sim1.6$.  $Z_0$ and $\gamma$ are statistically consistent between the two samples.  Therefore, the evolution of the MZ relation can be quantified by the evolution of the turnover mass $M_0$.  Motivated by this fact, we fit a model to our data with $Z_0$ and $\gamma$ fixed to the local values ($Z_0=9.086$ and $\gamma=0.56$; see Table \ref{tb:MZ}).  In this case, we find $M_0$ to be consistent with the case where $Z_0$ and $\gamma$ are free parameters.  Therefore, the evolution of the MZ relation since $z\sim1.6$ to the present-day is characterized by the decline of the turn-over stellar mass $M_0$ by about $0.6~\mathrm{dex}$.  Likewise, the model also fits well the local and high-$z$ MZ relations based on the \citet{2016Ap&SS.361...61D} metallicity indicator as shown in Figure \ref{fig:Dopita16}.  However, the best-fit values of $M_0$ for low- or high-$z$ samples are slightly different from those of the $N2$-based MZ relation and show less evolution in $M_0$ ($\Delta \log M_0=0.36~\mathrm{dex}$)  for the \citet{2016Ap&SS.361...61D} metallicity indicator.

\begin{figure}[htbp]
   \centering
   \includegraphics[width=3.6in]{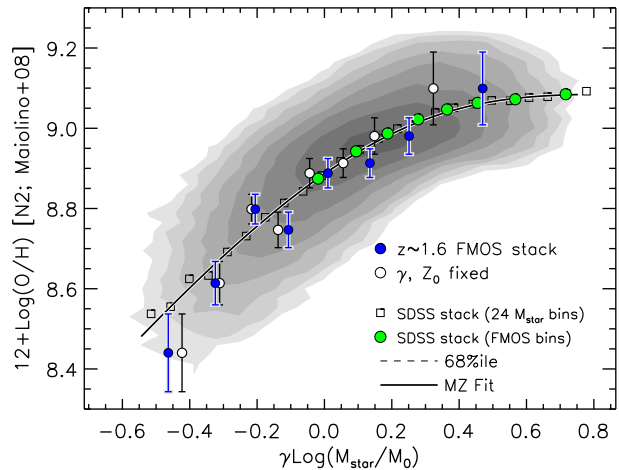} 
   \caption{Metallicity vs. $\gamma \log (M_\ast/M_0)$, a proxy of the stellar-to-gas mass ratio (see Section \ref{sec:MZev}), for the local and FMOS samples.  Blue circles show the FMOS stacked measurements for the best-fit $Z_0$ and $\gamma$; while circles for $Z_0$ and $\gamma$ fixed to the local values.  Contours and symbols representing the local sample are the same as in Figure \ref{fig:MZ}b.  A solid curve is given by Equation (\ref{eq:MZ}) with the best-fit parameters for the local sample.  Dashed lines indicate the 68 percentiles.  Excellent agreement between the two samples at $z\sim0$ and $z\sim1.6$ suggests the existence of a universal relation between metallicity and stellar-to-gas mass ratio.}
   \label{fig:uniMZ}
\end{figure}

Figure \ref{fig:uniMZ} shows the metallicity (from the [N\,{\sc ii}]/H$\alpha$ ratio) as a function of $\gamma \log (M_\ast / M_0)$ for the local sample and the FMOS sample (free $Z_0$, $\gamma$ -- filled circles; fixed $Z_0$, $\gamma$ -- open circles).  The measurements from our FMOS sample agree well with the local relation.  In the context of the model given in Equation (\ref{eq:MZ}), we have $(M_\ast / M_0)^\gamma \sim (M_\ast/M_\mathrm{gas})$ (see \citealt{2014ApJ...791..130Z} for details), thus our results support the existence of a universal relation between the metallicity and the stellar-to-gas mass ratio since $z\sim1.6$ to the present day.  The decline of $M_0$ with cosmic time is simply interpreted as a result of gas consumption and stellar mass assembly through star formation.  We also note that our results are consistent with a time-invariant slope of the $M_\ast$--$M_\mathrm{gas}$, while the errors on $\gamma$ are too large to constrain its evolution (if any).

However, selection effects may be present at low masses that impact our determination of $\gamma$.  Our sample is potentially biased towards a population having high SFRs and being less obscured by dust (see Section \ref{sec:SFRdet}).  This bias may be induced by the observational limit of detecting H$\alpha$ and the self-imposed selection on the predicted H$\alpha$ flux.  The amount of metals is commonly known to be correlated with the amount of dust (e.g., \citealt{1990A&A...236..237I,1993AN....314..361S,1998ApJ...496..145L}).  Therefore, such a bias may impact the shape of the MZ relation.  In particular, the effects may be considerable at low masses because low mass dusty galaxies, which are expected to have a low H$\alpha$ flux and to be more metal rich, could fall below our selection limit.  In addition, as naturally expected from the fact that high-$z$ galaxies have a higher ionization parameter, galaxies with higher sSFR may have a higher ionization parameter as compared to lower-sSFR counterparts at fixed $M_\ast$, although our data do not show any evidence for the positive correlation between these quantities.  If it is the case, the cut on predicted H$\alpha$ flux or the S/N makes the selected sample biased towards having higher ionization parameters, thus resulting in lower [N\,{\sc ii}]/H$\alpha$ ratios (see Section \ref{sec:MxN2}).  As a consequence of these possible biases, the measurements at the low-mass end could be biased towards lower metallicities, leading to a MZ relation with a steeper slope.  Indeed, such biases arise in the local relation as well where a high S/N or luminosity threshold of the H$\alpha$ line is applied.  A selection of local galaxies with an observed $L(\mathrm{H\alpha})\ge10^{41}~\mathrm{erg~s^{-1}}$ yields a slightly steeper slope $\gamma=0.61\pm0.03$ for the $N2$-based MZ relation.  In contrast, we find that the slope of the local MZ relation based on the \citet{2016Ap&SS.361...61D} indicator is almost insensitive to such luminosity threshold.

In Figure \ref{fig:MZ}{\it b}, we compare our result to MZ relations derived at higher redshifts ($z\gtrsim2$, \citealt{2006ApJ...644..813E,2014ApJ...795..165S}).  These MZ relations show lower metallicities as compared to our $z\sim1.6$ relation, while the difference is not significant at the lowest mass ($M_\ast\sim10^{9.8}M_\odot$).  The slope of the Steidel et al. MZ relation is found to be approximately $\gamma=0.34$, which is shallower than found in the local MZ relation, and no flattening feature appears over the proved stellar mass range.  In contrast, \citet{2014ApJ...789L..40W} report that the shape of the MZ relation does not change and the turnover mass $M_0$ keeps increasing up to $z\sim 2.3$ by applying the same formalism (Equation \ref{eq:MZ}) to a compilation of samples.  Beyond $z\sim2$, the shape of the MZ relation still poorly constrained \citep[see also][]{2015ApJ...799..138S}.  As mentioned above, selection effects probably impact the shape of the MZ relation.  In particular, there is the possibility that the $z\gtrsim2$ samples that are shown in Figure \ref{fig:MZ} miss some of the dusty galaxies, which tend to be metal rich, because of their UV-based selection \citep{2006ApJ...644..813E,2014ApJ...795..165S}.  Furthermore, effects from changes in the ionization parameter and/or the ionizing radiation field are likely to be considerable at higher redshift \citep{2016ApJ...826..159S}.

\section{Mass--metallicity--SFR relation}
\label{sec:M-Z-SFR}

Observations of an anti-correlation between metallicity and SFR at fixed stellar mass have advocated a simple picture, in which upward fluctuations in the inflow of metal-poor gas fuel star formation, while diluting the gas-phase metal abundance \citep[e.g.,][]{2008ApJ...672L.107E,2010MNRAS.408.2115M}.  \citet{2010MNRAS.408.2115M} proposed the so-called {\it fundamental metallicity relation} (FMR), a universal relation between stellar mass, metallicity, and SFR, holding for star-forming galaxies both locally and up to $z\sim2.5$.  However, the actual shape of the FMR is known to depend on the methodology with which the three parameters are measured \citep[see e.g.,][]{2013ApJ...765..140A,2016ApJ...823L..24K}.  While the anti-correlation between SFR and $Z$ appears to be characteristic of local low-mass galaxies ($M_\ast\lesssim 10^{10}~M_\odot$; see e.g.,  \citealt{2012MNRAS.422..215Y}), it remains to be demonstrated conclusively whether such a relation exists at $z>1$ and whether the FMR is truly fundamental (i.e., does not evolve with redshift).

\subsection{Anti-correlation between metallicity and SFR}
\label{sec:Z-SFR}

Most recent studies at $z\sim2$ find no significant (anti-) correlation between metallicity and SFR at fixed stellar mass \citep{2014ApJ...789L..40W,2014ApJ...795..165S,2015ApJ...799..138S}.  In contrast, we previously reported on an $Z$-SFR anti-correlation, by using an sBzK-selected subset of the current data \citep{2014ApJ...792...75Z}.  \citet{2015ApJ...808...25S} also find a weak anti-correlation at $z\sim2$ based on individual measurements, as well as \citet{2016ApJ...822..103G} do at $z\sim0.6$.  \citet{2015ApJ...808...25S} argue that binning by SFR (or sSFR) washes out a weak SFR dependence of metallicity when using data of low S/N.  With the current data, we examine again whether metallicity depends on SFR at given stellar mass.  Here, we use the [N\,{\sc ii}]/H$\alpha$ ratios to determine metallicity since the \citet{2016Ap&SS.361...61D} indicator is known to be less sensitive to instantaneous fluctuations of the metallicity due to infalling gas \citep{2016ApJ...823L..24K}.  

To investigate the relation between SFR and $Z$ at fixed $M_\ast$, we derive the excess both in $\log([\textrm{N\,{\sc ii}}]/\mathrm{H\alpha})$ and in SFR for individual FMOS galaxies.  The $N2$ excess is defined as the deviation from the average $M_\ast$--$N2$ relation that is converted from the best-fit MZ relation as shown in Figure \ref{fig:MZ}b.  Similarly, the H$\alpha$-based SFR excess, $\mathrm{SFR}/\left< \mathrm{SFR}\right>_\mathrm{MS}$, is defined as the distance from the star-forming main sequence (see Figure \ref{fig:sample}).  We note that our conclusions do not change if the absolute values of sSFR or $\log([\textrm{N\,{\sc ii}}]/\mathrm{H\alpha})$ are used instead of these quantities normalized by the corresponding average values.

\begin{figure*}[htbp]
   \centering
   \includegraphics[width=6.5in]{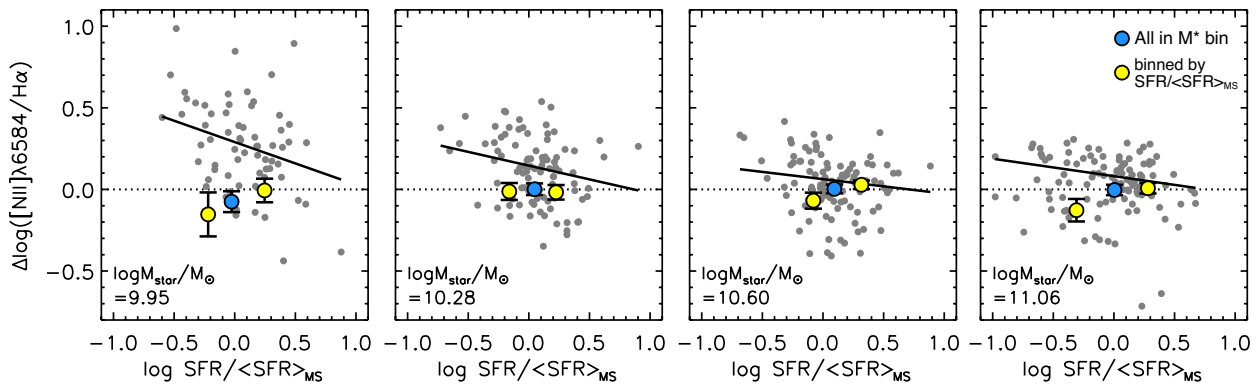} 
   \caption{Excess in the [N\,{\sc ii}]/H$\alpha$ ratio vs. the SFR excess in four bins of stellar mass (median $M_\ast$ is shown in each panel).  Individual FMOS galaxies with a [N\,{\sc ii}]/H$\alpha$ measurement are shown as gray circles.  In each bin, a solid line is a linear fit to individual points.  Stacked points of all galaxies in each $M_\ast$ bin and two subsets further separated by the SFR excess are shown by blue and yellow circles, respectively.  The individual points show a moderate anti-correlation between the two quantities in all bins, while the stacks do not.}
   \label{fig:SSFRxN2}
\end{figure*}

In Figure \ref{fig:SSFRxN2}, we present the $N2$ excess as a function of the SFR excess in four stellar mass bins. We find a moderate anti-correlation between SFR and $Z$ at all masses for the individual FMOS galaxies with a [N\,{\sc ii}]/H$\alpha$ measurement. From the lowest to the highest mass bins, the Spearman's rank correlation coefficients and the confidence levels are respectively $\rho=-0.32$ ($>99\%$), $-0.28$ ($>99\%$), $-0.22$ ($98\%$), $-0.11$ ($75\%$).  However, the individual points preferentially fall above $\Delta \log([\textrm{N\,{\sc ii}}]/\mathrm{H\alpha})=0$ (dotted lines in Figure \ref{fig:SSFRxN2}).  This clearly indicates that the bias concerning the detection of the faint [N\,{\sc ii}] line certainly exists in the individual measurements.  In particular, the [N\,{\sc ii}] line for low-$Z$ galaxies tends to be much fainter than H$\alpha$ and can only be detected when their SFRs are high.  Therefore, an SFR--$Z$ anti-correlation found with these individual measurements is likely caused by such a bias.  In fact, stacked measurements, based on two subsamples binned by SFR excess in each mass bin, fall below the individual points and do not show any significant dependence of $Z$ on SFR.  While this contradiction is possibly due to the effects of sample selection, it is still desirable to further study this issue with a larger sample and accurate SFR estimation to constrain a correlation (possibly weak if any) between metallicity and SFR.  

\subsection{Comparison with the local fundamental metallicity relation}
\label{sec:FMR}

We examine the cosmic evolution of the relation between metallicity, stellar mass, and SFR.  It is worth noting that a lack of a significant SFR--$Z$ correlation at a given single redshift (i.e., within a narrow sSFR range) does not mean that there is a lack of the correlation over a wider sSFR range with consideration of the redshift evolution of the main sequence.  In the context of the FMR, the redshift evolution of the MZ relation is expressed as a shift on the surface of the $M_\ast$--SFR--$Z$ space while these quantities vary with cosmic time.  \citet{2010MNRAS.408.2115M} defined a quantity combining $M_\ast$ and SFR as $\mu=\log M_\ast - \alpha \log \mathrm{SFR}$, where the projection parameter $\alpha$ is determined to minimize the dispersion in metallicity over small intervals on a grid of the $M_\ast$--SFR plane.  We choose $\alpha=0.32$ for comparison with the original FMR by \citet{2010MNRAS.408.2115M}, although $\alpha=0.30$ is derived for the [N\,{\sc ii}]/H$\alpha$-based metallicity by \citet{2013ApJ...765..140A} and \citet{2014ApJ...792...75Z}.  The use of $\alpha=0.30$ does not change our conclusions.  We note that \citet{2010MNRAS.408.2115M} use the metallicity calibration of \citet{2008A&A...488..463M}, the same as implemented here.

\begin{figure}[htbp]
   \centering
   \includegraphics[width=3.6in]{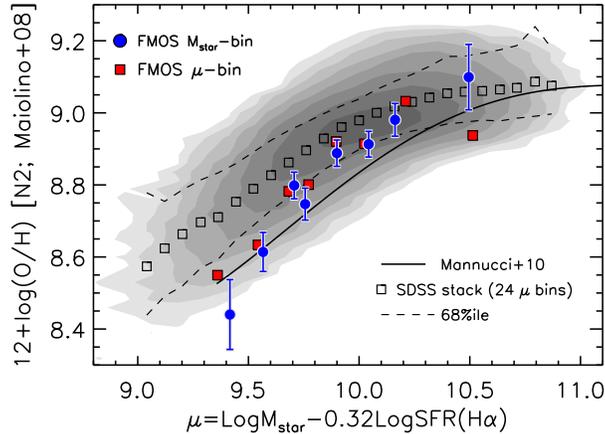} 
   \caption{Projected $M_\ast$-$Z$-SFR relation: metallicity versus $\mu_{\alpha=0.32} (M_\ast, \mathrm{SFR})$.  Metallicities are based on [N\,{\sc ii}]/H$\alpha$ and $\mu$ is calculated using H$\alpha$-based SFRs for both the local and FMOS samples.  Symbols indicate the stacked points in eight bins of $M_\ast$ (blue circles) or $\mu$ (red squares).  Shaded contours show local star-forming galaxies, with the stacked data points in equally-spaced 24 bins of $\mu$ (squares) and the central 68 percentiles (dotted lines).  A thin solid line indicates the original FMR derived by \citet{2010MNRAS.408.2115M} (based on in-fiber SFRs).}
   \label{fig:FMR}
\end{figure}

Figure \ref{fig:FMR} shows the metallicity derived from the [N\,{\sc ii}]/H$\alpha$ ratio as a function of the projected axis, $\mu$, with the FMR from \citet{2010MNRAS.408.2115M}.  The line ratios are measured on the composite spectra of the Sample-1 in eight bins of either the $\mu$ value or stellar mass.  In the latter case, the values of $\mu$ are calculated by median stellar mass and SFR in each bin.  Here, we show the results obtained using H$\alpha$-based SFRs; our conclusion does not change when using SED-based SFRs.  The measurements of two binned cases ($\mu$-bin or $M_\ast$-bin) are consistent within the uncertainties.  Our FMOS sample shows good agreement with the Mannucci et al. FMR (thin solid curve in Figure \ref{fig:FMR}) at $\mu\sim9.5$, while there is an offset towards higher metallicities at $\mu\gtrsim10$.  However, we note that \citet{2010MNRAS.408.2115M} derived the FMR using {\it in-fiber} SFRs, that are not corrected for light loss outside the fiber aperture, while we derive the {\it total} SFRs for our FMOS sample accounting for the fiber loss.  The use of a mixture of total and in-fiber SFRs may be misleading and result in an inaccurate physical understanding.  Therefore, we derive a new relation between $\mu$ and metallicity for local galaxies, using total SFRs, for which the aperture effect and dust extinction are taken into account (see Section \ref{sec:sdss}), as done in \citet{2014ApJ...792...75Z}.  We note that the metallicity is based on in-fiber emission-line fluxes, thus not a galaxy-wide value, which possibly causes a bias towards higher O/H.  To minimize the effect of such a bias, we select local galaxies with $z\ge0.04$ following \citet{2005PASP..117..227K}.  Our high-redshift measurements are in good agreement at $\mu>9.8$ with the local measurements, based on the median [N\,{\sc ii}]/H$\alpha$ ratios in twenty-four bins of $\mu$.  In contrast, there is a difference between the local and high-$z$ samples at $\mu<9.8$.  At the low-mass end, the offset is approximately $\Delta \log (\mathrm{O/H}) \sim 0.2~\mathrm{dex}$ in the FMR while about $0.4~\mathrm{dex}$ in the MZ relation (Figure \ref{fig:MZ}b).  Here, we highlight that there still remains a significant offset in the metallicity between the relations at these two epochs in spite of the fact that the potential bias towards high SFRs in our sample is mitigated by computing the projection axis $\mu$.  In conclusion, our data do not support an extrapolation of the FMR from local galaxies to $z\sim1.6$, as first reported in \citet{2014ApJ...792...75Z}, particularly at low $M_\ast$ and high SFR.

\subsection{Mass-$Z$-SFR relation with a gas regulation model}
\label{sec:FMR_Lilly13}

\citet{2014ApJ...792....3M} have studied the FMR at $z>2$ and found that high-$z$ galaxies in their sample are consistent with a universal FMR that is based on the physically motivated formulation by \citet{2013ApJ...772..119L} (see also \citealt{2016ApJ...822...42O}), but not with the Mannucci et al. FMR.  In the framework of \citet{2013ApJ...772..119L}, the equilibrium gas-phase metallicity, $Z_\mathrm{eq}$, is determined by the regulation of star formation by the mass of gas in a galaxy, and is given as a function of $M_\ast$ and SFR.  We follow the parametrization of Equation 3 in \citet{2014ApJ...792....3M}:
\begin{equation}
Z_\mathrm{eq} = Z_\mathrm{in}+\frac{y}{1+\lambda (1-R)^{-1}+\varepsilon^{-1}((1+\beta-b) \mathrm{SFR}/M_\ast - 0.15)}
\label{eq:FMR_Lilly13}
\end{equation}
where $Z_\mathrm{in}$ is the metallicity of the infalling gas, $y$ is the yield, $R = 0.27$ (for a Salpeter IMF) is the fraction of mass returned into ISM, $\lambda \propto M_\ast^a$ is the mass-loading factor, and $\varepsilon=\mathrm{SFR}/M_\mathrm{gas}\propto M_\ast^b$ is the star formation efficiency.  $\beta = -0.25$ is the slope of the relation between stellar mass and sSFR, i.e., $\mathrm{sSFR}\propto M_\ast^\beta$.  Although \citet{2013ApJ...772..119L} assume $R=0.4$ and $\beta=-0.1$, the use of such different values does not change our conclusions.  We note that Equation (\ref{eq:FMR_Lilly13}) is a form for $z=0$ and that the form for high redshifts will depend on the time variation of parameters, as discussed in \citet{2013ApJ...772..119L}.  In particular, the star formation efficiency is likely to increase with redshift \citep[e.g.,][]{2015ApJ...800...20G}.  Here, we compare our measurements at $z\sim1.6$ to this model without considering the time dependence of the parameters to examine whether our FMOS galaxies at $z\sim1.6$ follow a non-evolving $M_\ast$--$Z$--SFR relation.  

We determine the parameters from a model fitting for the local sample using total SFRs, as listed in Table \ref{tb:FMR_Lilly13}.  We use SDSS galaxies with $9.2\le \log M_\ast/M_\odot \le 10.6$ and $-1 \le \log \mathrm{SFR}/(M_\odot~\mathrm{yr}^{-1}) \le 1$, which are binned by $\Delta \log M_\ast = 0.1~\mathrm{dex}$ and $\Delta \log \mathrm{SFR} = 0.1~\mathrm{dex}$.  We utilize the {\sc mpfit2dfun} IDL procedure \citep{2009ASPC..411..251M} to fit the model to the median metallicity in each bin as a function of both $M_\ast$ and SFR.  $Z_\mathrm{in}$ is fixed to zero while the use of a different value between $0<Z_\mathrm{in}/y<0.1$ does not change our conclusions, alt0.hough the best-fit values of other parameters will slightly change.  The derived values are nearly equivalent to those given in \citet{2013ApJ...772..119L}.

\begin{deluxetable}{ccccc}
\tablecaption{Fit of Lilly et al. (2013) FMR model to the SDSS sample\label{tb:FMR_Lilly13}}
\tablehead{\colhead{$\log y$}&\colhead{$\lambda_{10}$}&\colhead{$a$}&\colhead{$\varepsilon_{10}^{-1}~\mathrm{[Gyr]}$}&\colhead{$b$}}
\startdata
$9.09\pm0.01$ & $0.38\pm0.05$ & $-0.61\pm0.03$ & $1.9\pm0.3$ & $0.48\pm0.05$
\enddata
\tablenotetext{ }{Parameters in Equation (\ref{eq:FMR_Lilly13}): $y$ -- yield parameter given in units of $12+\log (\mathrm{O/H})$; $\lambda_{10}$ -- mass-loading factor at $M_\ast=10^{10}~M_\odot$, $\varepsilon_{10}$ -- star formation efficiency at $M_\ast=10^{10}~M_\odot$.}
\end{deluxetable}

\begin{figure*}[htbp]
   \centering
   \includegraphics[width=6.8in]{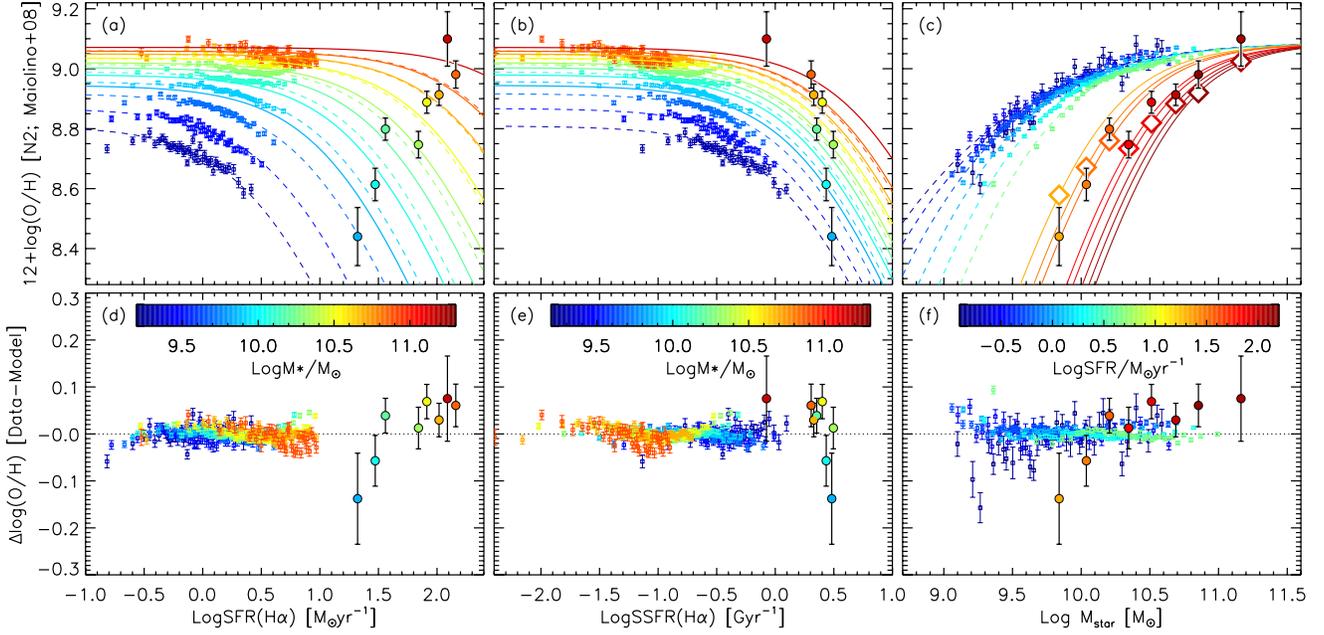} 
   \caption{Metallicity as a function of SFR ({\it Panel a}), sSFR ({\it b}) and stellar mass ({\it c}) as compared to the model of \citet{2013ApJ...772..119L}.  The FMOS measurements (Sample-1) based on composite spectra in eight mass bins are indicated by filled circles, color-coded by $M_\ast$ (panel {\it a} and {\it b}) or SFR ({\it c}).  Colored squares show median metallicities of local galaxies in bins of stellar mass and SFR.  Dashed lines show the model metallicity, $Z(M_\ast, \mathrm{SFR})$, at median $M_\ast$ ({\it a} and {\it b}) or SFR ({\it c}).  Solid lines indicate the model metallicities for median $M_\ast$ or SFR of our FMOS sample.  In {\it Panel c}, diamonds indicate the metallicities predicted by the best-fit $Z(M_\ast, \mathrm{SFR})$ at the median $M_\ast$ and SFR of each bin of our FMOS sample.  {\it Panels d, e, f:} Differences between observed metallicities and theoretical predictions from the model fit (related to panels {\it a}, {\it b}, and {\it c}, respectively.)}
   \label{fig:FMR_Lilly13}
\end{figure*}

Based on this gas regulation model, we find that our FMOS sample at $z\sim1.6$ is consistent with a non-evolving FMR relation.  Figure \ref{fig:FMR_Lilly13} shows average metallicities for galaxies in our sample and for local objects, as a function of SFR (panel {\it a}) and sSFR (panel {\it b}) for galaxies of different stellar masses, in comparison with the best-fit $Z(M_\ast, \mathrm{SFR})$ relation.  We first separate the local sources into $0.2~\mathrm{dex}$-wide mass bins, and divide into forty SFR bins within each mass bin.  The median metallicities are shown separately for each mass bin, as color-coded in the plot.  The functional form of $Z(M_\ast, \mathrm{SFR})$ (Equation \ref{eq:FMR_Lilly13}) reproduces the behavior of local galaxies in each mass bin (dashed curves).  The FMOS stacked data points are based on the co-added spectra of Sample-1 in eight mass bins.  Figure \ref{fig:FMR_Lilly13} shows good agreement between our $z\sim1.6$ measurements and the model prediction extrapolated from the local sample within $\Delta \log(\textrm{O/H})\sim0.1~\mathrm{dex}$ at each $M_\ast$.  This is more clearly seen in the lower panels {\it d} and {\it e}, in which we show the deviations of observed metallicities from the model values at corresponding $M_\ast$ and SFR.  

The agreement between our data and the model is also seen in panels {\it c} and {\it f} of Figure \ref{fig:FMR_Lilly13}, where we present the MZ relation and the deviation in metallicity from the model for the local (squares) and FMOS galaxies (filled circles), respectively.  We separate the local sample into $0.3~\mathrm{dex}$-wide SFR bins, then divide into forty $M_\ast$ bins. The median metallicities are presented as a function of $M_\ast$ separately for each SFR bin as color-coded.  We overlaid the predictions from the best-fit FMR at median SFR of each bin for the local (dashed lines) and our FMOS (solid lines) samples, and mark the expected metallicities at the median $M_\ast$ and SFR of each bin of the FMOS sample (open diamonds), which are in good agreement with the data points. 

Our measurements at low masses require a rapid decline of metallicity at higher SFRs (see Figure \ref{fig:FMR_Lilly13}{\it ab}).  This trend is naturally induced by considering the evolution of galaxies as regulated by the gas content with the consideration of gas inflow and mass loss in the model of \citet{2013ApJ...772..119L}.  Our finding is consistent with the conclusion in \citet{2014ApJ...792....3M}, claiming that whether a FMR is truly fundamental (i.e., independent of redshift) depends on the formulation of the model and its extrapolation to regions of the $M_\ast$--SFR parameter space different from that occupied by local galaxies.  Lastly, we note that, in this study, we do not take into account for the time evolution of model parameters.  

\section{Conclusions}
\label{sec:conclusion}
We investigate the physical conditions of the ionized gas in star-forming galaxies on the main sequence at $1.43 < z < 1.74$ using a sample of 701 galaxies with an H$\alpha$ detection from the FMOS-COSMOS survey.  This is the first time using such a large sample over a wide stellar mass range of $10^{9.6} \lesssim M_\ast / M_\odot \lesssim 10^{11.6}$ at these redshifts.  Based on the X-ray imaging data set from the {\it Chandra}-COSMOS Legacy survey, we carefully exclude the AGN contamination and make sure that ionization due to AGN does not impact our stacked measurements of the emission-line ratios.  Our main results are as follows:

\begin{enumerate}
	\item We confirm a clear offset of star-forming galaxies at $z\sim1.6$ from the sequence of local star-forming galaxies in the BPT diagram using both individual and stacked measurements (Figures \ref{fig:BPT} and \ref{fig:BPTst}).  This offset amounts to an increase in the [O\,{\sc iii}]/H$\beta$ ratio by a factor of $\sim 3$ in the mass--excitation diagram ($M_\ast$ vs. [O\,{\sc iii}]/H$\beta$; Figure \ref{fig:MEx}) between our FMOS sample and local SDSS galaxies.
		
	\item We measure the emission-line ratio [S\,{\sc ii}]$\lambda \lambda$6717, 6731/H$\alpha$ with co-added spectra and find lower values of [S\,{\sc ii}]/H$\alpha$ for the highest two mass bins ($M_\ast \gtrsim10^{10.6}~M_\odot$) as compared with local star-forming galaxies at fixed [O\,{\sc iii}]/H$\beta$ (Figure \ref{fig:BPT_S2}).  These lower [S\,{\sc ii}]/H$\alpha$ ratios are an indication of a higher ionization parameter of the ISM as compared to of local galaxies.
	
	\item The average electron density is measured based on the emission-line ratio $\text{[S\sc ii]}\lambda6717/ \text{[S\sc ii]}\lambda 6731$ using stacked spectra that gives a value of $n_\mathrm{e}=222^{+172}_{-128} ~\mathrm{cm^{-3}}$, which is higher than the average value of local star-forming galaxies ($n_\mathrm{e}\sim 10\text{--}10^2~\mathrm{cm^{-3}}$; Figure \ref{fig:Sii}).  
		
	\item Comparisons with theoretical models indicate a relatively high ionization parameter ($\log q_\mathrm{ion}/c \sim -3$) for H\,{\sc ii} regions in galaxies at $z\sim1.6$, which is approximately 0.3~dex higher as compared to local star-forming galaxies at a fixed stellar mass (Figure \ref{fig:D16Fig2}).  Given our measure of a relatively high gas density, a higher ionization parameter could reflect an increase in the efficiency of star formation expected for high-$z$ galaxies.
	
	\item We find that the enhancement in the ionization parameter {\it at fixed metallicity}, i.e., a higher $q_\mathrm{ion}$ than expected from the local $q_\mathrm{ion}$--$Z$ anti-correlation, is essential to produce the offset of high-$z$ galaxies with $M_\ast\gtrsim10^{10}~M_\odot$ from the loci of local galaxies in the emission-line diagrams. Furthermore, additional effects, attributed to a higher electron density and a hardening of ionizing radiation field, are likely important as well (Figure \ref{fig:model}).  In contrast, the changes in emission-line ratios in lower-mass, $z\sim1.6$ galaxies ($M_\ast<10^{10}~M_\odot$) could be explained by only the increase in $q_\mathrm{ion}$ accompanying the change in metallicity without invoking additional effects.

	\item We derive the mass--metallicity relation from the [N\,{\sc ii}]/H$\alpha$ ratio for our $z\sim1.6$ sample (Figure \ref{fig:MZ}).  Our results are consistent with the previous derivation for an sBzK-selected subsample in the FMOS-COSMOS survey \citep{2014ApJ...792...75Z}.  The current data strengthens our previous result: the most massive galaxies ($M_\ast \gtrsim 10^{11}~M_\odot$) are fully mature at a level similar to local galaxies, while, at the lowest masses ($M_\ast \sim10^{9.8}~M_\odot$), the metallicity is on average $\sim0.4$~dex smaller than local galaxies.  Our sample at $z\sim1.6$ is in good agreement with a universal metallicity relation proposed by \citet{2014ApJ...791..130Z}, which is founded on the fundamental relation between metallicity and stellar-to-gas mass ratio.  However, we highlight that the recipes for determining metallicity at high redshift may need further refinement due to changes in ionization parameter, hardness of stellar spectra, and in the N/O vs. O/H relation (if any).
	 
	  \item We evaluate metallicity adopting a new calibration by \citet{2016Ap&SS.361...61D}, which uses the line ratios [N\,{\sc ii}]/[S\,{\sc ii}] and [N\,{\sc ii}]/H$\alpha$ (Figure \ref{fig:Dopita16}) and assumes universality of the local N/O vs. O/H relation.  The novelty of this indicator is that it is less sensitive to changes in the ionization parameter.  The derived mass--metallicity relation shows a trend that is qualitatively equivalent to that seen in the relation based on the [N\,{\sc ii}]/H$\alpha$ ratio, while the absolute value of the metallicity is slightly different.  In addition, the MZ relations based on this new indicator show an offset, between the SDSS and FMOS samples, three times as large as the scatter of the local MZ relation at $M_\ast\sim10^{9.8}~M_\odot$, while the [N\,{\sc ii}]/H$\alpha$-based MZ relations show an offset five times the local scatter.   This difference between the two metallicity indicators may be caused by the fact that the \citet{2016Ap&SS.361...61D} indicator is relatively insensitive to variations of the ionization parameter.

	\item The evolution of the MZ relation since $z\sim1.6$ to the present-day is well characterized by the change of the turnover stellar mass at which the metallicity begins to saturate.  As a consequence, the MZ relations at both epochs are described by a universal relation between metallicity and the stellar mass normalized by the turnover mass, which is a proxy of the stellar-to-gas mass ratio in the context of a simple chemical evolution model (Figure \ref{fig:uniMZ}).
	
	\item We revise the FMR proposed by \citet{2010MNRAS.408.2115M} by using the total SFRs instead of in-fiber SFRs (as done in \citealt{2014ApJ...792...75Z}) and find that our FMOS sample shows lower metallicities than expected for their $M_\ast$ and SFR from the extrapolation of the FMR (Figure \ref{fig:FMR}).  In contrast, our data are consistent with a physically-motivated model from \citet{2013ApJ...772..119L}, in which the star formation is instantaneously regulated by the gas mass, gas inflow, and mass loss by outflows.  This model well reproduces a rapid decline in metallicity at high SFRs as shown in our data (Figure \ref{fig:FMR_Lilly13}).
\end{enumerate}

To conclude, our study has established the rest-frame optical emission-line properties of star-forming galaxies at $z\sim1.6$ and identified the processes likely responsible for differences with local samples such as shown with the BPT diagram.  The FMOS-COSMOS project and other related efforts have enabled us to take an important step forward in understanding the evolution of the interstellar medium in galaxies at the peak epoch of the cosmic star formation history that will be carried forward by the next-generation multi-object spectrographs (e.g., Subaru/PFS, VLT/MOONS).

\begin{acknowledgements}

We are grateful to the referee for careful reading and useful comments, M.~Dopita for kindly providing us with the photoionization model data, M.~Fukugita, K.~Yabe, T.~Kojima, R.~Shimakawa for useful discussions.  We thank the Subaru telescope staff, especially K.~Aoki, for their great help in the observations.  This paper is based on data collected at the Subaru Telescope, which is operated by the National Astronomical Observatory of Japan.  We appreciate the MPA/JHU team for making their catalog public.  Funding for SDSS-III has been provided by the Alfred P. Sloan Foundation, the Participating Institutions, the National Science Foundation, and the U.S. Department of Energy Office of Science. The SDSS-III web site is http://www.sdss3.org/.  T.~M. is supported by CONACyT Grants 179662, 252531 and UNAM-DGAPA PAPIIT IN104216.  A.~R. is grateful to the National Astronomical Observatory of Japan for its support and hospitality while this paper was set up. Partial support by the INAF-PRIN 2012 grant is also acknowledged.  This work was supported in part by KAKENHI (YT: 23244031 and 16H02166) through Japan Society for the Promotion of Science (JSPS).  J.~D.~S is supported by JSPS KAKENHI grant Number 26400221 and the World Premier International Research Center Initiative (WPI), MEXT, Japan.  D.~K. was supported through the Grant-in-Aid for JSPS Research Fellows (No.~26-3216). 

\end{acknowledgements}

\appendix

\section{Assessment of AGN contamination}
\label{sec:AGNcontami}

As described in Section \ref{sec:AGNrem}, we remove AGNs from our star-forming galaxy samples for all analyses in this paper.  Our data, as presented on line-ratio diagrams (Figures \ref{fig:BPT}, \ref{fig:BPT_S2} and \ref{fig:N2xS2}), shows no strong signature of ionizing radiation from AGN.  However, the possibility that weak and/or obscured AGNs affect the results cannot be completely ruled out.  Here, we assess the contribution of such hidden AGNs on the results based on the co-added spectra.

\begin{figure*}[htbp]
   \centering
   \includegraphics[width=6in]{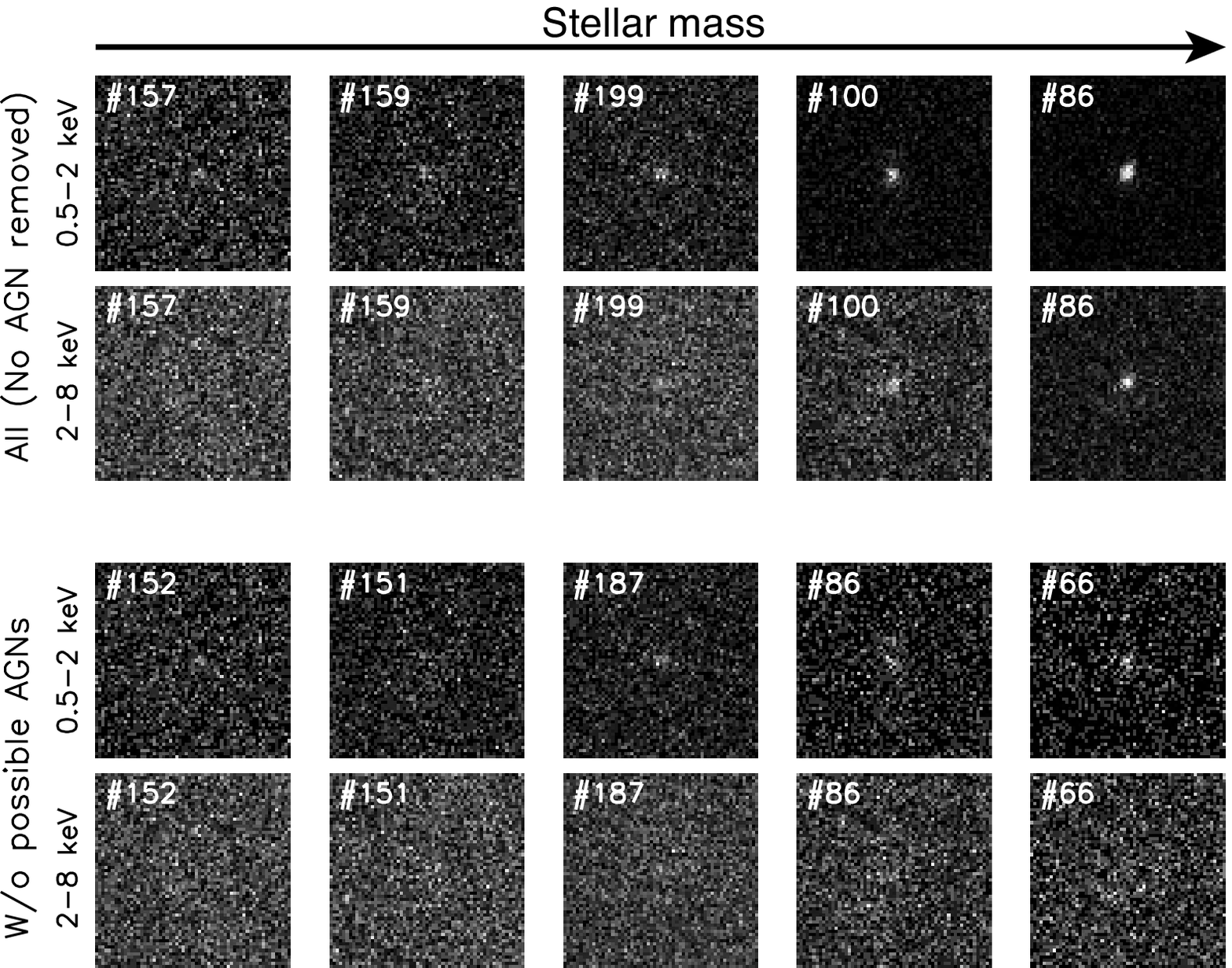} 
   \caption{{\it Chandra} stacked X-ray images (30 arcsec on each side) in two bandpasses (soft: 0.5--2 keV, hard: 2--8 keV) of our sample in five bins of stellar mass.   The number of individual images for stacking is indicated in each panel.  The upper group shows the stacked images for the sample including all SED-selected galaxies without any removal of possible AGNs (case 1).  The lower group shows for the sample in which all AGN candidates are removed (case 3) for the same stellar mass ranges as above.  The stellar mass increases from left to right (see Table \ref{tb:xrayflux}). }
   \label{fig:xray_imgs}
\end{figure*}

X-ray emission is an effective indicator of the existence of AGN.  We measure the average X-ray luminosities of our FMOS galaxies by stacking the {\it Chandra X-ray Observatory}/Advanced CCD Imaging Spectrometer (ACIS-I) images from the {\it Chandra}-COSMOS Legacy survey \citep{2009ApJS..184..158E,2016ApJ...819...62C} using the CSTACK tool\footnote{http://lambic.astrosen.unam.mx/cstack} (v4.3; \citealt{2008HEAD...10.0401M}).  To examine how the X-ray luminosity depends on the exclusion of individually-identified AGNs, we compare three samples with different criteria of the removal of possible AGNs; 1) no AGN cut is applied, 2) excluding individual X-ray detected sources, and 3) excluding all possible AGN candidates (X-ray sources, $\mathrm{FWHM(H\alpha)}>1000~\mathrm{km~s^{-1}}$, or identified with the BPT diagram; see Section \ref{sec:AGNrem}).  The last case is implemented for the results based on stacking analysis throughout the paper.  Figure \ref{fig:xray_imgs} shows the stacked X-ray images in five bins of stellar mass, separately in two bandpasses (soft: 0.5--2 keV, hard: 2--8 keV), from the two samples without (case 1) and with (case 3) AGN removal.

The average count rate is converted into flux assuming a spectrum with $\Gamma=1.7$ and a Galactic column density $N_\mathrm{H}=2.7\times 10^{20}~\mathrm{cm^{-2}}$ \citep{2009ApJS..184..158E,1990ARA&A..28..215D}.   The average rest-frame X-ray luminosity is calculated as Equation 4 in \citet{2008ApJ...681.1163L}, with a median redshift ($z\simeq1.6$).  Table \ref{tb:xrayflux} lists the X-ray luminosities and associated errors in each bin.  When no AGN cut is applied (case 1), the X-ray emission is significantly detected in the two most massive bins at $S/N>5$ in both the soft and hard bandpasses.  When the AGN cut is applied (case 2 and 3), only the most massive bin ($\log M_\ast / M_\odot>11.1$) presents a significant detection ($S/N>3$) in the X-ray band.  In the third and forth bins ($10.4<\log M_\ast/M_\odot<11.1$), the X-ray emission is detected at a significance of $S/N\sim2\textrm{--}3$.  There are cases with no detection ($S/N<1$) in the lowest two mass bins, for which there are at most a few objects that are individually detected in the X-ray band (see Table \ref{tb:xrayflux}).  

\onecolumngrid
\newpage
\begin{deluxetable}{lccccc}
\tablecaption{X-ray luminosity measurements\label{tb:xrayflux}}
\tablehead{\colhead{Median $\log M_\ast/M_\odot$ \tablenotemark{a}}&\colhead{9.93}&\colhead{10.26}&\colhead{10.60}&\colhead{10.96}&\colhead{11.25} \\
\colhead{Range}&\colhead{[9.1:10.1]}&\colhead{[10.1:10.4]}&\colhead{[10.4:10.8}&\colhead{[10.8:11.1]}&\colhead{[11.1:11.7]}}
\startdata
\multicolumn{6}{l}{Case 1: No AGN exclusion applied }\\
No. of images\tablenotemark{b} & 157 & 159 & 199 & 100 & 86 \\
Soft\tablenotemark{c} & $0.11\pm0.08$ & $0.32\pm0.09$ & $0.28\pm0.08$ & $1.20\pm0.13$ & $2.46\pm0.18$ \\
Hard\tablenotemark{d} & $<0.75\tablenotemark{f}$ & $0.58\pm0.50$ & $0.83\pm0.44$ & $3.26\pm0.66$ & $7.25\pm0.80$ \\
Full\tablenotemark{e} & $<0.76$ & $0.90\pm0.51$ & $1.11\pm0.45$ & $4.46\pm0.68$ & $9.71\pm0.82$ \\
\hline 
\multicolumn{6}{l}{Case 2: X-ray luminous objects removed}\\
No. of images & 157 & 156 & 195 & 93 & 78 \\
Soft & $0.11\pm0.08$ & $0.24\pm0.09$ & $0.23\pm0.08$ & $0.42\pm0.12$ & $0.48\pm0.13$ \\
Hard & $<0.75$ & $<0.76$ & $0.74\pm0.44$ & $1.57\pm0.66$ & $3.43\pm0.78$ \\
Full & $<0.76$ & $0.61\pm0.513$ & $0.97\pm0.45$ & $1.20\pm0.67$ & $3.92\pm0.79$ \\
\hline
\multicolumn{6}{l}{Case 3: All possible AGNs removed }\\
No. of images & 152 & 151 & 187 & 86 & 66 \\
Soft & $<0.12$ & $0.21\pm0.09$ & $0.20\pm0.08$ & $0.45\pm0.12$ & $0.50\pm0.15$ \\
Hard & $<0.77$ & $<0.77$ & $0.83\pm0.45$ & $1.48\pm0.68$ & $2.87\pm0.84$ \\
Full & $<0.78$ & $<0.79$ & $1.03\pm0.46$ & $1.93\pm0.69$ & $3.37\pm0.85$
\enddata
\tablenotetext{a}{Median stellar mass in each bin.}
\tablenotetext{b}{Number of stacked images in each bin.}
\tablenotetext{c}{Soft band (0.5--2~keV) luminosity in units of $10^{42}~\mathrm{erg~s^{-1}}$.}
\tablenotetext{d}{Hard band (2--8~keV) luminosity in units of $10^{42}~\mathrm{erg~s^{-1}}$.}
\tablenotetext{e}{Full band (0.5--8~keV) luminosity in units of $10^{42}~\mathrm{erg~s^{-1}}$.}
\tablenotetext{f}{Upper limit is given as 1.5 times of noise levels if $S/N<1$.}
\end{deluxetable}

\begin{figure}[htbp]
   \centering
   \includegraphics[width=3.5in]{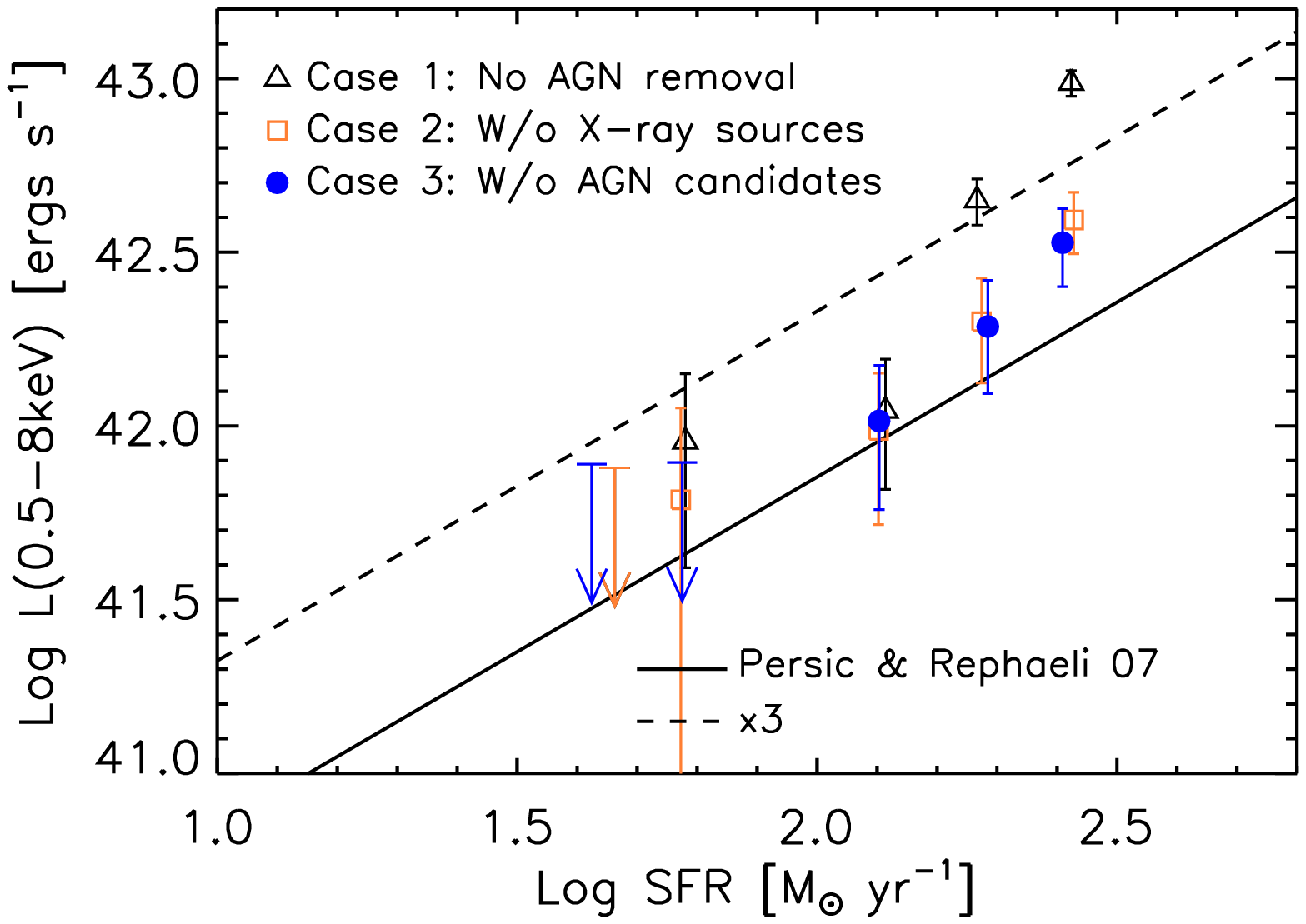} 
   \caption{Comparison between the mean SFR (from dust-corrected H$\alpha$ luminosity) and the 0.5--8 keV luminosity $L_\mathrm{0.5-8~keV}$ in each bin of stellar mass.  The X-ray luminosities $L_\mathrm{0.5-8~keV}$ are measured by stacking the {\it Chandra} ACIS-I images.  Different symbols shows the measurements for different cases of the AGN removal (triangles -- no AGN removed; squares -- without X-ray-detected sources; filled circles -- without all potential AGN candidates; see the text).  An SFR--$L_\mathrm{X}$ relation for star-forming galaxies from \citet{2007A&A...463..481P} and a relation with three times higher $L_\mathrm{0.5-8~keV}$ are shown by the solid and dashed lines, respectively.}
   \label{fig:Lx_SFR}
\end{figure}

To further investigate the impact of AGN, we compare the mean SFR and mean X-ray luminosity in the full {\it Chandra} bandpass ($L_\mathrm{0.5-8~keV}$) in each bin in Figure \ref{fig:Lx_SFR}.  The mean SFRs are derived from dust-corrected H$\alpha$ luminosity (see Section \ref{sec:SFRdet}).  Following \citet{2008ApJ...681.1163L}, we compare our data with an SFR--$L_\mathrm{X-ray}$ relation for star-forming galaxies derived by \citet{2007A&A...463..481P} and a relation with three times higher X-ray luminosity, which corresponds to $\sim2.5$ times the rms scatter of the $L_X$--SFR relation, to distinguish AGNs from non-AGN. We compare the measurements for the samples with the different X-ray exclusion criteria (cases 1, 2, and 3).  We see that $L_\mathrm{0.5-8~keV}$ is correlated with SFR at $\mathrm{SFR}\gtrsim100~M_\odot\mathrm{yr^{-1}}$ ($M_\ast \gtrsim 10^{10.4}M_\odot$) in all the cases and that the removal of individual X-ray detected sources primarily affects the average $X$-ray luminosity of the more massive bins.  As a result of the AGN exclusion, the X-ray luminosities are reduced down to the level where the residual contribution from AGN is likely to be negligible in three higher stellar mass bins.  For two lower mass bins, upper limits on $L_\mathrm{0.5-8~keV}$ are calculated as 1.5 times the noise level, as shown by downward arrows in Figure \ref{fig:Lx_SFR}.  

Figure \ref{fig:Lx_SFR} indicates that, once the individually-detected AGNs are removed, the contribution to the X-ray luminosities in the stack from unidentified AGNs is small, and that levels of the residual X-ray luminosity are comparable to the levels that are normally expected for star-forming galaxies (solid line in Figure \ref{fig:Lx_SFR}).  While the limit of the lowest mass bin does not reject a probability that a contribution from AGNs exists, the AGN contribution can be expected to be negligible in such a range of stellar mass.  Therefore, we conclude that the impact from AGN is small in our stacking analysis enough to appropriately assess the properties of star-forming H\,{\sc ii} regions over the entire stellar mass range spanned by our sample.

\end{document}